\documentclass[twocolumn, openany]{aastex62}

\newcommand{\RNum}[1]{\uppercase\expandafter{\romannumeral #1\relax}}

\usepackage[bookmarks=false]{hyperref}
\usepackage{booktabs}
\usepackage{booktabs,threeparttable}
\usepackage{natbib}
\usepackage{supertabular}
\usepackage{array}
\usepackage{afterpage}
\usepackage{multirow}
\usepackage{longtable}
\usepackage{lipsum}
\usepackage{comment}
\usepackage{tabu}
\usepackage{tabularx}
\usepackage{float}
\usepackage{stackengine}
\usepackage[caption = false]{subfig}
\usepackage{fixltx2e}
\usepackage{graphicx}
\nopagebreak
\graphicspath{{figs/}}
\received{November 27 2019}
\revised{\today}
\accepted{unkown}

\defcitealias{anderson2014}{Anderson2014}

\shortauthors{Chen et al.}

\begin{document}

\title{\LARGE A 4-6 GHz Radio Recombination Line Survey in the Milky Way\footnote{a machine readable catalog accompanies this paper}}

\correspondingauthor{X. Chen}
\email{chenxi@gzhu.edu.cn}

\author{Hong-Ying Chen}
\affiliation{Shanghai Astronomical Observatory, 80 Nandan Road, Xuhui District, Shanghai, 200030, China \\}
\affiliation{Center for Astrophysics, Guangzhou University, Guangzhou 510006, China}
\affiliation{University of Chinese Academy of Sciences, 19A Yuquanlu, Beijing 100049, China\\}
\affiliation{Jodrell Bank Centre for Astrophysics (JBCA), Department of Physics \& Astronomy, Alan Turing Building, The University of Manchester, M13 9PL, United Kingdom}

\author{Xi Chen}
\affiliation{Shanghai Astronomical Obervatory, 80 Nandan Road, Xuhui Districe, Shanghai, 200030, China \\}
\affiliation{Center for Astrophysics, Guangzhou University, Guangzhou 510006, China}
\affiliation{Key Laboratory of Radio Astronomy, Chinese Academy of Sciences, China}

\author{Jun-Zhi Wang}
\affiliation{Shanghai Astronomical Obervatory, 80 Nandan Road, Xuhui Districe, Shanghai, 200030, China \\}
\affiliation{Key Laboratory of Radio Astronomy, Chinese Academy of Sciences, China}

\author{Zhi-Qiang Shen}
\affiliation{Shanghai Astronomical Obervatory, 80 Nandan Road, Xuhui Districe, Shanghai, 200030, China \\}
\affiliation{Key Laboratory of Radio Astronomy, Chinese Academy of Sciences, China}

\author{Kai Yang}
\affiliation{Shanghai Astronomical Obervatory, 80 Nandan Road, Xuhui Districe, Shanghai, 200030, China \\}
\affiliation{University of Chinese Academy of Sciences, 19A Yuquanlu, Beijing 100049, China\\}

\begin{abstract}

We performed a radio recombination line (RRL) survey to construct a high-mass star-forming region (HMSFR) sample in the Milky Way based on the all-sky Wide-Field Infrared Survey Explorer (\textit{All-WISE}) point source catalog. The survey was observed with the Shanghai 65m Tianma radio telescope (TMRT) covering 10 hydrogen RRL transitions ranging from H98$\alpha$ to H113$\alpha$ (corresponding to the rest frequencies of 4.5$-$6.9 GHz) simultaneously. Out of 3348 selected targets, we identified an HMSFR sample consisting of 517 sources traced by RRLs, a large fraction of this sample (486) locate near the Galactic plane ($|$\textit{b}$|$ $<$ 2$\degr$). In addition to the hydrogen RRLs, we also detected helium and carbon RRLs towards 49 and 23 sources respectively. We cross-match the RRL detections with the 6.7 methanol maser sources built up in previous works for the same target sample, as a result, 103 HMSFR sources were found to harbor both emissions. In this paper, we present the HMSFR catalog accompanied by the measured RRL line properties and a correlation with our methanol maser sample, which is believed to tracer massive stars at earlier stages. The construction of an HMSFR sample consisting of sources in various evolutionary stages indicated by different tracers is fundamental for future studies of high-mass star formation in such regions.

\end{abstract}


\section{Introduction\label{1}}

Formation of high-mass stars in the giant molecular clouds, though intensively studied, remains mysterious (see review papers, e.g., \citealt{ZY2007, tan2014}). To reveal the intrinsic of high-mass star formation (HMSF) at the very early stage, the fundamental and vital step is to construct a complete sample of high-mass star-forming regions (HMSFRs). Ultra-compact H{\footnotesize \RNum{2}} regions (UCH{\footnotesize \RNum{2}}Rs) ($<$ 0.1 pc) are hot ionized gas surrounding an exciting central high-mass star. Such regions are excited by an early O$-$B star from which the ultra-violet photons are strong enough to ionize neutral hydrogen. H{\footnotesize \RNum{2}} regions spread widely at a Galactic scale and have strong luminosity across multiple wavebands (ultraviolet, visible, infra-red and radio), therefore, they are ideal tracers of HMSFRs. 

H{\footnotesize \RNum{2}} region surveys in the Milky Way were firstly studied in visible wavelengths \citep{sharpless1953,sharpless1959, gum1955, rodgers1960}. However, the extinction in the optical largely limited the capability of such researches. The dust-free radio observations are therefore needed to construct a more complete sample of Galactic H{\footnotesize \RNum{2}} regions.

In 1965, radio recombination line (RRL) was firstly detected by \citet{HM1965} from M 17 and Orion A. Its thin optical depth in centimeter wavelengths makes it an optimal tracer of H{\footnotesize \RNum{2}} regions. RRL surveys were then performed in the next few decades, e.g. \citet{MH1967, wilson1970, reifenstein1970, downes1980, CH1987} and \citet{lockman1989}. The properties of the Galactic RRLs, such as their spatial distribution, line widths, LSR velocities and intensities are probes of the morphological, chemical and dynamical information of the Milky Way (see \citealt{anderson2011}). Thus, RRL is important in a range of astrophysical topics, such as the Galactic structure (e.g. \citealt{HH2015, downes1980, AB2009}) and metallicity gradient across the Galactic disk which helps understanding the Galactic chemical evolution (GCE) \citep{wink1983, shaver1983, quireza2006, balser2011}.


More recent RRL surveys were performed with high-sensitivity facilities (e.g. \citealt{liu2013, alves2015, anderson2011, anderson2014}). In particular, the recent Green Bank Telescope (GBT) H{\footnotesize \RNum{2}} Region Discovery Survey (HRDS) detected 603 discrete RRL components from 448 targets which were considered to be H{\footnotesize \RNum{2}} regions, thus doubled the number of known Galactic H{\footnotesize \RNum{2}} regions \citep{anderson2011}. With the demonstration that H{\footnotesize \RNum{2}} regions can be reliably identified by their mid-infrared (MIR) morphology, \citet{anderson2014} extended the HRDS sample to $\sim~8000$ candidate sources based on the \textit{all-sky Wide-Field Infrared Survey Explorer} (\textit{WISE}) MIR images (hereafter the \citetalias{anderson2014} catalog). The \citetalias{anderson2014} catalog contains $\sim~1500$ confirmed H{\footnotesize \RNum{2}} regions with observed RRL data in the literature, it is the most complete sample of H{\footnotesize \RNum{2}} regions to date.

The \textit{WISE} data have four MIR bands: 3.4 $\mu$m, 4.6 $\mu$m, 12 $\mu$m and 22 $\mu$m, with angular resolutions of 6$\arcsec$.1, 6$\arcsec$.4, 6$\arcsec$.5 and 12$\arcsec$, respectively, which are sensitive to HMSFRs. Its complete sky coverage and up-to-date database provide an optimal target sample for identifying HMSFR candidates. To further extended the HMSFR sample traced by RRLs beyond the \citetalias{anderson2014} catalog, we conducted an RRL survey with the Shanghai 65m Tianma Radio Telescope (TMRT) based on the \textit{WISE} point source catalog rather than the \textit{WISE} MIR images. Since as H{\footnotesize \RNum{2}} regions form and evolve they will expand, selecting targets from the point source catalog will make our sample to include more compact, and therefore younger sources.

Comparing to other single-dish RRL surveys, we concentrate more on the correlation and association with methanol masers to signpost different periods of star-forming processes. Class \RNum{2} methanol maser is a powerful tracer of the hot molecular cloud phase of HMSFR \citep{minier2003, ellingsen2006, xu2008}, when there is significant mass accretion. H{\footnotesize \RNum{2}} region generally appears in more evolved phases of star formation, well before the main sequence \citep{walsh1998, BM1996}. As suggested by \citet{churchwell2002}, due to beam-blending and thick optical depth, the densest and earliest H{\footnotesize \RNum{2}} regions are heavily obscured, more extended detectable UCH{\footnotesize \RNum{2}} regions probably are only formed until the central star reaches main sequence, and no longer accreting significant mass. By cross-matching the RRL and class \RNum{2} methanol maser samples, the evolutionary stages of their hosts may be specified more accurately. Therefore, simultaneous observation for both the RRLs and 6.7 GHz methanol masers were conducted to investigate their associations. Notably, due to beam dilution, RRL emissions from dense UCH{\footnotesize \RNum{2}} regions at earlier stages will be undetectable, thus our detected RRL sources will trace more extended and evolved H{\footnotesize \RNum{2}} regions. Previous studies have demonstrated that 6.7 GHz methanol masers can be excited in the UCH{\footnotesize \RNum{2}} regions, including both extended and compact sources identified by radio continuum data (e.g \citealt{hu2016}). Since RRL-detected UCH{\footnotesize \RNum{2}} regions are generally more evolved than those without RRL emissions, RRL researches will be helpful for identifying which methanol maser sources are at more evolutionary stages.

In this paper, we report the RRL detections with the measured line parameters, as well as the results of a correlation with the 6.7 GHz methanol maser sample towards the same target sample built by \cite{yang2017, yang2019}. Section \ref{2} describes the sample selection and observations. Section \ref{f3} presents the results of the survey followed by a discussion in section \ref{4}. We summarize our main conclusions in Section \ref{5}.\\

\section{Observations and Data Reduction\label{2}}

\subsection{Source Selection\label{2.1}}


RRLs and 6.7 GHz methanol masers were observed simultaneously in our survey. The targets were selected with the following methodology: firstly, a cross-matching was applied between the 6.7 GHz methanol maser catalog created by the Methanol Multi beam (MMB) Survey conducted with the Parkes telescope \citep{caswell2010, caswell2011, green2010, green2012, breen2015}, and the \textit{All-WISE} point source catalog. As a result, there are 502 MMB maser sources which have a \textit{WISE} counterpart with a spatial offset within 7$\arcsec$. We only kept 473 sources with \textit{WISE} data available from all four bands. A magnitude and color-color analysis was then applied to those 473 sources (see \citealt{yang2017} for details). 73$\%$ of those sources fell in the color region with well-constrained \textit{WISE} color criteria: [3.4] $<$ 14 mag; [4.6] $<$ 12 mag; [12] $<$ 11 mag; [22] $<$ 5.5 mag; [3.4] - [4.6] $>$ 2, and [12] - [22] $>$ 2. To avoid repetition, we excluded sources locating in the MMB survey region ($20\degr < l < 186\degr$ and $|$\textit{b}$|$ $>$ 2$\degr$). Due to the limitation of observing range, we also excluded those with a declination below $-30 \degr$. In total, 3348 \textit{WISE} point sources were selected searching for RRLs and methanol maser emissions. In this sample, 1473 sources are located at a high Galactic latitude region with $|$\textit{b}$|$ $>$ 2$\degr$ and 1875 sources fall within $\pm 2\degr$ of the Galactic Plane.

Among the selected targets, \citet{yang2017, yang2019} detected 6.7 GHz methanol masers from 241 sources, 209 of them are near the Galactic Plane where $|$\textit{b}$|$ $<$ 1$\degr$.\\
\subsection{Observation \& Data Reduction\label{2.2}}

The observations were performed between 2015 September and 2018 January with the 65m TMRT in Shanghai, China \citep{yang2017, yang2019}. A cryogenically cooled C-band receiver (4-8 GHz) with two orthogonal polarizations was employed in this survey. We used an FPGA-based spectrometer Digital Backend System (DIBAS) (VEGAS; \citealt{bussa2012}) to receive and record the signals. A total of 16 spectral windows were applied in the observations, each has 16384 channels and a bandwidth of 23.4 MHz supplying a velocity resolution of $\sim$~0.1~km~s$^{-1}$ at 4.5 GHz. Ten of the spectral windows were set up to cover hydrogen RRL transitions spanning from H98$\alpha$ to H113$\alpha$ as specified in Table \ref{t1}. In addition to RRLs, the 6.7 GHz methanol maser (CH$_3$OH) emission line, 4.8 GHz H$_2$CO emission and absorption lines, as well as the 4.7 and 6.0 GHz excited-state OH maser transitions were also observed. The observed 6.7 GHz methanol maser results are reported in \citet{yang2017,yang2019}. This paper mainly focuses on RRL detection and a cross-match between RRLs and methanol masers. In the observations, the system temperature is about 20 $\sim$~ 30 K and the main beam efficiency of the TMRT is $\sim~60\%$. The beam has a full width at half-maximum ({\tt\string FWHM}) of $\sim~3\arcmin - 4\arcmin$ at the frequencies of RRLs. There is an uncertainty of $<~20\%$ in the detected flux densities for the sources estimated from the observed variation of the calibrators. 

\begin{table*}[ht!]
\caption{The information of the spectral windows of the TMRT receiver we applied in this survey.\label{t1}}
\centering
\begin{tabular*}{500pt}[c]{>{\centering}p{2.5cm}|>{\centering}p{1.3cm}>{\centering}p{1.3cm}>{\centering}p{1.3cm}>{\centering}p{1.3cm}>{\centering}p{1.3cm}>{\centering}p{1.3cm}>{\centering}p{2.7cm}>{\centering\arraybackslash}p{1.3cm}}
\toprule
Window Number& 1 & 2 & 3 & 4 & 5 & 6 & 7 & 8\\
\hline
$\nu_o$ (MHz) & 4497.78 & 4593.09 & 4618.79 & 4758.11 & 4829.66 & 4874.16 & 5008.92 & 5148.7\\
Line Name & H113$\alpha$ & H$_2^{13}$CO & H112$\alpha$ & OH & H$_2$CO & H110$\alpha$ & H109$\alpha$ & H108$\alpha$\\
\hline
Window Number& 9 & 10 & 11 & 12 & 13 & 14 & 15 & 16\\
\hline
$\nu_o$ (MHz) & 6016.75 & 6032.92 & 6049.08 & 6106.85 & 6289.14 & 6478.76 & 6672.30 (6676.08)\tablenotemark{*} & 6881.49\\
Line Name & OH & OH & OH & H102$\alpha$ & H101$\alpha$ & H100$\alpha$ & CH$_3$OH $\&$ H99$\alpha$ & H98$\alpha$\\
\bottomrule
\end{tabular*}\\
\tablenotetext{*}{The 15$^{th}$ spectral window covers both the 6.7 GHz methanol maser and H99$\alpha$ line, we set the central frequency of the window to the medium value of the two lines.}
\end{table*}

\begin{deluxetable*}{cccccccc}
\tablecaption{Overview of the main detection results.\label{t2}}
\tablehead{
\colhead{Total HMSFRs} & \colhead{HMSFRs traced by RRLs} & \colhead{Maser associations} & \colhead{Maser-only sources} & \colhead{He$\alpha$ lines} & \colhead{C$\alpha$ lines} & \colhead{SNR} & \colhead{PNe} \\
\colhead{(1)} & \colhead{(2)} & \colhead{(3)} & \colhead{(4)} & \colhead{(5)} & \colhead{(6)} & \colhead{(7)} & \colhead{(8)}}

\startdata
654 & 517 & 103 & 137 & 49 & 23 & 5 & 5 \\
\enddata
\tablecomments{Overview of the main results: (1) the final HMSFR sample combing those traced by RRL and/or 6.7 GHz methanol maser reported in \citet{yang2017, yang2019}, (2) HMSFRs traced by RRL detections excluding potential PNe and SNRs, (3) HMSFR sources associated with both RRL and 6.7 GHz methanol maser, (4) HMSFRs associated only with 6.7 GHz methanol maser, (5) helium recombination lines and (6) carbon recombination lines, (7) SNR and (8) PNe candidates catalogd in {\tt\string SIMBAD}.}
\end{deluxetable*}

A position-switching mode was applied in the observations. Each source was observed with two ON-OFF cycles, both the ON- and OFF-positions in each cycle take $\sim~2$ minutes. To ensure data reliability, sources with bad data quality (eg. radio frequency interferences (RFI), high noise level or abnormal signals) were re-observed with many more cycles. For each source, we primarily set the OFF-point to (0.0$\degr$, $-0.4\degr$) away from the ON-position in (R.A., Decl.). Sources with OFF-positions showing RRL emission were re-observed with a different OFF-position to exclude the background emissions. In total, the observations took $\sim$ 700 hours of observing time excluding calibrations.

The data were processed with the {\tt\string GILDAS/CLASS}\footnote{https://www.iram.fr/IRAMFR/GILDAS/} software package \citep{pety2005,gildas2013}. Adjacent Hn$\alpha$ transitions with large quantum numbers (n $> 50$) have similar line properties, such as line intensity and {\tt\string FWHM} line width (see \citealt{balser2006}). Therefore, the spectra of the observed 10 transitions through H98$\alpha$ to H113$\alpha$ can be averaged to achieve a higher S/N. After averaging over the 10 RRL transitions at two polarizations, a typical 3-$\sigma$ sensitivity of $\sim$~7 mK per channel was achieved for the majority of our sources. 


Due to different rest frequencies, the 10 RRL transitions have different velocity resolutions in their individual spectrum. When averaging over the spectral windows, the {\tt\string GILDAS/CLASS} software will resample the data with the coarsest resolution of them. Depending on the spectral quality, we typically smooth the averaged spectrum over 5 to 30 channels to have a spectral resolution of $\sim$~0.4 to $\sim$~2.4 km~s$^{-1}$. After that, depending on the background fluctuations, the spectral baselines were subtracted by a first or multi-order ($<$ 4) polynomial fitting. Then the line profiles were fitted with a Gaussian model. For the multi-component sources, the number of Gaussian components was decided via a visual inspection. Notably, it is somehow hard to disentangle the blended line components which have very close central frequencies, so some of the blended sources may be identified as single emission with wide line width (see Section \ref{3.1}). The Gaussian fitting results are reported in Section \ref{3}. \\

\section{Result\label{3}}

\subsection{RRL Detections\label{3.1}}
A summary of RRL detections in this work and 6.7 GHz methanol maser detections from \citet{yang2017, yang2019} are presented in Table \ref{t2}. Out of the 3348 targets, we detected hydrogen (H) RRL emissions from 527 positions, corresponding to a detection rate of 15.7\%. Excluding the potential Planetary Nebula (PNe) and Supernova Remnant (SNR) (see Sec. \ref{3.2}), we built a sample of 517 HMSFRs based on the RRL detections. The derived line parameters and source information for the H RRLs from the HMSFR sample are listed in Table \ref{t3}. The spectra for all the detected RRLs (including helium (He) and carbon (C) RRLs; see Section \ref{4.3}) are given in Appendix A. Amongst the 517 HMSFR candidates, 488 of them reside within $|b| < 2 \degr$, only 28 are from higher Galactic latitude regions ($|b| > 2 \degr$). Combining with the 240 methanol maser sample in \citet{yang2017, yang2019}, excluding one associated with a potential SNR, there are 654 HMSFR sources traced by RRLs and/or methanol maser listed in Table \ref{t4}.

\begin{figure}
\gridline{\fig{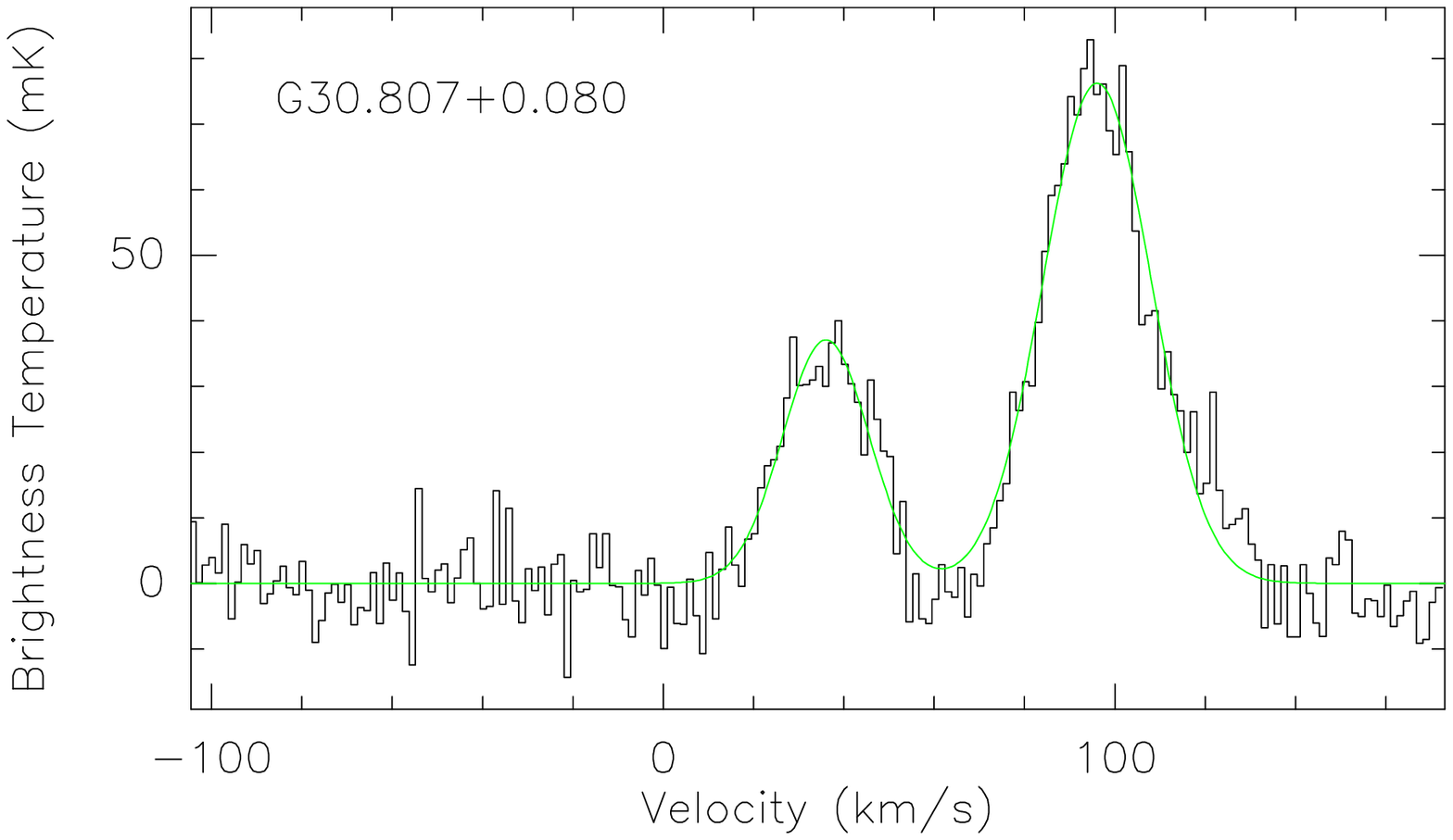}{0.45\textwidth}{(a)}}
\gridline{\fig{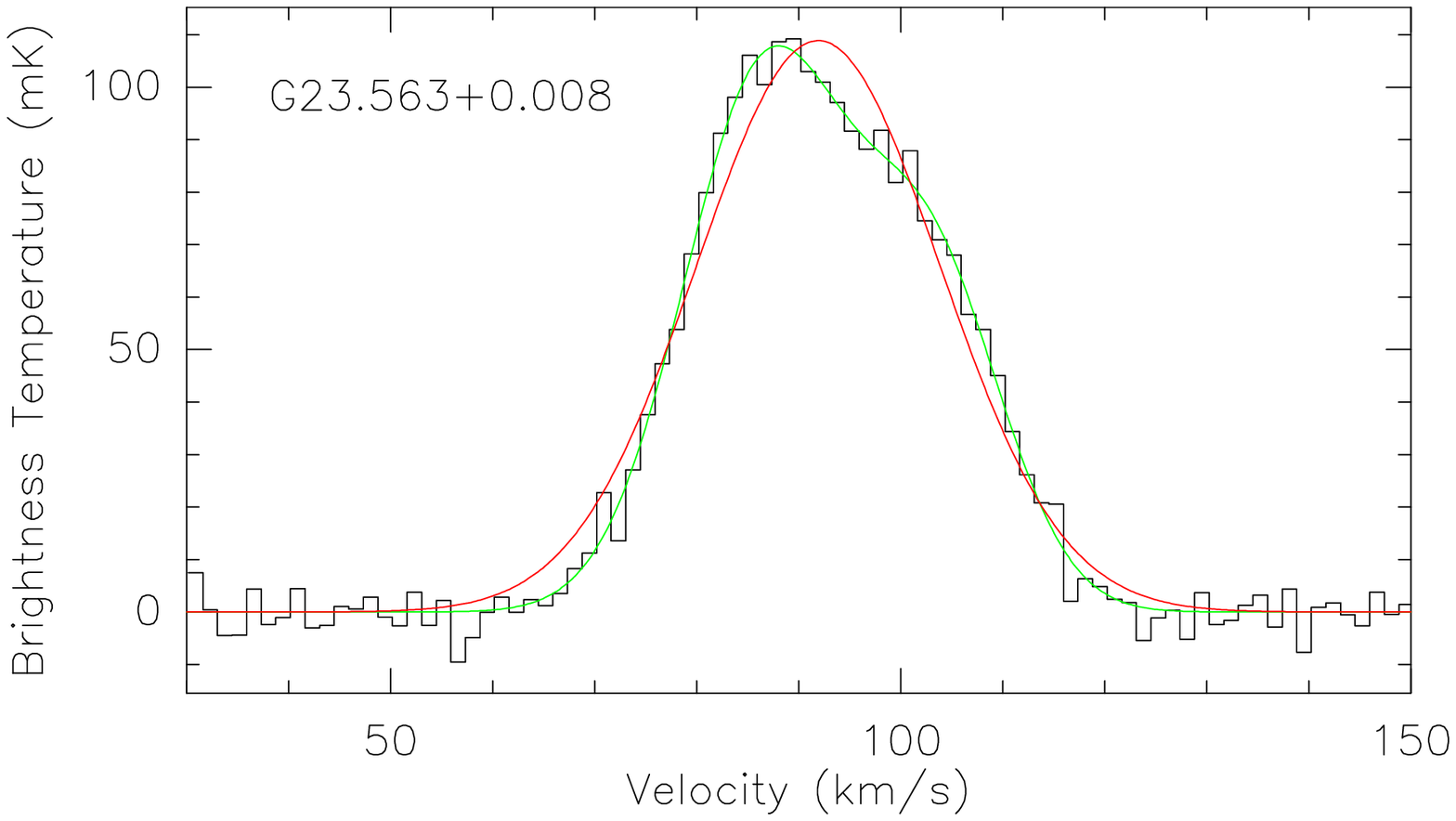}{0.45\textwidth}{(b)}}
\gridline{\fig{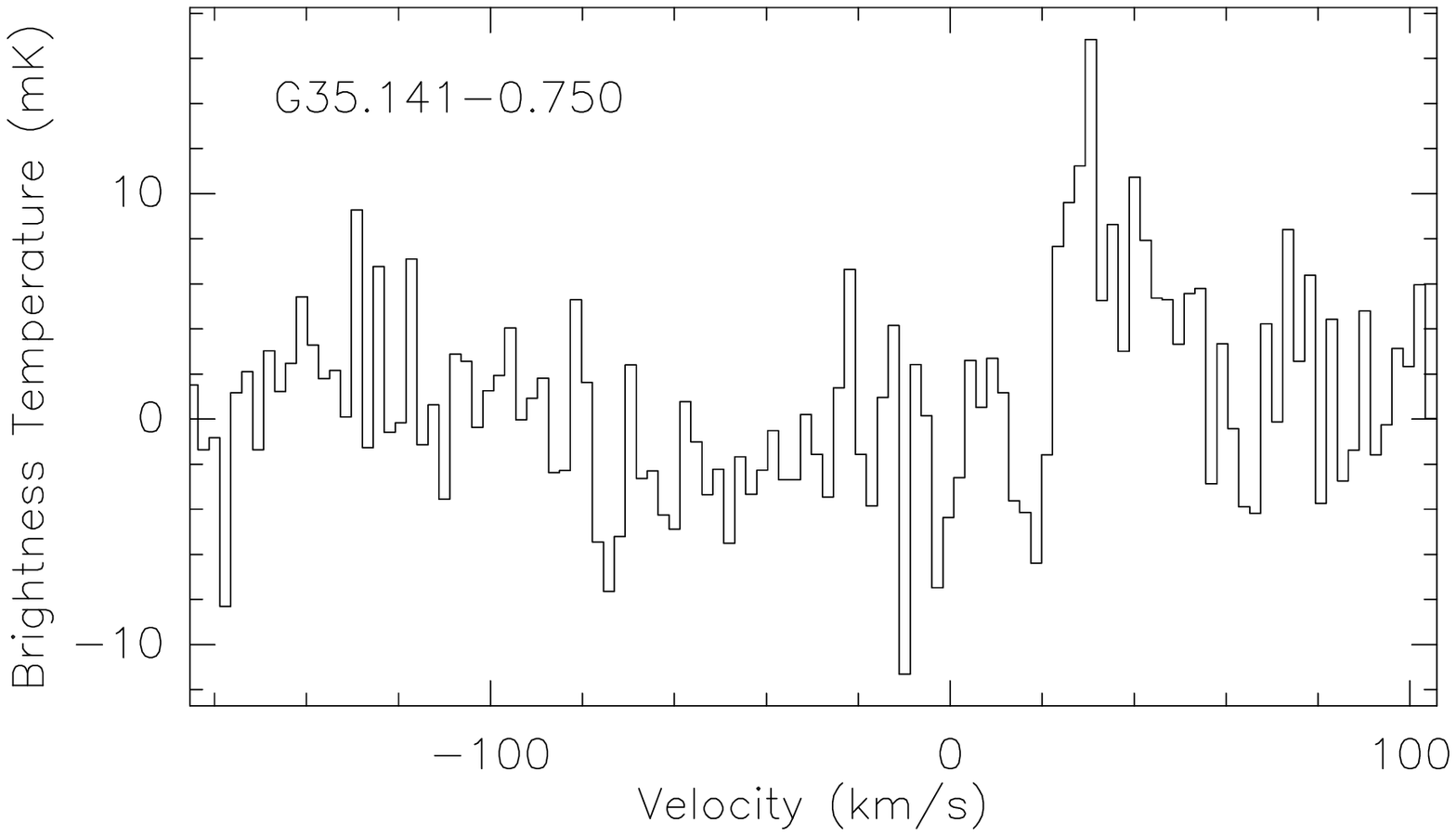}{0.45\textwidth}{(c)}}
\caption{Example spectra of the detected hydrogen RRLs: (a) a source with two explicit Gaussian line components; (b) a non-Gaussian line component possibly caused by two very close RRL emissions with a velocity offset smaller than their width; and (c) a weak RRL signal. The spectra for the whole RRL sample can be found in the Appendix A.\label{f1}}
\end{figure}

We cross-matched our detected sources with the \citetalias{anderson2014} catalog consisting of $\sim$~8000 sources. Due to the lack of radio observations, only $\sim$~1500 sources in their sample were confirmed to be ``Known'' H{\footnotesize \RNum{2}} regions (denoted as ``K" sources in Table 2 of \citealt{anderson2014}). For targets which spatially associate with multiple sources in the \citetalias{anderson2014} catalog, ``Known'' sources would be designated preferentially. For targets associate with multiple ``Known'' sources or other types of \citetalias{anderson2014} sources, the closest \citetalias{anderson2014} counterpart would be designated. There are 467 HMSFR candidates in our sample which are associated with at least one of \citetalias{anderson2014} sources within a separation of 3$\arcmin$ (corresponding to the beam size of TMRT) plus the radius of the \citetalias{anderson2014} source. Amongst them, 425 were classified as ``known'' H{\footnotesize \RNum{2}} regions, we thus confirmed the other 42 sources being H{\footnotesize \RNum{2}} regions. For the sources included in the \citetalias{anderson2014} catalog, we label their type accordingly in Column (11) of Table \ref{t3}. There are also 3 PNe and 4 SNR candidate sources associated with \citetalias{anderson2014} sources, due to the extended morphology of H{\footnotesize \RNum{2}} regions, these targets may be overlapped with the \citetalias{anderson2014} sources along the line of sight (LOS).

There are 133 sources showing multiple (typically two or three) H RRL emission components. Figure \ref{f1}a shows an example spectrum for such sources. The multiple RRL components may arise from nearby H{\footnotesize \RNum{2}} regions within the TMRT beam or overlapped H{\footnotesize \RNum{2}} regions along the LOS. Diffused ionized gas leaked from nearby HMSFRs may also cause multi-components in the RRL spectra \citep{zavagno2007, anderson2010, OK1997}. Some components have line emissions with close peak velocities causing confusion with a wide line width or non-Gaussian profile. An example of this is presented in Figure \ref{f1}b.

Notably, RRL detections at adjacent observing positions with similar peak velocities are possibly from the same extended H{\footnotesize \RNum{2}} region. Moreover, H{\footnotesize \RNum{2}} regions with large angular size may have leaked RRL emission detected by the side lobe of the telescope when observing its nearby target points. Those RRL components have similar line velocities but much weaker line intensities comparing to the emission from the central position of the source. There are 5 RRL sources which were potentially detected by the side lobe, we label those emissions as possible duplicated sources by ``SL" and the name of the real source in Column (11) in Table \ref{t3}.

Our RRL detections have peak intensities ranging from 0.01 to 2 K with an average value of 0.07 K, and an integrated intensity from 0.1 to 58 K$\cdot$km~s$^{-1}$ with an average of 1.9 K$\cdot$km~s$^{-1}$. Among the 517 HMSFR candidates, there are 12 weak sources, labeled with ``?" in column (11) in Table \ref{t3}, which have a line intensity only slightly above our 3$\sigma$ detection threshold, no accurate Gaussian fitting results can be achieved for them. For those weak sources, we only give their peak intensity in Table \ref{t3}. Further observations with longer integration time are required for them to get a higher S/N. Figure \ref{f1}c illustrates an example of these weak sources. In addition to the 12 weak sources, for multi-component sources containing such weak line component(s), we only provide Gaussian fitting results for the stronger components.

\begin{figure}[!tb]
\epsscale{1.1}
\plotone{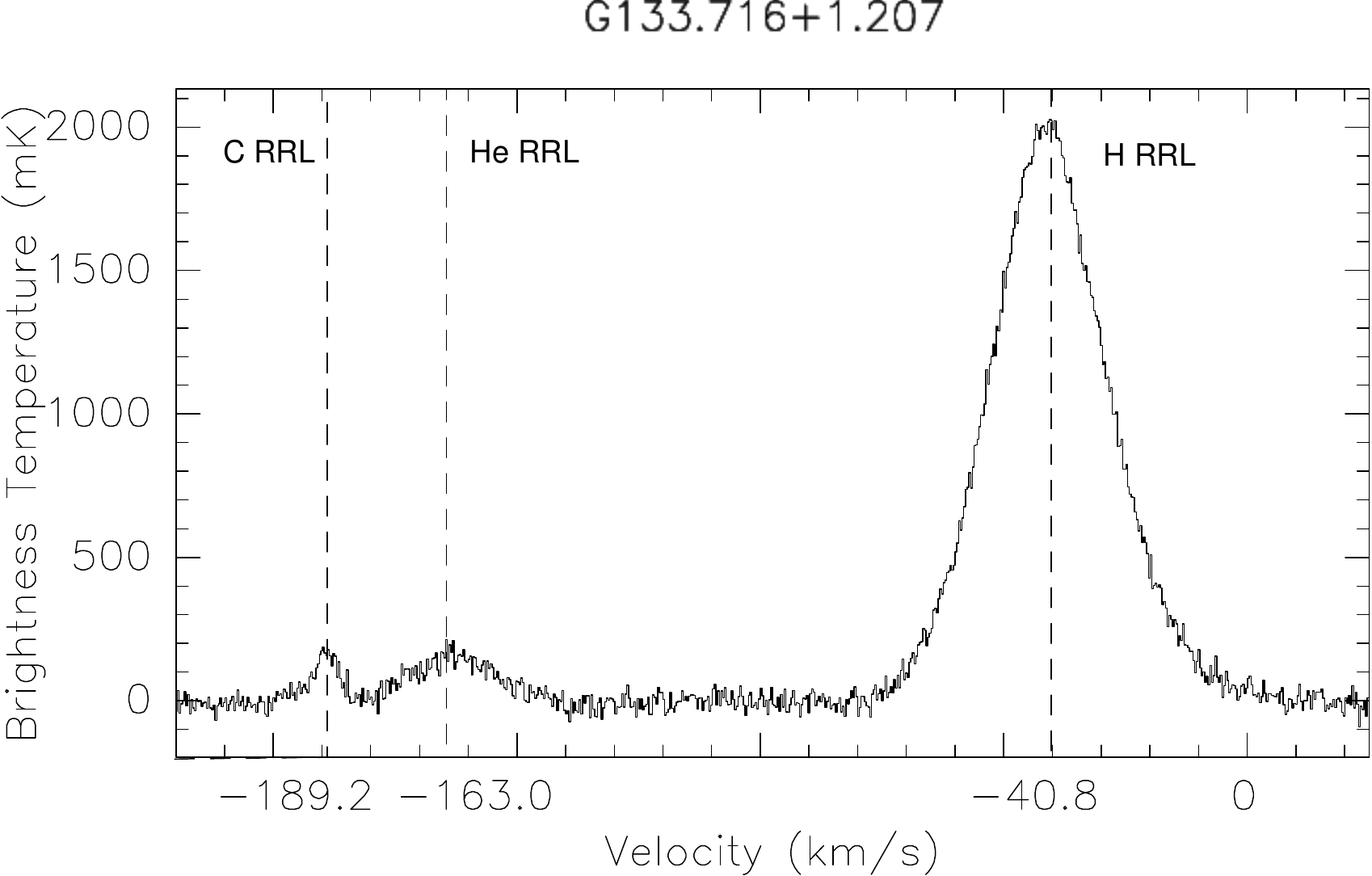}
\caption{An example source with the three atomic RRL detections. Helium and carbon RRLs generally have a constant velocity offset with respect to hydrogen RRLs of $-$ 122~km~s$^{-1}$ and $-$149~km~s$^{-1}$ respectively. The spectra for the whole RRL sample can be found in the Appendix A.\label{f2}}
\end{figure}

In addition to H RRL transitions, the 23.4 MHz spectral windows (see Section \ref{2.2}) also simultaneously cover the rest frequencies of the nearby He and C RRLs, they have the same intrinsic quantum numbers with the H RRLs within each spectral window. In the velocity domain, He and C RRLs typically have a constant velocity offset with respect to H RRLs of $-$122 km~s$^{-1}$ and $-$149~km~s$^{-1}$, respectively. The derived line parameters of He and C RRLs are given in Table \ref{t5}. Figure \ref{f2} shows an example source with all the three atomic RRLs. In total, we found 49 He RRLs and 23 C RRLs in the observed sample (see Section \ref{4.3} for further discussions).\\

\subsection{Other Sources\label{3.2}}

In addition to HMSFRs, ionized gas associated with other astrophysical objects such as planetary Nebula (PNe) and Supernova Remnant (SNR) can also produce RRLs. We performed a matching analysis for our sample with the {\tt\string SIMBAD}\footnote{http://simbad.u-strasbg.fr/simbad/} catalog to exclude previously known PNe and SNR sources with a positional criterion of $< 3 \arcmin$. We mark sources with an explicit PNe or SNR identifier as ``PNe" or ``SNR" in Tables \ref{t6} and \ref{t7}, respectively. To maximize the reliability of our HMSFR sample, for those candidate PNe/SNR sources, which may be associated with both PNe/SNR and H{\footnotesize \RNum{2}} regions or with PNe/SNR located near the edge of the detecting beam ($\sim3 \arcmin$ away), we remove them from the final HMSFR sample and denote them as ``PNe?"/``SNR?" in Table \ref{t6} / \ref{t7}. Notably, though spatially associated with PNe or SNR, the RRLs detected from those target positions may originated from H{\footnotesize \RNum{2}} regions along the LOS.

\begin{figure}[!tb]
\epsscale{1.1}
\plotone{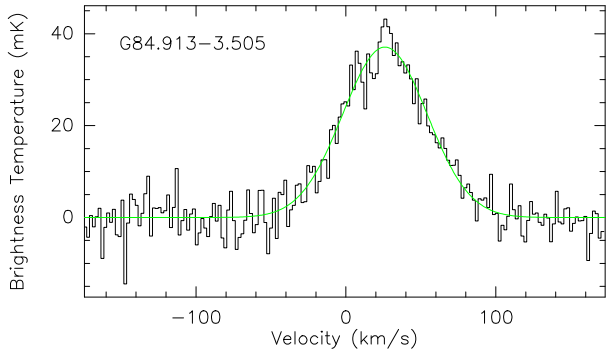}
\caption{An example RRL from a PNe source with a line width of 58.7 km s$^{-1}$. The nominal position of this source is $\sim~68\arcsec$ away from the well-known PNe NGC7027. The spectra for the whole PNe sample can be found in the Appendix B.\label{f3}}
\end{figure}

PNe usually has an expanding shell of ionized gas ejected from red giant stars. It is the main contamination of H{\footnotesize \RNum{2}} region samples traced by RRLs. RRLs from PNe usually have wider line width (typically 30 $\sim$~ 50 km s$^{-1}$) than those from H{\footnotesize \RNum{2}} regions (typical line width of 20 $\sim$~ 30 km s$^{-1}$) due to their expansion (see \citealt{garay1989, balser1997}). A typical RRL spectrum from PNe with a line width of $\sim$ 60 km s$^{-1}$ is shown in Figure \ref{f3}. The existence of unknown PNe may cause a bias in the statistical analysis of the line width distribution for H{\footnotesize \RNum{2}} regions (see Section \ref{4.2.2}). There are 2 sources (G84.913$-$3.505; G85.946$-$3.488) which are explicitly associated with the well-built PNe NGC 7027, as well as 3 possible PNe sources. The full list of PNe candidates is given in Table \ref{t6}.

\begin{figure}[!tb]
\epsscale{1.1}
\plotone{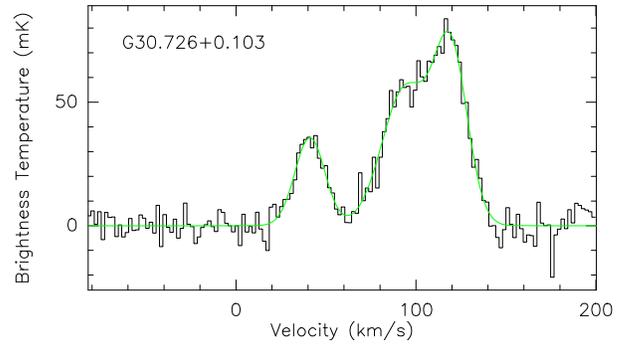}
\caption{An example SNR source with three Gaussian RRL components. This source is $\sim$ 71$\arcsec$ away from the previously identified SNR G030.3+00.7. The spectra for the whole SNR sample can be found in the Appendix C. \label{f4}}
\end{figure}

SNRs are non-thermal radio sources, and generally have only weak RRLs with wide line width. \citet{liu2019} suggested that the broad line width (mostly $>$ 50 km s$^{-1}$) of the RRLs toward SNRs implies high temperature or turbulent motions of the plasma. Former studies indicated that stimulated emission may be a possible origin for RRLs from SNRs (see \citealt{liu2019} and references therein). The potentially similar radio morphology of SNR and H{\footnotesize \RNum{2}} regions is the major confusion when disentangling these two samples \citep{anderson2017}. There are 5 potential SNR candidates in our sample, all these sources are spatially associated with multiple sources thus no clear identifier can be designated. We list the potential SNR sources in Table \ref{t7}. Figure \ref{f4} shows an example of RRL spectra from SNR. There is one potential SNR source (G28.532+0.129) which also exhibit methanol maser emission, however, it has a small spatial offset (57.41$\arcsec$) to a possible SNR source G28.5167+0.1333, and has a very wide RRL line width of 70.2 km s$^{-1}$, thus there may have both HMSFR and SNR sources along the LOS of this target. To minimize contamination, this source was excluded from the HMSFR sample and classified as a potential SNR.\\

\section{Discussion\label{4}}


\subsection{Association and Correlation with Methanol Masers\label{4.1}}

In addition to our RRL emissions, there are 241 6.7-GHz class \RNum{2} methanol maser sources detected towards the same sample with TMRT \citep{yang2017,yang2019} including one towards a potential SNR (G28.532+0.129; see Section \ref{3.2}). The majority of the maser sample (224/241) are close to the Galactic Plane ($|\textit{b}| < 2\degr$). 

6.7 GHz methanol masers are believed to appear at an earlier stage of star formation, while H{\footnotesize \RNum{2}} regions typically exist at more evolved stages (see Section \ref{2.1}). Thus the association of 6.7 GHz methanol masers and RRLs would be helpful on discriminating HMSFRs at different evolutionary stages \citep{walsh1998, jordan2017}.

Our survey observed RRLs and methanol masers simultaneously with the same pointing positions, providing the most accurate results of cross-matching the signals. In addition, combing data from different surveys would bring systematic error in the cross-matching caused by inconsistent sensitivities and resolutions. Since we mainly focus on building up a cross-matched HMSFR sample, to reach higher reliability, we do not combine our date with the literature for the analysis in this paper. The 6.7-GHz class \RNum{2} methanol maser signal is observed and received by the 15$^{th}$ window in our survey (see Table \ref{t1}). Out of the 517 HMSFR sample with RRL detections, there are 103 (20.1$\%$) sources associated with methanol masers, meanwhile, 43.2$\%$ (103/240) methanol maser sources (excluding one potential SNR) exhibit RRL emissions. The remaining 137 methanol maser sources without RRLs may correspond to a younger evolutionary stage compared to those with RRLs. We label the sources showing both emission features in Column (12) in Table \ref{t3}. Notably, selecting targets from the point source catalog will bias the sample to include younger sources, and therefore may increase the number of sources associated with both RRL and maser. As a comparison, \citet{anderson2011} found that only $\sim$ 10\% (46/448) of their H{\footnotesize \RNum{2}} region sample associated with methanol maser.

\begin{figure}[!tb]
\epsscale{1.1}
\plotone{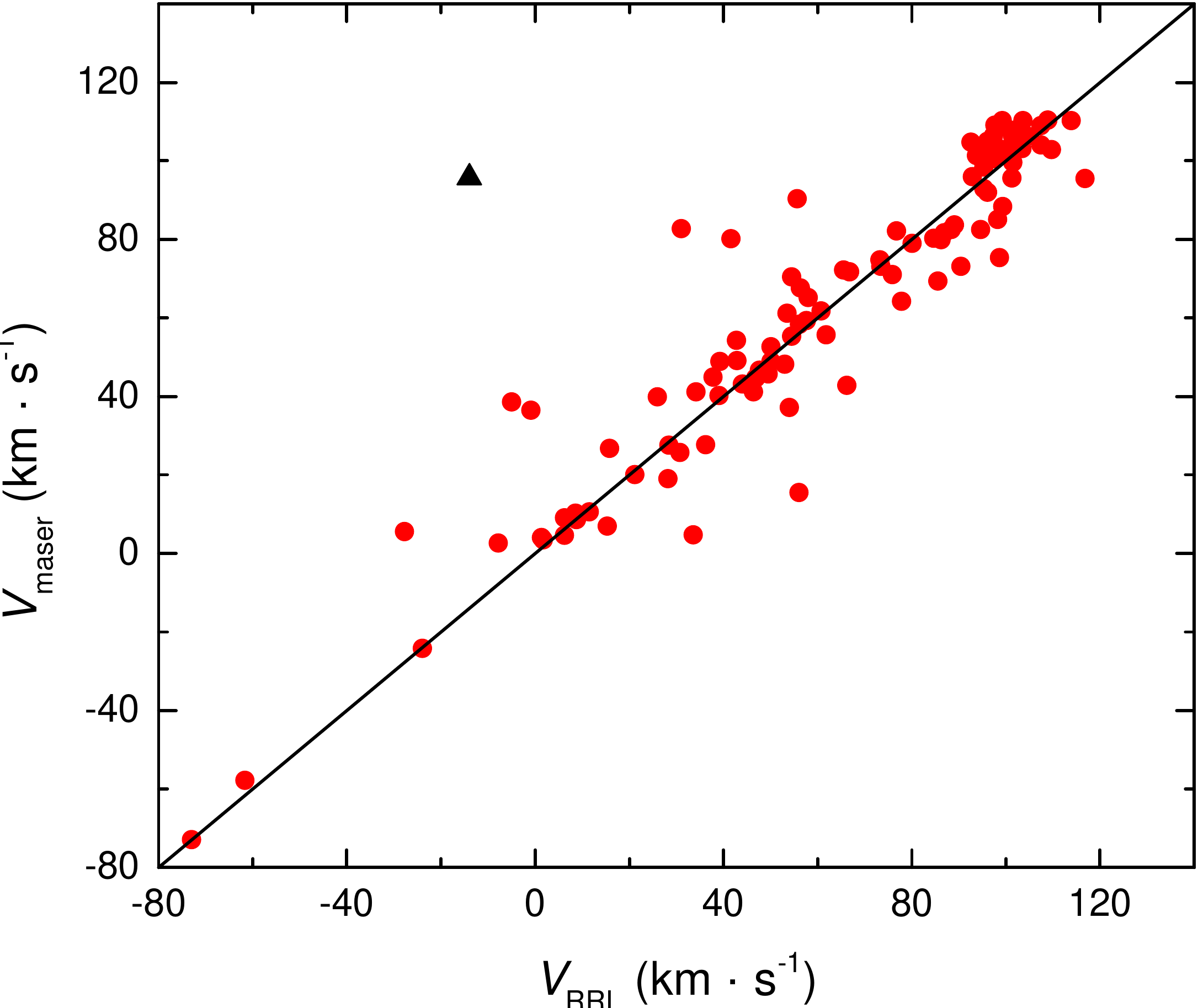}
\caption{The correlation between radio $V_{{\rm LSR}}$ at peaks of RRL and methanol maser spectra. The solid line denotes where $V_{{\rm maser}}$ = $V_{{\rm RRL}}$. There is one source marked with black triangles (G25.395+0.033) significantly departs from the rest of the sources.\label{f5}}
\end{figure}

The correlation of LSR velocities between RRLs and methanol masers is shown in Figure \ref{f5}. The majority of the associated RRL and methanol emissions seem to have fairly similar velocities, which intrinsically represent the systemic motion of the sources. As shown in Figure \ref{f5}, there is one source (G25.395+0.033) which has a large offset between the $V_{\rm LSR}$ of RRL and maser. Due to the commonly extended morphology of H{\footnotesize \RNum{2}} regions, RRL and maser emissions detected from the same target position but with a large velocity offset may from different sources along the LOS, or different regions in the same extended cluster in our $\sim 3~\arcmin$ beam.

As the processes of star formation at early stages are poorly understood, comparing the physical properties of HMSFR in different evolutionary stages will help us to study the formation and early evolution of massive star in such sources. We briefly discuss the spatial and intensity distributions of the three sub-sampled sources in the following sections, a more detailed discussion on their physical properties will be given in our future works. \\

\begin{figure*}[!tb]
\gridline{\fig{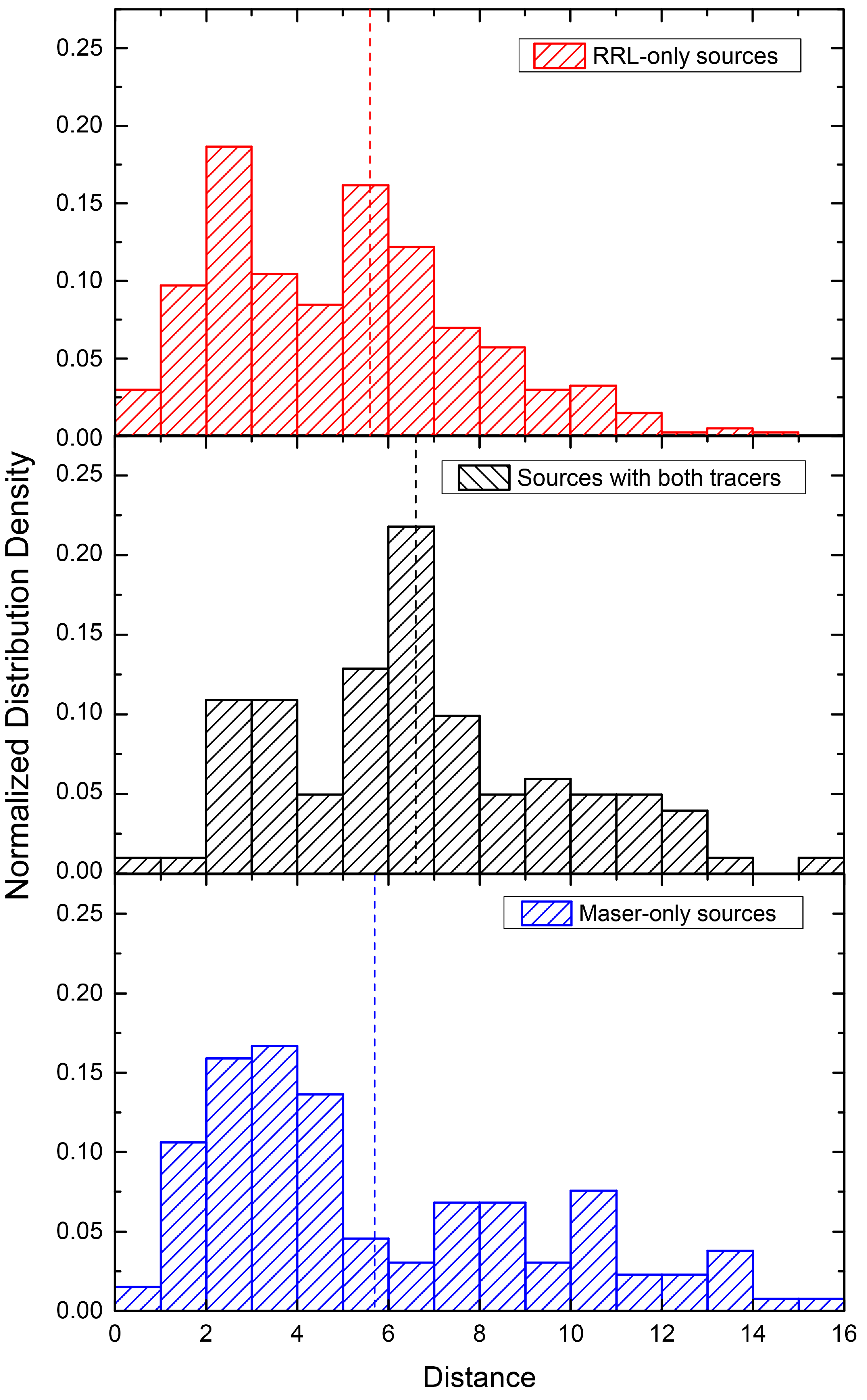}{0.45\textwidth}{(a)}\fig{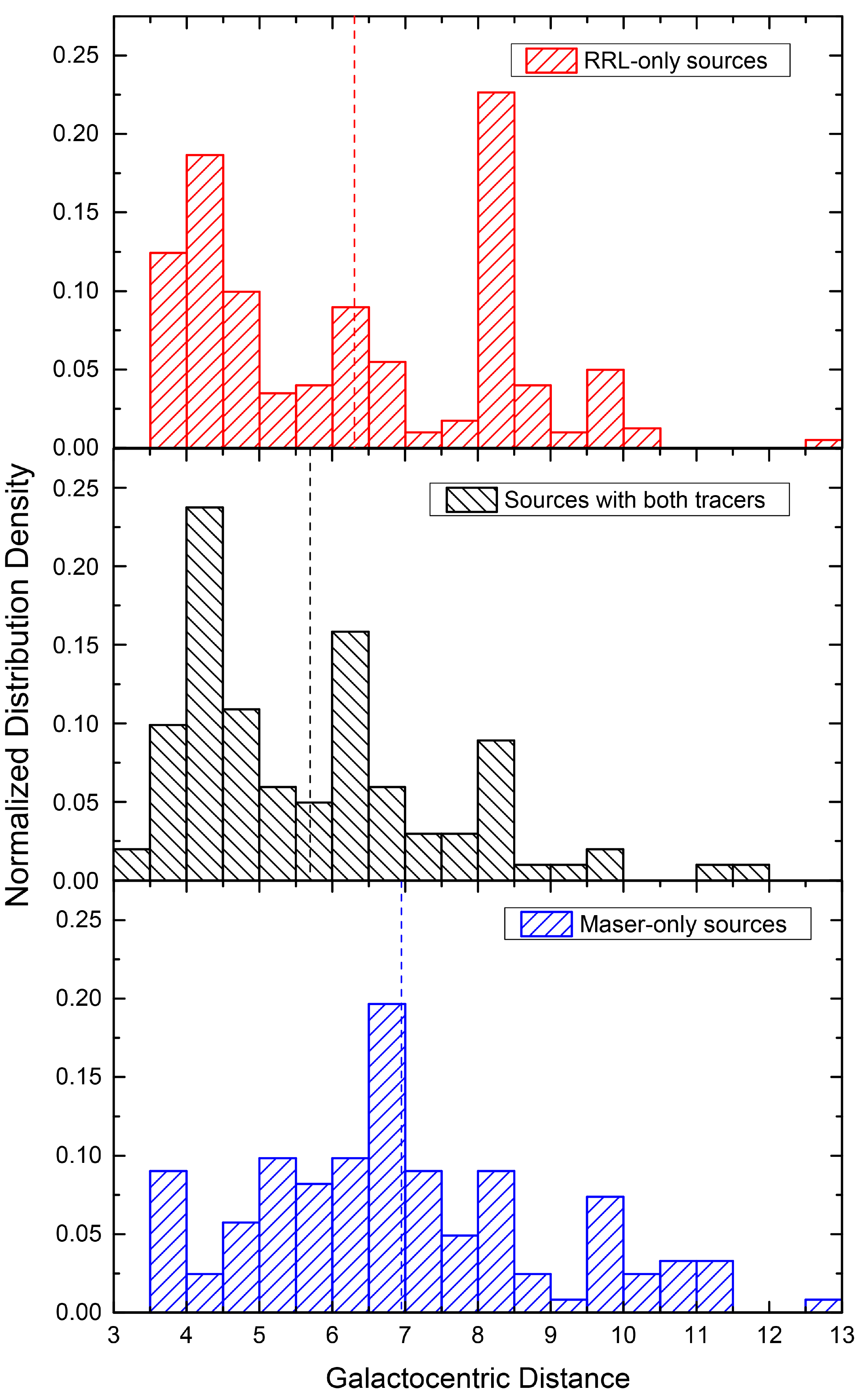}{0.45\textwidth}{(b)}}
\caption{(a) Normalized distance distribution densities of the 414 RRL-only sources (upper panel), 103 sources with both RRL and maser (middle panel) and 137 maser-only sources (lower panel), as shown by the dotted lines, the average distances with standard deviations of the three samples are 5.62 $\pm$ 2.25, 6.63 $\pm$ 3.03 and 5.72 $\pm$ 3.69 kpc, respectively. (b) Normalized galactocentric distance distribution densities of 414 RRL-only sources (upper panel), 103 sources associated with both tracers (middle panel) and 137 maser-only sources (lower panel), the dotted lines show the average galactocentric distances with standard deviations of 6.25 $\pm$ 2.09, 5.70 $\pm$ 1.77 and 6.95 $\pm$ 1.86 kpc, respectively.\label{f6}}
\end{figure*}

\subsubsection{Distance and Galactocentric Distance Distributions\label{4.1.1}}

Figure \ref{f6} shows the normalized distance and galactocentric distance ($R_{{\rm Gal}}$) distributions of RRL-only sources, maser-only sources, and sources associated with both tracers. The average distances with standard deviations of the three samples are 5.62 $\pm$ 2.25, 5.72 $\pm$ 3.69 kpc, and 6.63 $\pm$ 3.03 respectively. As shown in Figure \ref{f6}a, the majority of our sources are located at $\lesssim$ 8kpc, and there is no significant difference between the distance distribution of the three samples. This is consistent with previous studies that HMSFRs at different stages are distributed similarly with distance (e.g. \citealt{urquhart2014}).

The sources are more dispersed in the term of $R_{{\rm Gal}}$. As shown in Figure \ref{f6}b, although the three sub-samples have similar average $R_{{\rm Gal}}$ values (6.25 $\pm$ 2.09, 6.95 $\pm$ 1.86 and 5.70 $\pm$ 1.77 kpc for RRL-only sources, maser-only sources, and sources with both RRL and maser, respectively), both RRL-only sources and sources associated with both tracers seem to be more adequate near the Galactic center. This can be explained by the fact that the thin gas at the outer Galaxy makes ionized hydrogen hard to be formed, and vice versa.

As illustrated by Figure \ref{f6}b, three peaks can be seen in the distribution of $R_{{\rm Gal}}$ at $\sim$ 3$-$4, 6$-$7 and 8 kpc. The peak at 3$-$4 kpc for RRL-only sources and sources with both tracers may associate with the 4-kpc molecular ring \citep{dame2001}. The peak at 6$-$7 kpc appears in the $R_{{\rm Gal}}$ distribution for all sources may coincident with the northern segment of Sagittarius arm. The sources at the 8-kpc peak may be mostly associated with the local Sagittarius–Carina arm.\\


\subsubsection{RRL and Maser Line Intensity Distributions\label{4.1.2}}

\begin{figure}[!tb]
\gridline{\fig{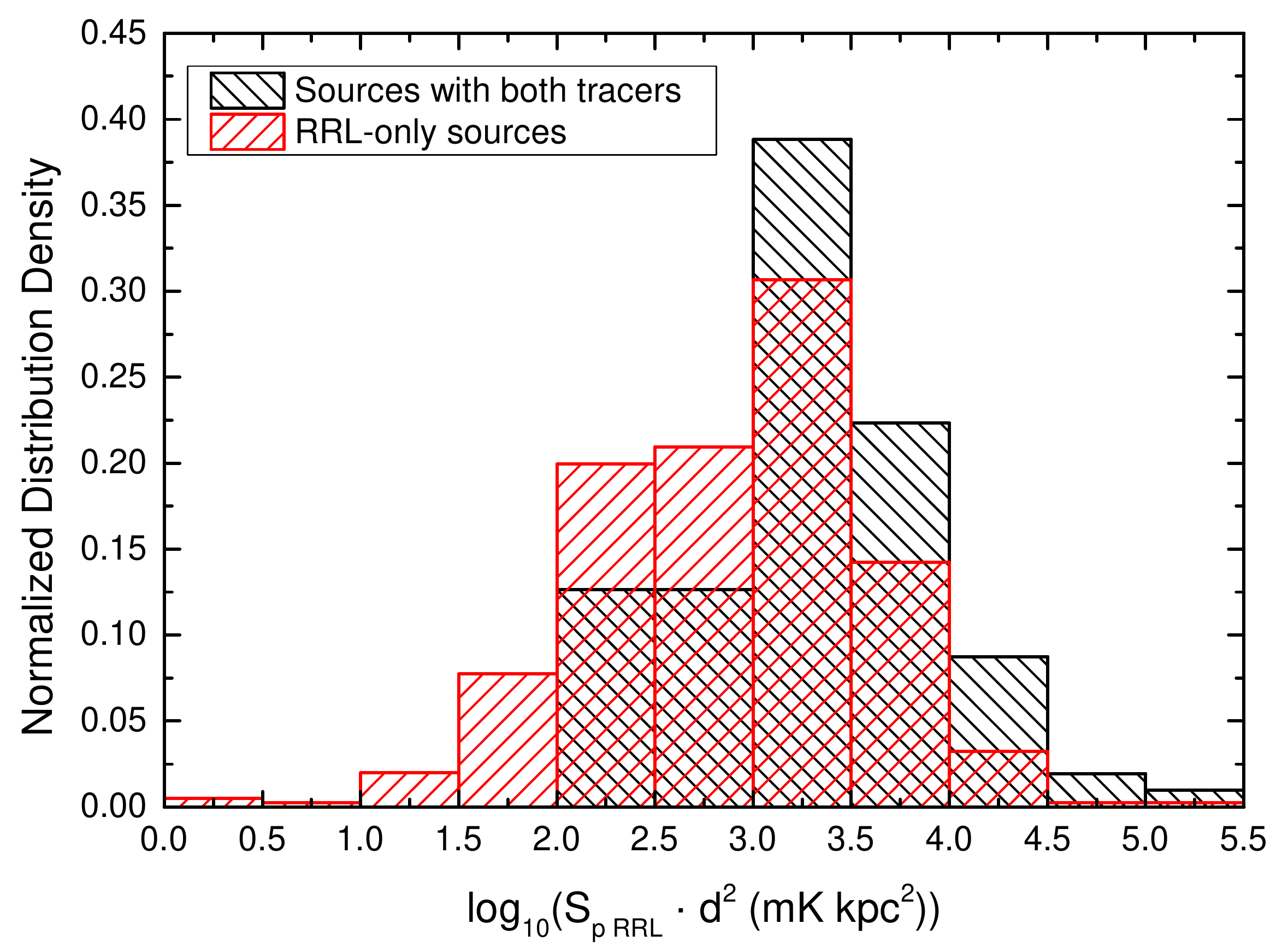}{0.45\textwidth}{(a)}}
\gridline{\fig{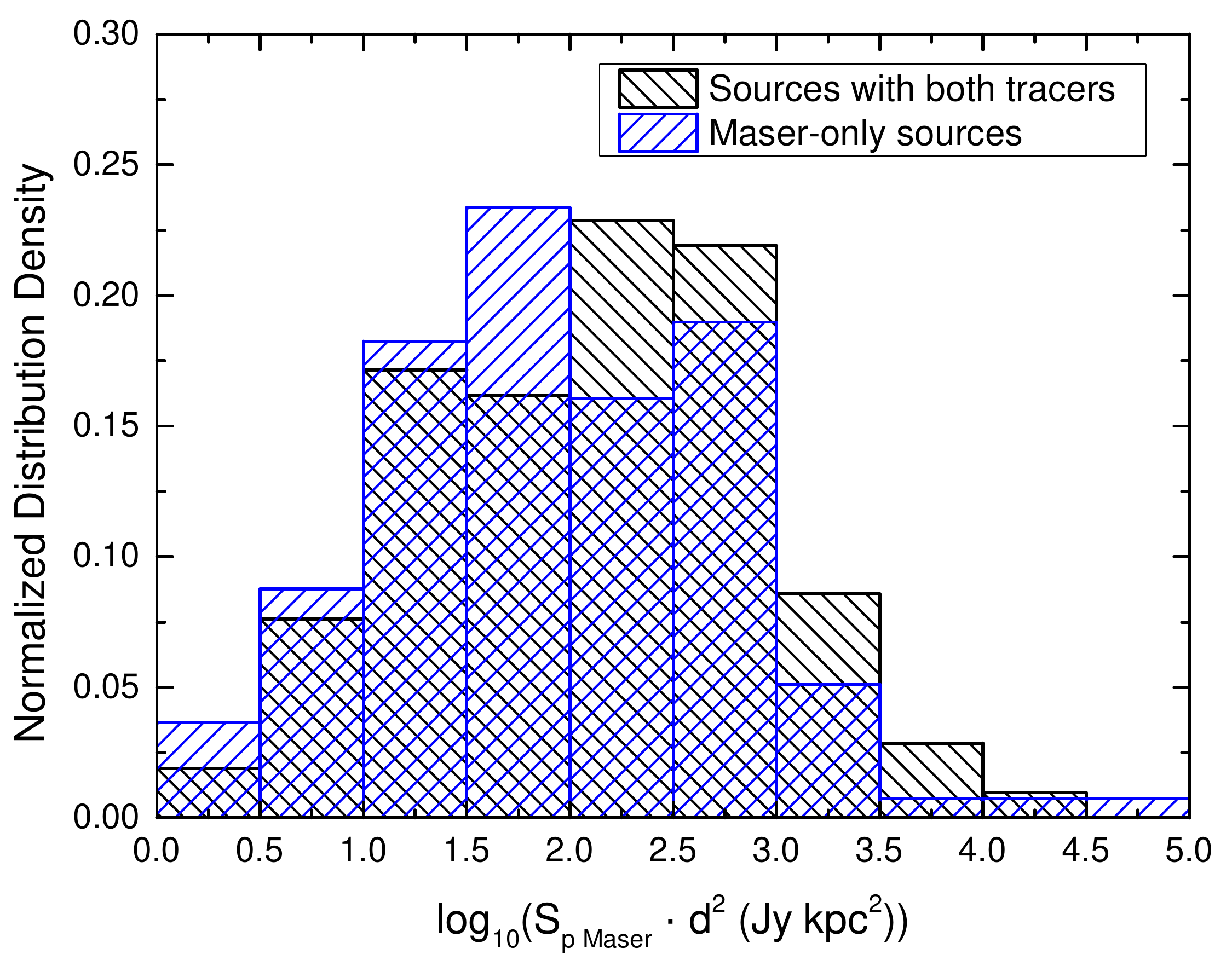}{0.45\textwidth}{(b)}}
\caption{(a) The normalized distributions of RRL peak line intensity corrected by distance for the 414 RRL-only sources and 103 sources associated with both RRL and methanol maser, (b) the normalized distribution of maser peak line intensity corrected by distance for the 137 methanol-maser-only sources comparing to that of the 103 sources associated with both RRL and methanol maser. \label{f7}}
\end{figure}

Figure \ref{f7} presents the normalized distributions of peak line intensity corrected by distance of our samples, which is defined by the peak line intensity times the square of Bayesian distance (see Section \ref{4.2.1} for more information). For the RRL sources, the peak line intensity used here are the main beam temperature (T$_{\rm mb}$) of the RRLs. The RRL-only sources have a mean RRL intensity of 2.88 $\pm$ 0.70 mK kpc$^2$, which is lower than the mean intensity of 3.29 $\pm$ 0.60 mK kpc$^2$ of the sources associated with both RRL and maser. Meanwhile, the maser-only sources and sources associated with both tracers have similar mean maser intensity of 1.95 $\pm$ 0.83 Jy kpc$^2$ and 2.11 $\pm$ 0.8 Jy kpc$^2$, respectively.

\citet{urquhart2014} performed similar analyses and found that clumps with methanol maser sources have lower bolometric luminosity than those with H{\footnotesize \RNum{2}} regions. The authors suggested that this is because earlier stage massive stars with methanol masers are more heavily embedded than those at later stages without methanol maser. Although there is no significant difference between our maser line intensity distribution of maser-only and sources associated with both RRL and maser (Figure \ref{f7}b), as shown in Figure \ref{f7}a, RRL-only sources have lower RRL line intensities than that of sources exhibiting methanol maser on average, and there is a lack of maser detection for RRL sources below 2.0 mK kcp$^2$. As illustrated by \citet{ouyang2019}, H{\footnotesize \RNum{2}} regions with methanol masers appear to have higher electron temperature and emission measure than those without, therefore methanol masers are more likely to be produced in regions with high gas densities and hence have a higher detection rate at more luminous H{\footnotesize \RNum{2}} regions.\\


\subsection{Galactic Distribution}

\subsubsection{RRL Spatial Distribution\label{4.2.1}}

\begin{figure*}[!thb]
\epsscale{1.2}
\plotone{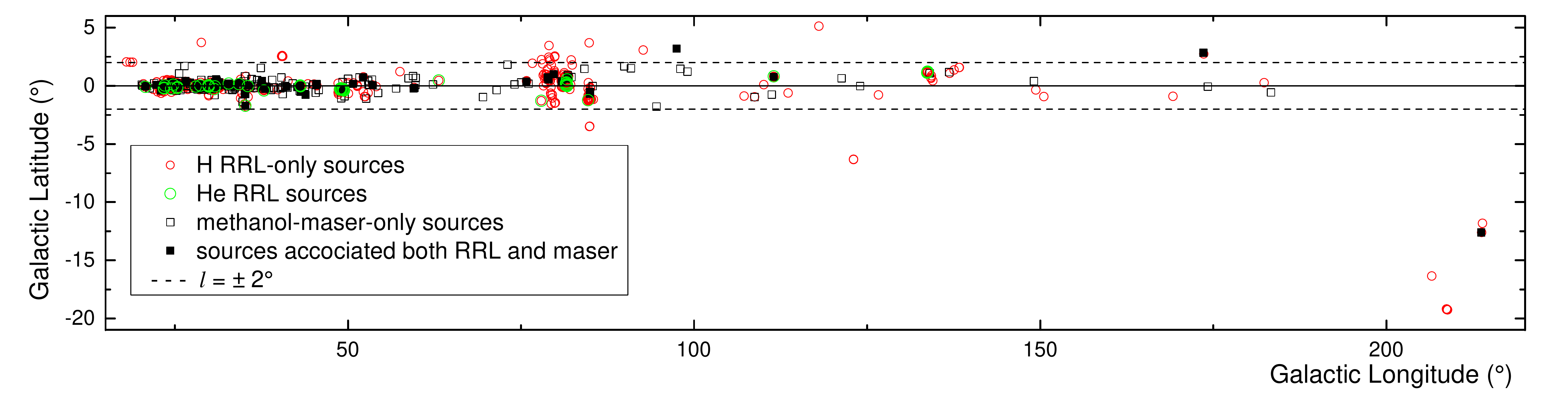}
\caption{Positional distribution of sources with only H RRL, only methanol maser, He RRL, and both RRL and methanol maser. The dashed lines denote where $| b |$ = 2$\degr$.\label{f8}\\}
\end{figure*}

\begin{figure*}[!thb]
\plotone{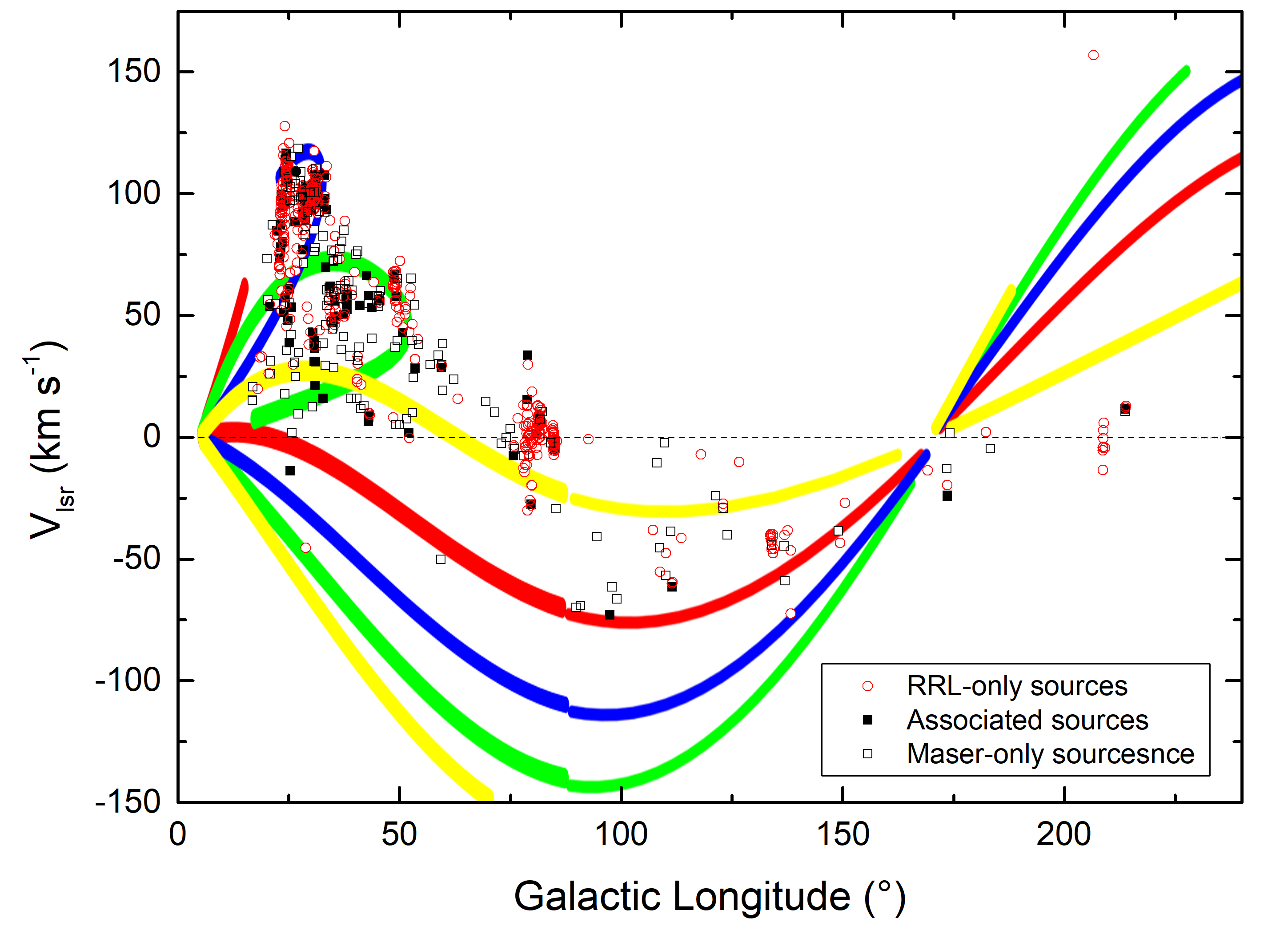}
\caption{LSR velocities of HMSFRs along Galactic longitude. The background curves showing trajectories of spiral arms’ radial velocities as a function of Galactic longitude were taken from \citet{vallee2008}. The spiral arms are colored by: red (Norma–Cygnus), blue (Scutum–Crux); green (Sagittarius–Carina) and yellow (Perseus).\label{f9}}
\end{figure*}

Figure \ref{f8} represents the Galactic latitude and longitude distributions of the H-RRL-only sources, methanol-maser-only sources, sources associated with both tracers, and He RRL sources (see Section \ref{4.3} for more details). As shown in this figure, a large fraction of the sources associated with both signals locate near the Galactic Plane, only 3 of them with a Galactic latitude larger than $\pm$ 2$\degr$, reflecting a sparse abundance in such regions.

The LSR velocities of the RRLs can be used to calculate the distance of its host H{\footnotesize \RNum{2}} regions using the Bayesian distance calculator built by \citet{reid2016}\footnote{http://bessel.vlbi-astrometry.org/bayesian}. \citet{reid2016} combines various types of distance information (spiral arm mode, kinematic distance, Galactic latitude, and parallax source) in a Bayesian approach, and fits the combined probability density function with multiple Gaussian components. The distance and error are estimated by the peak probability density and width of the Gaussian fitted component with maximum integrated probability density. The kinematic distances augmented by H{\footnotesize \RNum{1}} absorption spectra were used to resolve the near/far ambiguity for sources within the Solar circle. A user-adjustable prior probability (from 0 to 1) that the source is beyond the tangent point was set to the default value of 0.5. The result distances with error are presented in Column (9) and (10) in Table \ref{t3}.

In Figure \ref{f9}, we compare the velocity distribution along Galactic longitude of our HMSFRs to the longitude-velocity diagram retrieved from \citet{vallee2008}. For sources exhibiting multiple RRL components with different peak velocities, we only use the peak velocity of the strongest component to do the analysis. As shown in this figure, most of the sources located in the first quadrant and our data align with the model curves well in the first and second quadrants.\\

\subsubsection{Line Width Distribution\label{4.2.2}}
\begin{figure}[!tb]
\epsscale{1.1}
\plotone{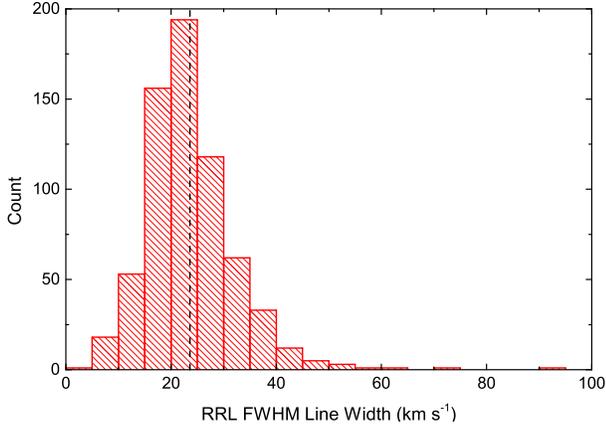}
\caption{Line width distribution of the detected H RRL components. The dashed line denotes the mean value of 23.6 km~s$^{-1}$\label{f10}}
\end{figure}

Figure \ref{f10} shows the RRL line width distribution of our H{\footnotesize \RNum{2}} region sample excluding those potential PNe, SNR and weak sources. These sources have an average line width with a standard deviation of 23.6 $\pm$ 2.0 km~s$^{-1}$, which is very close to that of the \citetalias{anderson2014} catalog (22.3 $\pm$ 5.3 km~s$^{-1}$).

Thermal broadened RRL has a line profile with an {\tt\string FWHM} width proportional to T$^{1/2}\nu_o$ \citep{BS1972}, where T is the temperature of the host H{\footnotesize \RNum{2}} region, $\nu_o$ is the rest central frequency. For a typical RRL-hosting source with T $\sim$ 5000 to 13000 K across the Galaxy \citep{balser2015}, the corresponding thermal-broadened line width is $\sim$ 15.3 to 24.6 km s$^{-1}$. In addition to thermal broadening, pressure broadening is also significant for RRLs at centimeter wavelengths \citep{keto2008}. The ratio between pressure-broadened and thermal broadened line width is proportional to $n_e N^7$, where $n_e$ is the electron density and $N$ is the principal quantum number of the RRL transition \citep{BS1972, grim1974, keto1995, keto2008}. As illustrated by \citet{keto2008}, since pressure broadening is proportional to the density of host H{\footnotesize \RNum{2}} region and less significant at high frequencies, comparing the line widths of RRLs measured with high-resolution interferometer in various wavelengths can be used to measure the electron density of the natal H{\footnotesize \RNum{2}} region.


RRLs with a line width narrower than 15 km s$^{-1}$ are likely from comparably cold, sparse H{\footnotesize \RNum{2}} regions and are broadened purely by thermal, thus they can be used to probe the temperature of the host H{\footnotesize \RNum{2}} regions. There are 69 sources in our sample containing H RRLs with a line width of $<$ 15 km~s$^{-1}$, for thermally broadened RRL source with such narrow line width, the host H{\footnotesize \RNum{2}} regions have an upper limit to the temperature of $\lesssim 5000$ K. According to \citet{shaver1970} and \citet{shaver1979}, RRLs with such narrow line width are from cold nebulae. Another possible explanation for the narrow line width is that the observed RRL line width is underestimated due to radiative transfer effect caused by strong free-free emission, which typically has a larger optical depth.


We detected 51 sources with a very broad line width ($>$ 35~km~s$^{-1}$). Out of these sources, 22 have a line width broader than 40 km~s$^{-1}$. RRL with a line width $>$ 40 km~s$^{-1}$ may be very dense H{\footnotesize \RNum{2}} regions. For these RRLs, there may also exist large-scale motions around the central young stars (\citealt{sewilo2004, ridge2001, RB2001} and references therein) such as nebular expansion, rotation of the inner parts of the accretion disk, infall of matter, shocks, bipolar jets or photo evaporating flows.

Some of those broad line sources are possibly previously unknown PNe or SNR sources. As mentioned in Section \ref{3.2}, since PNe generally have larger line width due to expansion, their existence may affect the RRL line width distribution of our sample. For example, one of the sources, G28.393+0.085, with the widest line width in our sample was identified as a known H{\footnotesize \RNum{2}} region (G028.394+00.076) in \citet{anderson2011}. However, its RRL emission with extremely large line width is potentially from an unknown PNe or SNR source along the LOS. In addition, blended multiple RRL components with very close peak velocities may also cause confusion with a single wide line component.\\




\subsection{Helium and Carbon RRLs\label{4.3}}
\subsubsection{Helium RRL Detections\label{4.3.1}}

\begin{figure}[!tb]
\epsscale{1.1}
\plotone{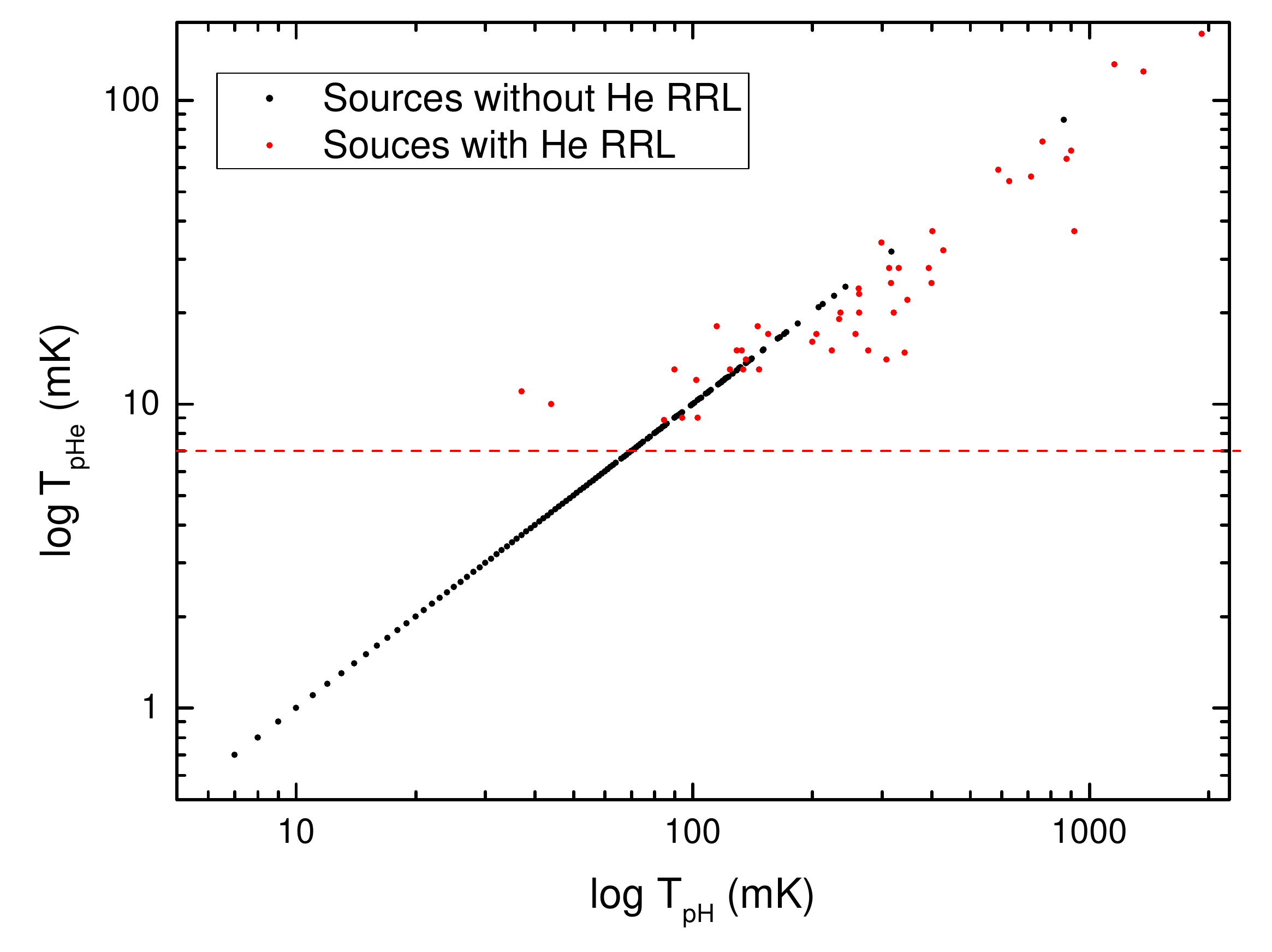}
\caption{The peak temperature distribution of He RRLs versus the peak temperature of H RRLs. The red and black dots represent sources with and without He RRL detections. For sources without He RRL detections, the expected $T_{p {\rm He}}$ values are calculated by multiplying their $T_{p {\rm H}}$ by the mean $T_{p {\rm He}}$ to $T_{p {\rm H}}$ ratio of 0.1. The dashed red line denotes the detection threshold of 7 mK. H-RRL-only sources with $T_{p {\rm H}}$ $>$ 70 mK are likely to have no or very weak He RRL emissions. \label{f11}}
\end{figure}

In addition to H RRLs, there are 49 sources also exhibit He RRLs. Their spatial distribution is shown in Figure \ref{f8}. He RRLs are believed to have the same origin with H RRLs. The atomic heliums are ionized by the UV photons emitted by a central O6 or hotter star \citep{mezger1978, roshi2017}. Since higher ionizing energy is needed for He RRLs, they are generally from stars that are more massive than those emit H RRL only. He RRLs are usually weaker than H RRLs, they can only be detected from sources with strong H RRL intensities. In our sample, the mean peak temperature of H RRL from the sources with He RRL emissions is 374.7 mK, which is $\sim$ 8 times brighter than those without He RRL detections. The detected He RRLs have a mean peak temperature of 31.6 mK and an average value of $T_{p {\rm He}}$/$T_{p {\rm H}}$ $\sim$ 0.1, where $T_{p {\rm He}}$ and $T_{p {\rm H}}$ denote the peak temperatures of He and H RRLs. Due to sensitivity limitation, He RRLs with peak temperature lower than $\sim$ 7 mK are below our detection threshold, this results in a lower limit to the $T_{p {\rm H}}$ of 70 mK for He RRL to be detected. Figure \ref{f11} shows the plot between $T_{p {\rm He}}$ and $T_{p {\rm H}}$. For sources without He RRL detections, the expected $T_{p {\rm He}}$ values are calculated by multiplying their $T_{p {\rm H}}$ by 0.1. There are 76 sources which have an H RRL peak temperature above 70 mK (black dots above the red dashed line in Figure \ref{f11}) but had no He RRL detections, showing a low abundance of He$^+$.

\begin{figure}[!tb]
\gridline{\fig{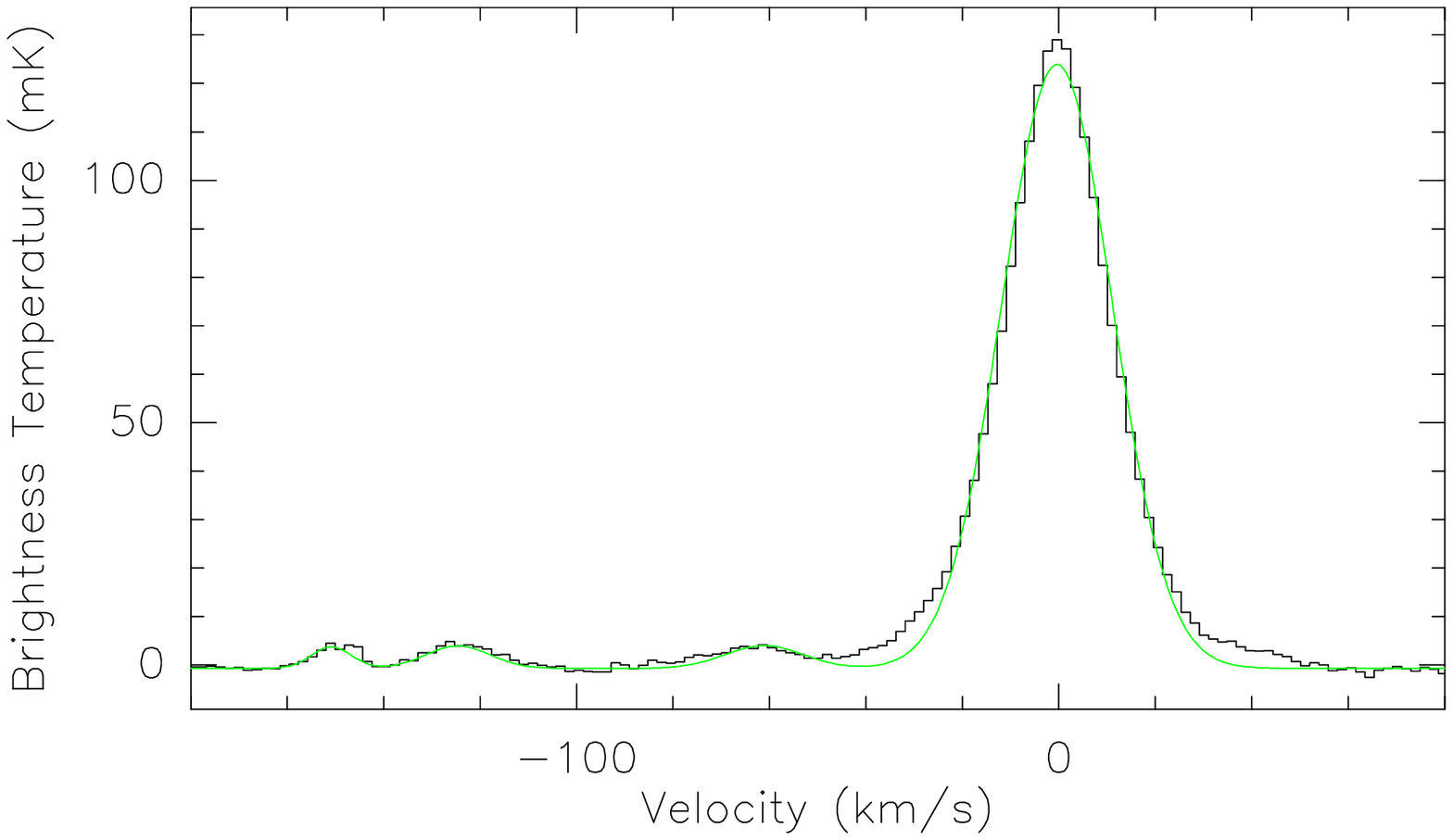}{0.45\textwidth}{(a)}}
\gridline{\fig{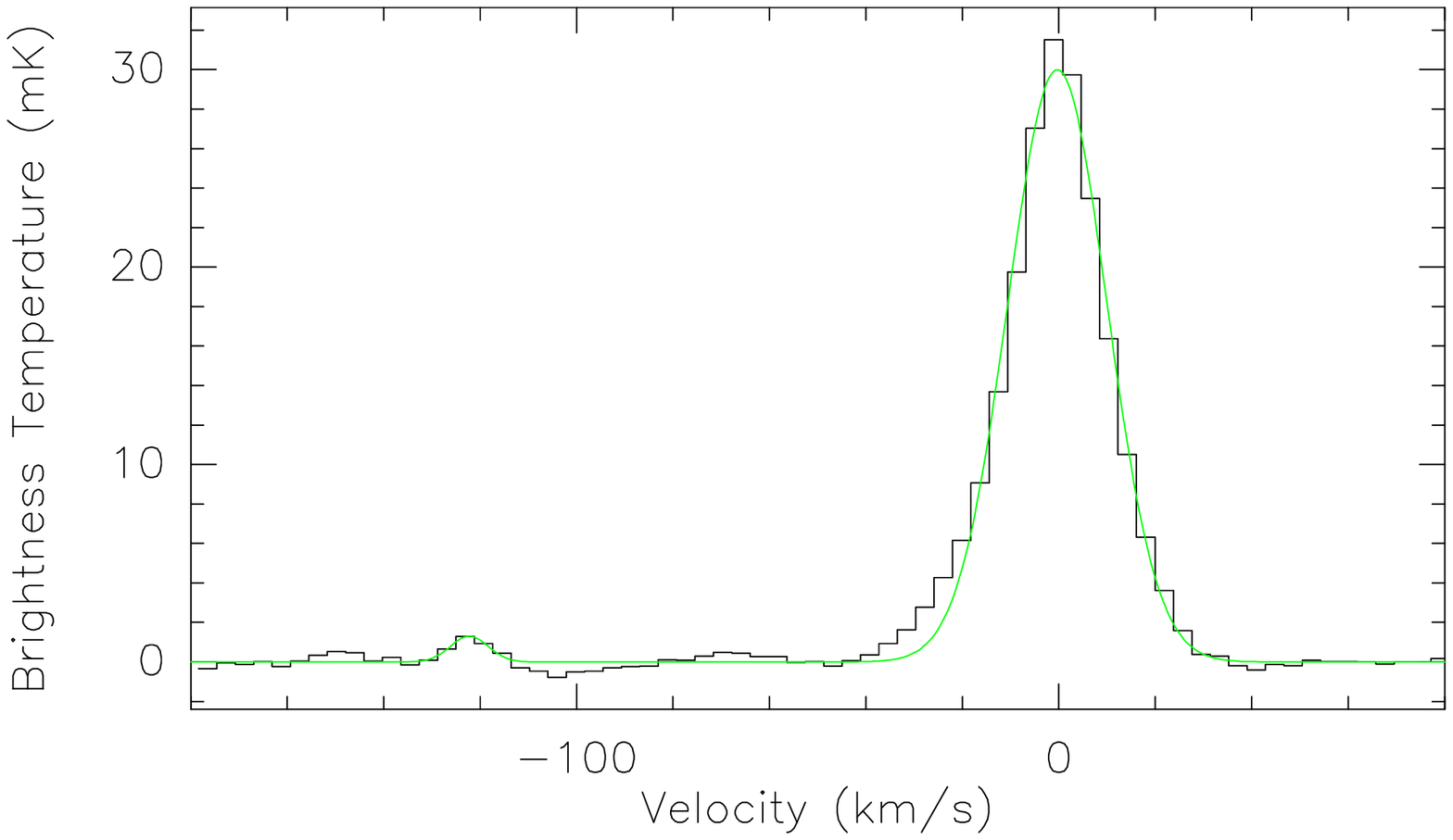}{0.45\textwidth}{(b)}}
\caption{The averaged spectra over sources without He RRL detection with (a) $T_{p {\rm H}}$ $>$ 70 mK and (b) $T_{p {\rm H}}$ $<$ 70 mK. The peak velocity of each individual spectrum was aligned and set to zero. He RRLs in (a) and (b) appear at $-$122 km s$^{-1}$. The C RRL ($-$149 km s$^{-1}$) and H$\beta$ line ($-$60 km s$^{-1}$) can also be seen clearly in (a).\label{f12}}
\end{figure}

We averaged over the H-RRL-only sources with $T_{p {\rm H}}$ above and below 70 mK, as shown in Figure \ref{f12}. After averaging, He RRLs can be seen with a $S_{i {\rm He}}$ to $S_{i {\rm H}}$ ratio of $\sim$ 0.01 and $\sim$ 0.02 for sources above and below 70 mK, respectively, where $S_{i {\rm He}}$ and $S_{i {\rm H}}$ denote the integrated intensities of He and H RRLs. These values are much lower than the mean $S_{i {\rm He}}$ to $S_{i {\rm H}}$ ratio of 0.06 for sources with He RRL detections. This fact may suggest that in addition to the selecting effect caused by the sensitivity limit, the non-detections of He RRLs for the H-RRL-only sources may be mostly originated from the low abundance of He$^+$. Those sources are possibly being ionized by a less massive, later OB-type star. That is, helium is under ionized with respect to hydrogen.\\

\begin{figure}[!tb]
\gridline{\fig{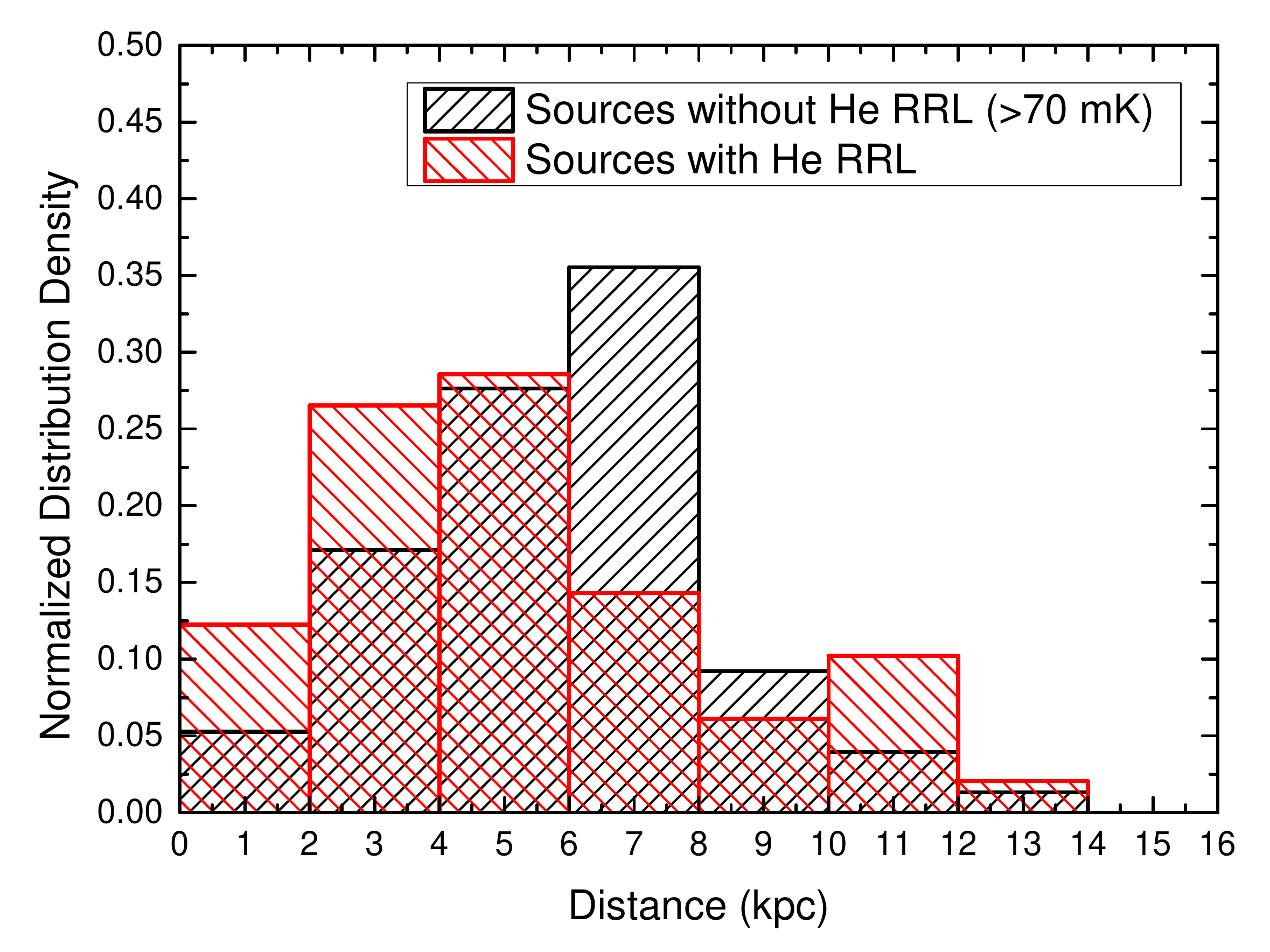}{0.45\textwidth}{(a)}}
\gridline{\fig{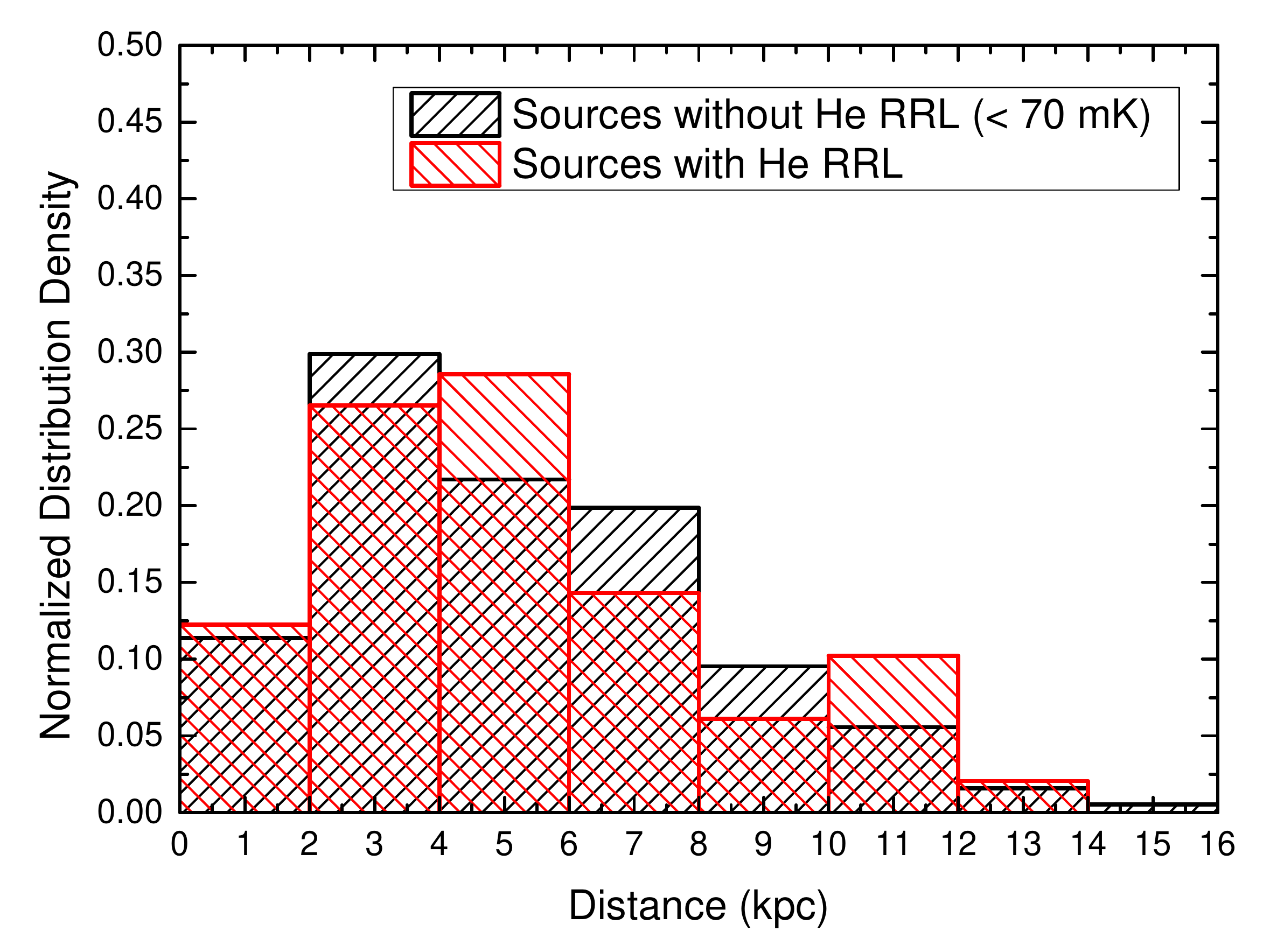}{0.45\textwidth}{(b)}}
\caption{The distance distribution of (a) 76 H-RRL-only sources with $T_{p {\rm H}}$ $>$ 70 mK and (b) 380 sources with $T_{p {\rm H}}$ $<$ 70 mK with respect to the distribution of 49 He RRL sources. There is a slight higher distance distribution on the H-RRL-only sources in (a) and no significant difference found in (b).\label{f13}}
\end{figure}

\begin{figure}[!tb]
\epsscale{1.1}
\plotone{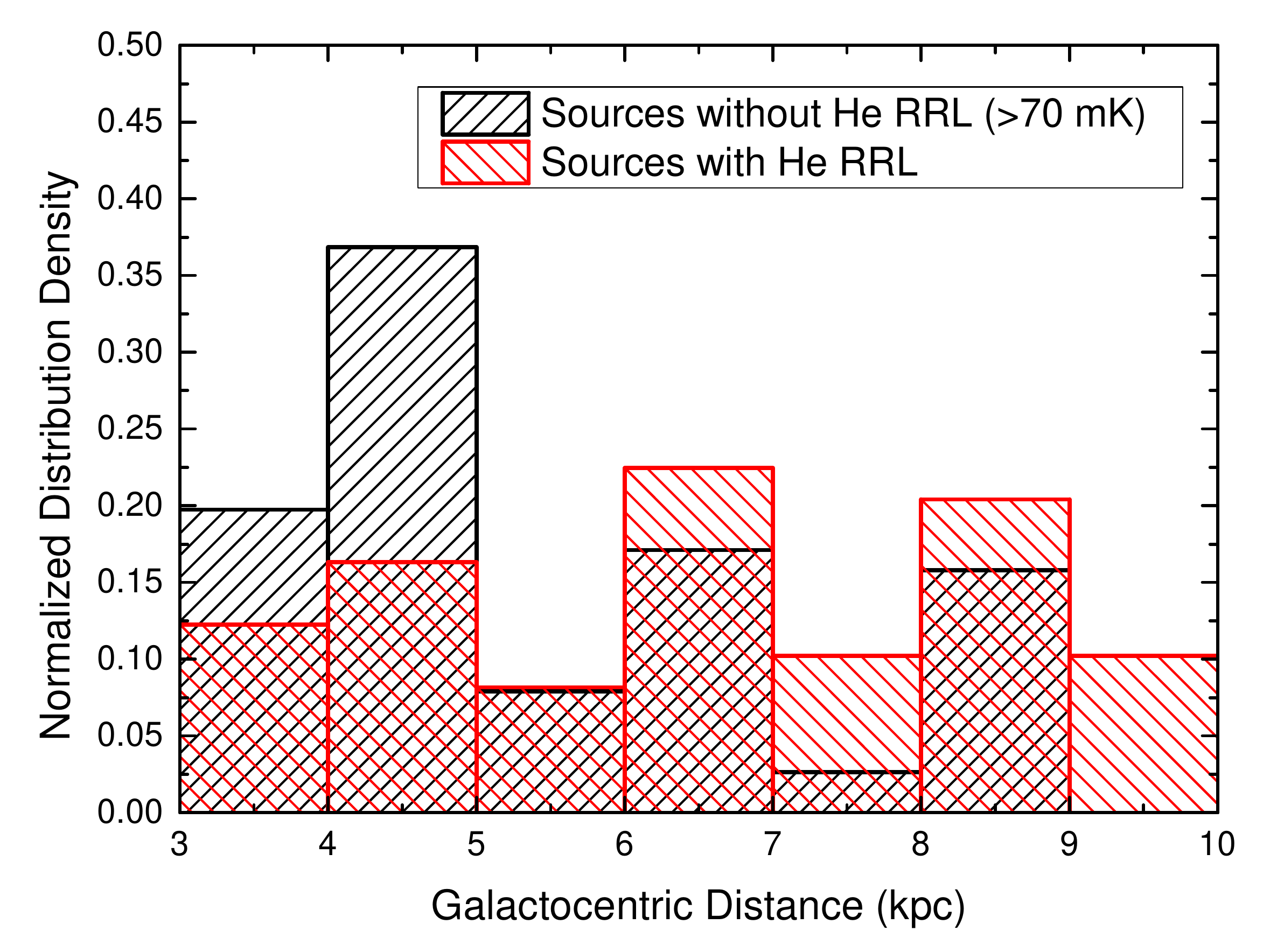}
\caption{Normalized galactocentric distance distribution density of the 49 He sources and the 76 H-RRL-only sources with $T_{p {\rm H}}$ above 70 mK.\label{f14}}
\end{figure}

\subsubsection{Distance and Galactocentric Distance Distribution}

Figure \ref{f13} shows the normalized distance distribution of the 76 H-RRL-only sources with $T_{p {\rm H}}$ $>$ 70 mK and 380 sources (excluding the weak sources, and for multi-component sources, only the strongest emission were taken into consideration.) with $T_{p {\rm H}}$ $<$ 70 mK comparing to the 49 sources with He RRL. No significant difference in the distance distributions were found for the two sub-samples in this figure. This fact may support the argument that the non-detections of He RRLs are mainly caused by the low abundance of He$^+$ rather than sensitivity limit caused by distance.

Figure \ref{f14} shows the normalized $R_{{\rm Gal}}$ distribution density of the 49 He RRL sources and the 76 H-RRL-only sources with $T_{p {\rm H}}$ above 70 mK. As shown in this figure, sources without He RRL emissions locate nearer to the Galactic center than those with He RRLs on average. Under the above assumption that He RRL sources are being ionized by a more massive star with respect to those without He RRL, we suggest that more massive SFRs locate at longer distances on average. This is contradicting to previous studies that due to higher gas density, there is a concentration of mass in the inner Galaxy (e.g., \citealt{green2011, casassus2000, lepine2011}). A more plausible explanation for the lower abundance of ionized He is line-blanketing effect caused by higher metallicity in such regions. There is a known negative metallicity radial gradient along the Galactic disk \citep{HW1999}, the higher metal content in the atmosphere of OB-type stars near the Galactic Center will cause line-blanketing and reduce the number of He-ionizing photons that escape the star. This may result in a lower abundance of ionized He in such regions. Nevertheless, both Figures \ref{f13} and \ref{f14} may suffer from statistical bias due to limited sample size.\\

\subsubsection{He$^+$ Abundance along the Galactic Plane\label{4.3.3}}

H and He RRLs with high principal quantum numbers in the radio act similarly, so their line intensity ratio y$^+$ can be used to diagnose the abundance ratio between $^4$He$^+$ and H$^+$ \citep{wenger2013, balser2006}. y$^+$ is defined as the following:

\begin{equation}
{\rm y}^+ = \frac{T_{p {\rm He}^+}\Delta \nu_{{\rm He}^+}}{T_{p {\rm H}^+}\Delta \nu_{{\rm H}^+}}
\end{equation}

where $T_{p {\rm He}^+}$ and $T_{p {\rm H}^+}$ are the peak line temperatures of He and H RRLs, $\Delta \nu_{{\rm He^+}}$ and $\Delta \nu_{{\rm H^+}}$ are the line widths of them, respectively. 


\begin{figure}[!tb]
\epsscale{1.1}
\plotone{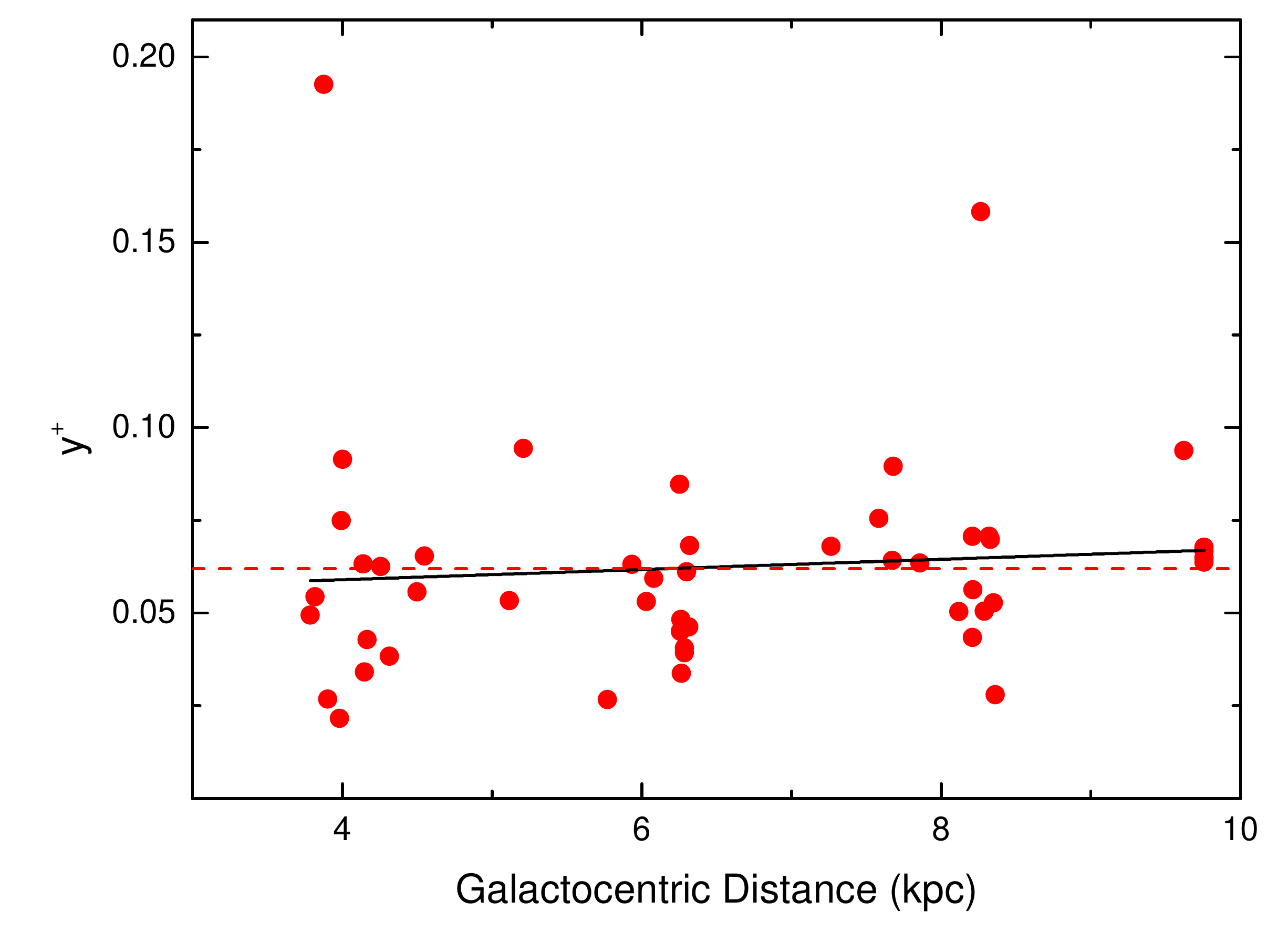}
\caption{The distribution of y$^+$ values along the galactocentric distance for the sources which exhibit He RRLs. The solid black line shows a linear fit to the sample. The dashed red line mark the location of the mean y$^+$ of 0.062 $\pm$ 0.029.\label{f15}}
\end{figure}

The distribution of y$^+$ values along the Galactic Plane for the sources exhibiting He RRLs is presented in Figure \ref{f15}. Our sample has a mean y$^+$ value with a standard deviation of 0.062 $\pm$ 0.029, this value is similar to that of the \citetalias{anderson2014} catalog (0.068 $\pm$ 0.023) measured in \cite{wenger2013}. Our sample shows no significant trend on y$^+$ with $R_{{\rm Gal}}$, as presented by the black fitting line in Figure \ref{f15}, a small positive slope of 0.002 $\pm$ 0.015 is derived for the sources. There is a very low correlation coefficient of $\sim$ 0.09 between y$^+$ and $R_{{\rm Gal}}$, and a large standard error of $\pm$ 0.015 to the fitting line slope, both showing weak dependence of y$^+$ on $R_{{\rm Gal}}$. This is consistent with earlier studies that y$^+$ has a negative or no obvious gradient with $R_{{\rm Gal}}$ through our Galaxy (see \citealt{balser2001} for a review). A weak increasing trend was also found on y$^+$ (y$^+$ = 0.0035 $\pm$ 0.0016 kpc$^{-1}$) in \cite{wenger2013}, however, they pointed out that this result only weakly constrains the actual y$^+$ gradient due to the large uncertainties in their data. Due to the limited sample size, our analysis may also suffer from statistical bias.


There are two sources with prominently high y$^+$ values ($>$ 0.15) (G23.563+0.008 and G84.722$-$1.248; see Figure \ref{f15}), as concluded by \citet{balser2001}, possible origins for the high y$^+$ value are mass loss of helium near the surface of the massive star or overestimated abundance due to the radiative transfer effect.\\



\subsubsection{MIR Color Distribution\label{4.3.4}}
\begin{figure}[!thb]
\gridline{\fig{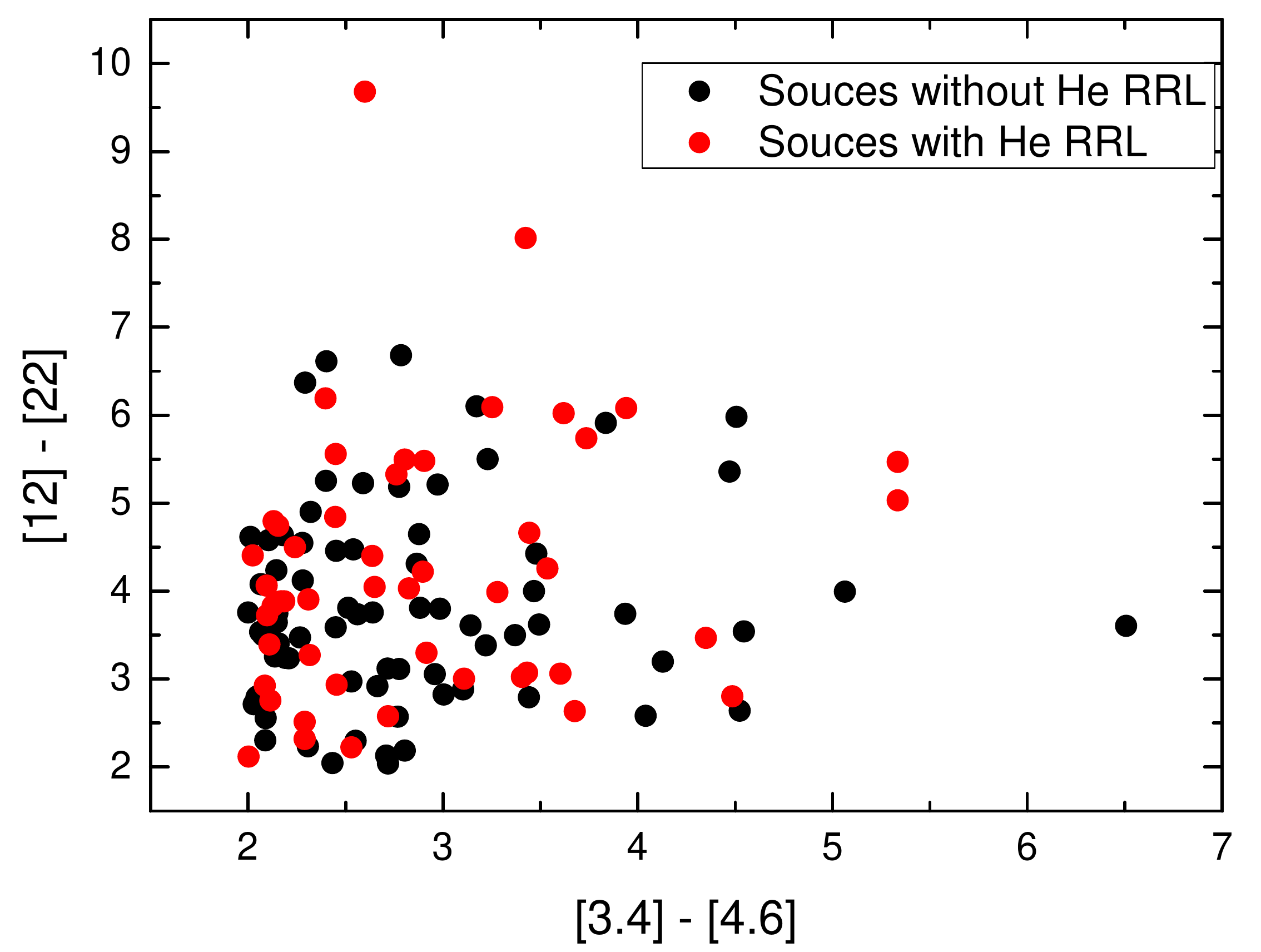}{0.45\textwidth}{(a)}}
\gridline{\fig{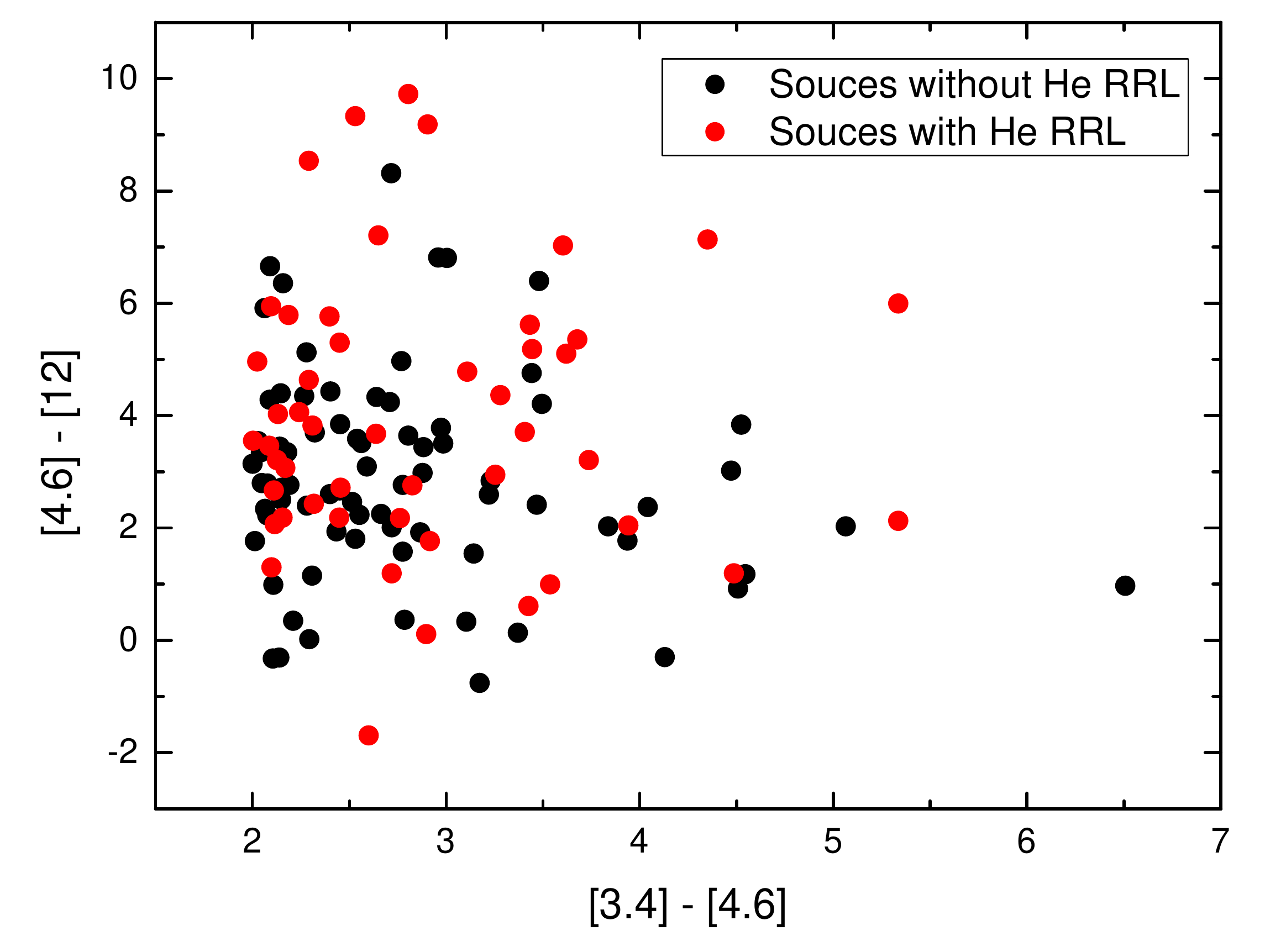}{0.45\textwidth}{(b)}}
\caption{The \textit{All-WISE} (a) [12] $-$ [22] vs. [3.4] $-$ [4.6] and (b) [4.6] $-$ [12] vs. [3.4] $-$ [4.6] colors of sources with and without He RRL emission. For the He-undetected sources, only those with $T_{p {\rm H}}$ above 70 mK are included in this figure. Both (a) and (b) show no significant difference in the color distribution of the two sub-samples.\label{f16}}
\end{figure}

Due to the emission from the dust heated by the central star, one would expect higher luminosities at longer MIR wavelength from more massive SFRs, thus there may exist a difference between the color-color distribution of H{\footnotesize \RNum{2}} regions with different masses of their exciting stars. Figure \ref{f16} shows the \textit{WISE} [12] $-$ [22] vs. [3.4] $-$ [4.6] and [4.6] $-$ [12] vs. [3.4] $-$ [4.6] color distributions of sources with and without He RRL detections (above 70 mK). However, there is no significant separation found between the color distributions of these two sub-samples. This may suggest that \textit{WISE} colors are insensitive to mass variation of UCH{\footnotesize \RNum{2}} regions.\\

\subsubsection{Carbon RRLs\label{4.3.5}}



Carbon RRLs are from cooler gas in photo-dissociation regions (PDRs) or diffuse gas ionized by the interstellar UV radiation (see \citealt{alves2015} and references therein). For a target position with both H and C RRLs, if the C RRL is emitted from a cold gas with a different velocity to the H{\footnotesize \RNum{2}} region, it can have a shifted velocity offset with respect to the H RRLs \citep{alves2012}. There are 23 sources showing C RRL in our sample, 3 of them are from the nearby Orion Molecular Cloud Complex (G208.894$-$19.313; G213.706$-$12.602; G213.885$-$11.832). The line properties of the 23 C RRLs are presented in Table \ref{t5}.


C RRLs are essential to the studies of morphology and physical properties of its host PDR (e.g., \citealt{RA2001, roshi2002, wenger2013}). For example, the non-thermal component of the carbon RRL line width can be used to diagnose the magnetic field in the host PDR \citep{roshi2007, balser2016}. For a thermally broadened C RRL, assuming a typical PDR temperature of $\sim$ 10$^3$ K, a line width of $\sim$ km~s$^{-1}$ is expected. However, the C RRLs in our sample have an average line width of 8.9 km~s$^{-1}$. The large average line width of C RRLs may indicate that there may exist non-thermal turbulence in the PDRs in our sample (\citealt{roshi2007} and \citealt{barrett1964}).\\



\section{Summary\label{5}} 

(1) Using the TMRT, we performed a Galactic RRL survey in C band toward 3348 targets, selected from the \textit{WISE} point source catalog. Excluding 5 potential PNe and 5 potential SNR candidates, we built a sample of 517 HMSFRs traced by RRL. The peak flux densities of the detected hydrogen RRLs are in a range of $\sim$ 10 to 1900 mK. Though the majority of the sources have a line width within 15 $\sim$ 35 km~s$^{-1}$, there are 82 of them show very narrow line width characteristic and 30 of them have line width larger than 35 km~s$^{-1}$. Our sample further expanded the H{\footnotesize \RNum{2}} region sample on the basis of previous surveys.

(2) Within the detected H{\footnotesize \RNum{2}} region sample, 103 sources also harbor 6.7 GHz methanol masers emissions. Combining the sources traced by RRL and/or maser, we built up a sample of 654 HMSFRs, providing fundamental information to study the high-mass star formation evolutionary stages. According to the argument that methanol maser appears earlier than the formation of H{\footnotesize \RNum{2}} region, our sources may be associated with HMSFR at various star-forming sequences. By comparing the physical properties of the RRL-only sources, maser-only sources and sources associated with both tracers, we found no significant difference in distance and $R_{{\rm Gal}}$ distribution of the three sub-samples. A slightly higher maser association rate was found for more luminous RRL sources.

(3) In addition to H RRL, we also detected He RRLs from 49 sources which may associate with more massive HMSFRs, no significant gradient on the He/H abundance along the galactocentric distance was found from the 49 He RRL sources. 

(4) A sample of 23 C RRLs were also built in this survey, which provides a promising sample for future studies on the physical properties of PDRs surrounding H{\footnotesize{\RNum{2}}} regions, the wide average line width of C RRLs may indicate non-thermal turbulence in their host PDRs.\\

\section*{Acknowledgement}

We are thankful for the assistance from the operators of the TMRT during the observations, the funding and support from China Scholarship Council (CSC) (File No.201704910999), Science and Technology Facilities Council (STFC) and the University of Manchester. This work was supported by the National Natural Science Foundation of China (11590781, 11590783, 11590784 and 11873002) , and Guangdong Province Universities and Colleges Pearl River Scholar Funded Scheme (2019).

\software{GILDAS/CLASS\citep{pety2005,gildas2013}}

\bibliography{ref}
\bibliographystyle{apj}


\floattable
\begin{longrotatetable}
\floattable
\begin{deluxetable}{p{1.8cm}p{1.6cm}p{1.5cm}p{1cm}p{0.6cm}p{1.6cm}p{1.6cm}p{1.6cm}p{0.5cm}p{0.5cm}p{1.5cm}p{0.6cm}p{3cm}}
\centering
\tablecaption{Characteristics of the Hydrogen RRL emission sample.\label{t3}}
\tablewidth{700pt}
\tabletypesize{\scriptsize}
\tablehead{
\colhead{Name} & \colhead{R.A.} & 
\colhead{Decl.} & \colhead{Epoch} & 
\colhead{T$_b$} & \colhead{S$_i$} & 
\colhead{$\Delta$V} & \colhead{$\nu$$_p$} & 
\colhead{d} & \colhead{$\sigma$$_d$} & \colhead{Notes} & \colhead{Maser} & \colhead{Other Name(s)}\\ 
\colhead{(l,b)} & \colhead{(J2000)} & \colhead{(J2000)} & \colhead{(dd/mm/yy)} & 
\colhead{} & \colhead{} & \colhead{} &
\colhead{} & \colhead{} & \colhead{}& \colhead{} & \colhead{} & \colhead{}\\
\colhead{($\degr$,$\degr$)} & \colhead{(h m s)} & \colhead{($\degr$ $\arcmin$ $\arcsec$)} & \colhead{} & 
\colhead{(mK)} & \colhead{(K km s$^{-1}$)} & \colhead{(km s$^{-1}$)} &
\colhead{(km s$^{-1}$)} & \colhead{(kpc)} & \colhead{(kpc)} & \colhead{} & \colhead{} & \colhead{}\\
} 
\colnumbers
\startdata
G18.059+2.035 & 18:16:27.60 & $-$12:14:45.7 & 22/04/16 & 26 & 0.53(0.05) & 19.4(2.0) & 19.7(0.8) & 1.69 & 0.24  &  &  & (K) G018.426+01.922\\
G18.559+2.029 & 18:17:26.97 & $-$11:48:32.0 & 15/12/17 & 49 & 1.26(0.05) & 24.4(1.1) & 32.7(0.5) & 1.77  & 0.26  &  &  & (K) G018.426+01.922 / IRAS 18146-1148\\
G18.915+2.027 & 18:18:08.75 & $-$11:29:45.2 & 15/12/17 & 31 & 0.42(0.04) & 12.7(1.3) & 33.1(0.5) & 1.77  & 0.26  &  &  & (K) G018.426+01.922 / IRAS 18153-1131\\
G20.495+0.157 & 18:27:54.06 & $-$10:58:37.3 & 04/07/16 & 52 & 1.10(0.04) & 19.7(0.9) & 26.1(0.4) & 14.01  & 0.31  &  &  & (K) G020.481+00.168\\
G20.749$-$0.112 & 18:29:21.28 & $-$10:52:38.5 & 04/07/16 & 263 & 7.54(0.06) & 26.9(0.2) & 54.5(0.1) & 3.46  & 0.20  &  &  & (K) G020.728-00.105 / Kes 68\\
G20.762$-$0.064 & 18:29:12.19 & $-$10:50:36.0 & 04/07/16 & 243 & 5.52(0.06) & 21.3(0.2) & 53.5(0.1) & 3.45  & 0.20  &  & yes & (K) G020.728-00.105 / Kes 68\\
G21.919$-$0.324 & 18:32:19.23 & $-$09:56:19.9 & 04/07/16 & 41 & 0.85(0.05) & 19.5(1.4) & 83.2(0.6) & 5.27  & 0.32  &  &  & (K) G021.884-00.318 / IRAS 18294-0959\\
G22.355+0.066 & 18:31:44.05 & $-$09:22:17.3 & 15/07/16 & 39 & 0.80(0.05) & 19.1(1.3) & 84.7(0.6) & 5.41  & 0.37  &  & yes & (K) G022.357+00.064\\
G22.396+0.334 & 18:30:50.75 & $-$09:12:39.8 & 15/07/16 & 21 & 0.27(0.03) & 12.1(1.6) & 82.9(0.7) & 5.38  & 0.40  &  &  & (C) G022.424+00.337\\
\enddata
\tablecomments{Properties of the H RRL emissions from the HMSFR sample with (1) source name, (2)(3) the pointing center coordinates, (4) epoch of observations, (5) measured peak antenna temperature of a Gaussian fitting, (6) integral flux of the Gaussian fitting with error, (7) ({\tt\string FWHM}) of the Gaussian fitting with error, (8) radio defined LSR velocity at the peak value with error, (9)(10) Bayesian distance and error calculated following \citet{reid2016} with a default equal weighting ($P_{far}$ = 0.5) on near and far distance probability, (11) complementary information: ``SL" for side-lobe detected sources, ``?" for weak sources without accurate Gaussian line fitting, (12) association with 6.7 GHz methanol maser, (13) other name(s) of the sources, for sources associated with an \citetalias{anderson2014} source, we label the type of the \citetalias{anderson2014} counterpart: ``K" for ``Known" sources, ``C" for ``candidate" sources, ``G" for ``Grouped" sources and ``Q" for ``Radio quiet" sources. For multi-component sources, the components are sorted by peak line intensity.\\}
\tablecomments{(This table is available in its entirety in a machine-readable form in the online journal and the source files on arXiv.org. A portion is shown here for guidance regarding its form and content.)}
\end{deluxetable}
\end{longrotatetable}%

\begin{longrotatetable}
\floattable
\begin{deluxetable*}{lllllllll}
\centering
\tablecaption{List of HMSFR sources traced by RRL and/or methanol maser.\label{t4}}
\tablewidth{700pt}
\tabletypesize{\scriptsize}
\tablehead{
\colhead{l($\degr$), b($\degr$)} & \colhead{l($\degr$), b($\degr$)} & \colhead{l($\degr$), b($\degr$)} & \colhead{l($\degr$), b($\degr$)} & \colhead{l($\degr$), b($\degr$)} & \colhead{l($\degr$), b($\degr$)} & \colhead{l($\degr$), b($\degr$)} & \colhead{l($\degr$), b($\degr$)} & \colhead{l($\degr$), b($\degr$)} 
} 
\startdata
\multicolumn{9}{c}{414 RRL-only sources}\\
\hline
G18.059+2.035 & G18.559+2.029 & G18.915+2.027 & G20.495+0.157 & G20.749$-$0.112 & G21.919$-$0.324 & G22.396+0.334 & G22.397+0.300 & G22.551$-$0.522 \\ 
G22.835$-$0.438 & G22.873$-$0.258 & G22.877$-$0.432 & G22.953$-$0.358 & G23.009$-$0.379 & G23.039$-$0.641 & G23.096$-$0.413 & G23.172$-$0.183 & G23.240$-$0.114 \\ 
G23.241$-$0.481 & G23.271$-$0.139 & G23.315$-$0.184 & G23.323$-$0.294 & G23.338$-$0.213 & G23.351$-$0.139 & G23.386$-$0.130 & G23.402+0.450 & G23.416$-$0.108 \\ 
G23.428$-$0.231 & G23.431$-$0.519 & G23.458+0.066 & G23.458$-$0.016 & G23.465+0.115 & G23.473$-$0.212 & G23.490$-$0.028 & G23.538$-$0.004 & G23.563+0.008 \\ 
G23.601$-$0.015 & G23.696+0.167 & G23.740+0.157 & G23.771+0.149 & G23.823+0.135 & G23.868$-$0.117 & G23.873+0.086 & G23.899$-$0.268 & G23.900+0.520 \\ 
G23.929+0.499 & G23.959+0.405 & G23.964+0.168 & G23.995$-$0.097 & G24.010+0.503 & G24.113$-$0.176 & G24.132+0.123 & G24.191$-$0.036 & G24.235$-$0.223 \\ 
G24.273$-$0.137 & G24.283$-$0.009 & G24.323+0.047 & G24.349+0.020 & G24.351$-$0.269 & G24.359+0.127 & G24.393+0.013 & G24.425+0.243 & G24.426+0.351 \\ 
G24.427+0.122 & G24.443$-$0.228 & G24.470+0.464 & G24.479$-$0.250 & G24.499+0.390 & G24.519$-$0.111 & G24.520$-$0.565 & G24.546$-$0.245 & G24.554+0.503 \\ 
G24.564$-$0.308 & G24.615+0.421 & G24.626$-$0.101 & G24.639$-$0.030 & G24.730+0.153 & G24.775+0.118 & G24.811+0.056 & G24.818$-$0.108 & G24.820+0.158 \\ 
G24.826$-$0.073 & G24.848$-$0.102 & G24.865+0.145 & G24.910+0.037 & G25.141$-$0.400 & G25.156$-$0.273 & G25.229+0.175 & G25.356+0.263 & G25.383$-$0.377 \\ 
G25.392$-$0.131 & G25.664$-$0.120 & G26.087$-$0.055 & G26.327+0.307 & G26.331+0.134 & G26.353+0.010 & G26.374+0.246 & G26.526+0.381 & G26.579$-$0.120 \\ 
G26.580+0.080 & G26.854$-$0.077 & G26.902$-$0.306 & G26.983$-$0.228 & G27.028+0.283 & G27.069+0.072 & G27.180$-$0.004 & G27.185$-$0.082 & G27.977+0.078 \\ 
G28.063$-$0.085 & G28.150+0.169 & G28.188$-$0.212 & G28.231+0.039 & G28.264$-$0.182 & G28.280$-$0.154 & G28.291+0.010 & G28.328$-$0.075 & G28.342+0.101 \\ 
G28.393+0.085 & G28.413+0.145 & G28.439+0.035 & G28.452+0.002 & G28.577$-$0.333 & G28.579+0.144 & G28.585+3.712 & G28.688$-$0.278 & G28.692+0.028 \\ 
G28.747+0.270 & G28.757+0.059 & G28.852+3.701 & G28.855$-$0.219 & G28.920$-$0.228 & G28.928+0.019 & G29.119+0.029 & G29.414+0.185 & G29.609+0.197 \\ 
G29.725+0.074 & G29.780$-$0.260 & G29.833$-$0.261 & G29.873+0.032 & G29.887$-$0.005 & G29.887$-$0.779 & G29.939$-$0.870 & G29.941$-$0.071 & G30.030$-$0.383 \\ 
G30.103$-$0.079 & G30.136$-$0.228 & G30.145$-$0.067 & G30.197+0.309 & G30.301$-$0.203 & G30.338$-$0.251 & G30.339$-$0.174 & G30.365+0.288 & G30.392+0.121 \\ 
G30.446$-$0.359 & G30.464+0.033 & G30.587$-$0.125 & G30.604+0.176 & G30.610+0.235 & G30.624$-$0.107 & G30.652$-$0.204 & G30.667$-$0.332 & G30.668+0.063 \\ 
G30.672+0.014 & G30.693+0.228 & G30.735$-$0.295 & G30.741$-$0.195 & G30.769+0.105 & G30.810+0.046 & G30.810+0.314 & G30.846$-$0.075 & G30.857+0.004 \\ 
G30.902$-$0.035 & G30.912+0.020 & G30.920+0.088 & G30.927+0.351 & G30.945+0.158 & G31.036+0.236 & G31.041$-$0.232 & G31.101+0.265 & G31.122+0.063 \\ 
G31.375+0.483 & G31.496+0.177 & G31.508$-$0.164 & G31.544$-$0.043 & G31.554$-$0.101 & G31.611+0.151 & G31.677+0.245 & G32.010$-$0.323 & G33.031+0.084 \\ 
G33.086+0.001 & G33.265+0.066 & G33.430$-$0.016 & G33.548+0.021 & G34.033$-$0.024 & G34.185+0.114 & G34.286+0.129 & G34.366$-$0.058 & G34.515+0.066 \\ 
G34.530$-$1.087 & G34.546+0.535 & G34.712$-$0.595 & G34.719$-$0.678 & G35.067$-$1.569 & G35.291+0.808 & G35.360$-$1.781 & G35.442$-$0.018 & G35.467+0.138 \\ 
G35.500$-$0.021 & G35.579$-$0.031 & G35.603$-$0.203 & G35.615$-$0.951 & G35.681$-$0.176 & G35.823$-$0.202 & G36.454$-$0.187 & G37.359$-$0.074 & G37.593$-$0.124 \\ 
G37.659+0.119 & G37.669$-$0.093 & G37.769$-$0.263 & G37.800$-$0.372 & G39.312$-$0.216 & G39.537$-$0.378 & G39.882$-$0.346 & G40.445+2.528 & G40.495+2.570 \\ 
"G40.545+2.596 & G40.592+	2.509 & G41.355+0.406 & G43.177$-$0.008 & G43.181$-$0.056 & G43.262$-$0.045 & G44.241+0.152 & G45.124+0.136 & G45.525+0.012 \\ "
G45.541$-$0.016 & G48.628+0.214 & G48.632$-$0.587 & G48.652$-$0.315 & G48.655$-$0.728 & G48.742$-$0.512 & G48.888$-$0.410 & G48.923$-$0.445 & G48.946$-$0.331 \\ 
G48.961$-$0.396 & G49.025$-$0.526 & G49.028$-$0.217 & G49.072$-$0.327 & G49.224$-$0.334 & G49.268$-$0.337 & G49.341$-$0.337 & G49.368$-$0.303 & G49.391$-$0.235 \\ 
G49.406$-$0.372 & G49.461$-$0.551 & G49.958+0.126 & G50.042+0.260 & G50.094$-$0.677 & G50.817+0.242 & G51.341+0.065 & G51.371$-$0.045 & G51.383$-$0.007 \\ 
G52.234+0.759 & G52.260$-$0.521 & G52.355$-$0.588 & G52.399$-$0.936 & G52.540$-$0.927 & G52.921$-$0.621 & G53.575+0.069 & G54.110$-$0.081 & G59.474$-$0.185 \\ 
G59.582$-$0.147 & G63.153+0.442 & G75.840+0.367 & G75.841+0.425 & G76.659+1.922 & G77.973+2.236 & G78.161+1.871 & G78.231+0.905 & G78.259$-$0.017 \\ 
G78.377+1.020 & G78.405+0.609 & G78.633+0.979 & G78.641+0.672 & G78.662+0.266 & G78.670+0.184 & G78.697+1.234 & G78.728+0.946 & G78.840+0.695 \\ 
G78.873+0.754 & G78.881+1.427 & G78.901+0.661 & G79.024+2.449 & G79.027+0.436 & G79.075+3.462 & G79.128+2.278 & G79.170+0.396 & G79.207+2.146 \\ 
G79.246+0.451 & G79.312$-$0.654 & G79.330$-$0.800 & G79.362$-$0.131 & G79.385$-$1.564 & G79.393+1.782 & G79.545$-$1.057 & G79.561$-$0.766 & G79.699+1.019 \\ 
G79.843+0.890 & G79.854$-$1.495 & G79.877+2.476 & G79.886+2.552 & G79.887$-$1.481 & G79.900+1.111 & G79.998$-$1.454 & G80.820+0.405 & G80.859$-$0.083 \\ 
G80.865+0.342 & G80.865+0.420 & G80.939$-$0.127 & G81.111$-$0.145 & G81.250+1.123 & G81.252+0.982 & G81.264$-$0.136 & G81.266+0.931 & G81.337+0.824 \\ 
G81.341+0.759 & G81.435+0.704 & G81.469+0.023 & G81.512+0.030 & G81.525+0.218 & G81.548+0.095 & G81.582+0.103 & G81.601+0.291 & G81.663+0.465 \\ 
G81.683+0.541 & G81.685$-$0.040 & G81.722+0.021 & G81.840+0.917 & G81.876+0.734 & G81.898+0.809 & G81.918$-$0.010 & G82.069$-$0.309 & G82.186+0.100 \\ 
G82.278+2.209 & G82.434+1.785 & G84.586$-$1.111 & G84.638$-$1.140 & G84.649$-$1.089 & G84.707$-$0.270 & G84.708$-$1.285 & G84.716$-$0.848 & G84.722$-$1.248 \\ 
G84.724$-$1.138 & G84.753+0.253 & G84.773$-$1.046 & G84.826$-$1.137 & G84.835$-$1.187 & G84.841$-$1.085 & G84.852+3.697 & G84.854$-$0.744 & G84.856$-$0.500 \\ 
G84.870$-$1.073 & G84.929$-$1.095 & G84.941$-$1.126 & G84.941$-$1.162 & G85.019$-$1.131 & G85.021$-$0.157 & G85.081$-$0.215 & G85.082$-$1.159 & G85.112$-$1.207 \\ 
G85.171$-$1.169 & G85.481$-$1.176 & G92.670+3.072 & G94.442$-$5.478 & G107.222$-$0.893 & G108.763$-$0.948 & G110.081+0.081 & G111.526+0.803 & G111.567+0.752 \\ 
G113.603$-$0.616 & G118.038+5.108 & G123.035$-$6.355 & G123.050$-$6.310 & G126.645$-$0.786 & G133.690+1.113 & G133.716+1.207 & G133.718+1.137 & G133.750+1.198 \\ 
G133.948+1.065 & G134.004+1.144 & G134.219+0.721 & G134.239+0.639 & G134.279+0.856 & G134.469+0.431 & G136.918+1.067 & G137.585+1.351 & G138.297+1.556 \\ 
G138.327+1.570 & G149.383$-$0.361 & G150.525$-$0.930 & G169.174$-$0.921 & G173.615+2.732 & G182.339+0.249 & G206.573$-$16.362 & G208.675$-$19.191 & G208.724$-$19.192 \\ 
G208.726$-$19.232 & G208.760$-$19.216 & G208.792$-$19.243 & G208.824$-$19.256 & G208.894$-$19.313 & G209.184$-$19.494 & G213.752$-$12.616 & G213.885$-$11.832 &  \\
\hline
\multicolumn{9}{c}{103 sources associated with both RRL and methanol maser}\\
\hline
G20.762$-$0.064 & G22.355+0.066 & G23.010$-$0.410 & G23.185$-$0.380 & G23.271$-$0.256 & G23.389+0.185 & G23.436$-$0.184 & G23.653$-$0.143 & G23.680$-$0.189 \\ 
G23.899+0.065 & G23.965$-$0.110 & G24.313$-$0.154 & G24.328+0.144 & G24.362$-$0.146 & G24.485+0.180 & G24.528+0.337 & G24.633+0.153 & G24.790+0.084 \\ 
G24.943+0.074 & G25.177+0.211 & G25.346$-$0.189 & G25.395+0.033 & G25.709+0.044 & G26.545+0.423 & G28.147$-$0.004 & G28.287$-$0.348 & G28.320$-$0.012 \\ 
G28.609+0.017 & G28.804$-$0.023 & G28.832$-$0.250 & G28.862+0.066 & G29.320$-$0.162 & G29.835$-$0.012 & G29.927+0.054 & G30.004$-$0.265 & G30.250$-$0.232 \\ 
G30.403$-$0.297 & G30.419$-$0.232 & G30.536$-$0.004 & G30.589$-$0.043 & G30.662$-$0.139 & G30.789+0.232 & G30.807+0.080 & G30.810$-$0.050 & G30.823+0.134 \\ 
G30.866+0.114 & G30.897+0.163 & G30.959+0.086 & G30.973+0.562 & G30.980+0.216 & G31.076+0.458 & G31.159+0.058 & G31.221+0.020 & G31.237+0.067 \\ 
G31.413+0.308 & G31.579+0.076 & G32.118+0.090 & G32.798+0.190 & G32.992+0.034 & G33.092$-$0.073 & G33.143$-$0.088 & G33.393+0.010 & G33.638$-$0.035 \\ 
G34.411+0.235 & G35.141$-$0.750 & G35.194$-$1.725 & G35.398+0.025 & G35.578+0.048 & G37.479$-$0.105 & G37.602+0.428 & G38.076$-$0.266 & G38.119$-$0.229 \\ 
G38.202$-$0.068 & G38.255$-$0.200 & G38.258$-$0.074 & G41.121$-$0.107 & G42.692$-$0.129 & G43.076$-$0.078 & G43.089$-$0.011 & G43.148+0.013 & G43.178$-$0.519 \\ 
G43.890$-$0.790 & G45.454+0.060 & G48.905$-$0.261 & G48.991$-$0.299 & G49.466$-$0.408 & G50.779+0.152 & G52.199+0.723 & G53.618+0.036 & G59.498$-$0.236 \\ 
G75.770+0.344 & G78.882+0.723 & G78.969+0.541 & G79.736+0.991 & G80.862+0.383 & G81.752+0.591 & G81.871+0.779 & G84.951$-$0.691 & G84.984$-$0.529 \\ 
G97.527+3.184 & G111.532+0.759 & G173.596+2.823 & G213.706$-$12.602 &  &  &  &  &  \\ 
\hline
\multicolumn{9}{c}{137 methanol-maser-only sources}\\
\hline
G16.872$-$2.154 & G16.883$-$2.186 & G17.021$-$2.402 & G173.482+2.446 & G173.617+2.883 & G174.205$-$0.069 & G183.349$-$0.575 & G20.234+0.085 & G20.363$-$0.014 \\ 
G20.926$-$0.050 & G21.023$-$0.063 & G21.370$-$0.226 & G213.752$-$12.615 & G22.050+0.211 & G24.148$-$0.009 & G24.634$-$0.323 & G25.256$-$0.446 & G25.410+0.105 \\ 
G25.498+0.069 & G25.613+0.226 & G25.649+1.050 & G25.837$-$0.378 & G26.421+1.686 & G26.598$-$0.024 & G26.623$-$0.259 & G26.645+0.021 & G27.220+0.261 \\ 
G27.222+0.136 & G27.287+0.154 & G27.725+0.037 & G27.784+0.057 & G27.795$-$0.277 & G28.180$-$0.093 & G28.393+0.085 & G28.843+0.494 & G29.281$-$0.330 \\ 
G29.941$-$0.070 & G30.370+0.483 & G30.770$-$0.804 & G30.788+0.203(R) & G30.819+0.273 & G30.972$-$0.141 & G31.253+0.003(L) & G31.253+0.003(R) & G32.045+0.059 \\ 
G32.773$-$0.059 & G32.828$-$0.315 & G33.229$-$0.018 & G33.322$-$0.364 & G33.425$-$0.315 & G33.641$-$0.228 & G33.726$-$0.119 & G34.096+0.018 & G34.229+0.133 \\ 
G34.757+0.025 & G34.789$-$1.392 & G34.974+0.365 & G35.149+0.809 & G35.197$-$0.729 & G35.225$-$0.360 & G35.247$-$0.237 & G35.792$-$0.174 & G36.137+0.564 \\ 
G36.634$-$0.203 & G36.705+0.096 & G36.833$-$0.031 & G36.919+0.483 & G37.043$-$0.035 & G37.430+1.517 & G37.554+0.201 & G37.763$-$0.215 & G38.598$-$0.213 \\ 
G38.933$-$0.361 & G39.100+0.491 & G39.387$-$0.141 & G40.282$-$0.220 & G40.425+0.700 & G40.597$-$0.719 & G40.622$-$0.138 & G40.964$-$0.025 & G41.307$-$0.169 \\ 
G42.035+0.191 & G43.037$-$0.453 & G43.808$-$0.080 & G45.070+0.124 & G45.360$-$0.598 & G45.493+0.126 & G45.804$-$0.356 & G49.043$-$1.079 & G49.265+0.311 \\ 
G49.537$-$0.904 & G49.599$-$0.249 & G50.034+0.581 & G51.678+0.719 & G52.663$-$1.092 & G52.922+0.414 & G53.022+0.100 & G53.141+0.071 & G53.485+0.521 \\ 
G54.371$-$0.613 & G56.963$-$0.234 & G58.775+0.647 & G59.436+0.820 & G59.634$-$0.192 & G59.785+0.068 & G59.833+0.672 & G62.310+0.114 & G69.543$-$0.973 \\ 
G71.522$-$0.385 & G73.063+1.796 & G74.098+0.110 & G75.010+0.274 & G76.093+0.158 & G78.122+3.633 & G81.794+0.911 & G82.308+0.729 & G84.193+1.439 \\ 
G85.394$-$0.023 & G89.930+1.669 & G90.921+1.487 & G94.609$-$1.790 & G98.036+1.446 & G99.070+1.200 & G108.184+5.518 & G108.758$-$0.986 & G109.839+2.134 \\ 
G109.868+2.119 & G110.196+2.476 & G111.256$-$0.770 & G121.329+0.639 & G123.035$-$6.355 & G123.050$-$6.310 & G124.015$-$0.027 & G134.029+1.072 & G136.859+1.165 \\ 
G137.068+3.002 & G149.076+0.397 &  &  &  &  &  &  &  \\ 
\enddata
\tablecomments{For sources showing multiple methanol emissions, ``R'' and ``L'' label the right/left hand side of the components (see \citealt{yang2017, yang2019}\\}
\end{deluxetable*}
\end{longrotatetable}

\begin{longrotatetable}
\floattable
\begin{deluxetable*}{p{1.9cm}p{1.7cm}p{1.6cm}p{1.2cm}p{0.7cm}p{1.7cm}p{1.7cm}p{1.7cm}p{0.6cm}p{0.6cm}p{0.8cm}p{0.6cm}p{3cm}}
\centering
\tablecaption{Characteristics of Helium and Carbon RRL emissions.\label{t5}}
\tablewidth{700pt}
\tabletypesize{\scriptsize}
\tablehead{
\colhead{Name} & \colhead{R.A.} & 
\colhead{Decl.} & \colhead{Epoch} & 
\colhead{T$_b$} & \colhead{S$_i$} & 
\colhead{$\Delta$V} & \colhead{$\nu$$_p$} & 
\colhead{d} & \colhead{$\sigma$$_d$} & \colhead{Type} & \colhead{Maser} & \colhead{Other Name(s)}\\ 
\colhead{(l,b)} & \colhead{(J2000)} & \colhead{(J2000)} & \colhead{(dd/mm/yy)} & 
\colhead{} & \colhead{} & \colhead{} &
\colhead{} & \colhead{} & \colhead{}& \colhead{} & \colhead{} & \colhead{}\\
\colhead{($\degr$,$\degr$)} & \colhead{(h m s)} & \colhead{($\degr$ $\arcmin$ $\arcsec$)} & \colhead{} & 
\colhead{(mK)} & \colhead{(K km s$^{-1}$)} & \colhead{(km s$^{-1}$)} &
\colhead{(km s$^{-1}$)} & \colhead{(kpc)} & \colhead{(kpc)} & \colhead{} & \colhead{} & \colhead{}\\
} 
\colnumbers
\startdata
G20.749$-$0.112 & 18:29:21.28 & $-$10:52:38.5 & 04/07/16 & 20  & 0.70(0.06) & 33.4(3.5) & $-$68.4(1.3) & 3.46  & 0.20  & He &  & (K) G020.728-00.105 / Kes 68\\
G23.428$-$0.231 & 18:34:48.38 & $-$08:33:22.0 & 21/08/16 & 15  & 0.24(0.06) & 15.2(3.8) & $-$20.0(1.9) & 5.93  & 0.23  & He &  & (K) G023.423-00.216\\
G23.436$-$0.184 & 18:34:39.21 & $-$08:31:40.4 & 21/08/16 & 14  & 0.17(0.04) & 10.8(2.1) & $-$24.6(1.3) & 5.87  & 0.24  & He & yes & (K) G023.458-00.179\\
G23.473$-$0.212 & 18:34:49.22 & $-$08:30:28.2 & 21/08/16 & 9  & 0.17(0.04) & 17.6(4.6) & $-$19.8(2.0) & 5.86  & 0.24  & He &  & (K) G023.458-00.179\\
G23.563+0.008 & 18:34:12.06 & $-$08:19:36.6 & 06/08/16 & 13  & 0.30(0.04) & 21.6(3.6) & $-$34.1(1.3) & 5.60  & 0.40  & He &  & (K) G023.572-00.020\\
G24.479$-$0.250 & 18:36:49.63 & $-$07:37:55.3 & 31/08/16 & 15  & 0.16(0.03) & 10.0(1.7) & $-$26.7(1.1) & 5.83  & 0.24  & He &  & (K) G024.493-00.219\\
G24.790+0.084 & 18:36:12.46 & $-$07:12:10.8 & 05/09/16 & 28  & 0.50(0.06) & 16.4(2.1) & $-$15.1(0.9) & 6.04  & 0.23  & He & yes & (K) G024.844+00.093\\
G25.346$-$0.189 & 18:38:12.83 & $-$06:50:00.8 & 05/09/16 & 17  & 0.40(0.05) & 22.4(3.5) & $-$64.9(1.7) & 3.83  & 0.29  & He & yes & (K) G025.382-00.151 / IRAS 18354-0652\\
G25.392$-$0.131 & 18:38:05.54 & $-$06:45:58.3 & 06/09/16 & 22  & 0.61(0.11) & 26.5(6.8) & $-$22.1(2.1) & 8.88  & 0.31  & He &  & (K) G025.382-00.151\\
 &  &  &  & 23  & 0.20(0.06) & 7.9(3.3) & $-$55.7(1.0) &  &  & C &  & \\
\enddata
\tablecomments{Properties of the He and C RRL components with (1) source name, (2)(3) the pointing center coordinates, (4) epoch of observations, (5) measured peak antenna temperature of a Gaussian fitting, (6) integral flux of the Gaussian fitting with error, (7) ({\tt\string FWHM}) of the Gaussian fitting with error, (8) radio defined LSR velocity at the peak value with error, (9)(10) Bayesian distance and error calculated using peak velocity of corresponding H RRLs. The distances were derived following \citet{reid2016} with a default equal weighting ($P_{far}$ = 0.5) on near and far distance probability, (11) emission type: ``He" and ``C" label the He and C RRLs respectively, (12) association with 6.7 GHz methanol maser, (13)other name(s) of the sources, for sources associated with an \citetalias{anderson2014} source, we label the type of the \citetalias{anderson2014} counterpart: ``K" for ``Known" sources, ``C" for ``candidate" sources, ``G" for ``Grouped" sources and ``Q" for ``Radio quiet" sources. \\}
\tablecomments{(This table is available in its entirety in a machine-readable form in the online journal and the source files on arXiv.org. A portion is shown here for guidance regarding its form and content.)}
\end{deluxetable*}
\end{longrotatetable}

\begin{longrotatetable}
\begin{deluxetable*}{p{1.9cm}p{1.7cm}p{1.6cm}p{1.2cm}p{0.7cm}p{1.7cm}p{1.7cm}p{1.7cm}p{0.6cm}p{0.6cm}p{0.8cm}p{3.5cm}}
\centering
\tablecaption{Characteristics of the PNe candidates\label{t6}}
\tablewidth{700pt}
\tabletypesize{\scriptsize}
\tablehead{
\colhead{Name} & \colhead{R.A.} & 
\colhead{Decl.} & \colhead{Epoch} & 
\colhead{T$_b$} & \colhead{S$_i$} & 
\colhead{$\Delta$V} & \colhead{$\nu$$_p$} & 
\colhead{d} & \colhead{$\sigma$$_d$} & \colhead{Notes} & \colhead{Other Name(s)}\\ 
\colhead{(l,b)} & \colhead{(J2000)} & \colhead{(J2000)} & \colhead{(dd/mm/yy)} & 
\colhead{} & \colhead{} & \colhead{} &
\colhead{} & \colhead{} & \colhead{}& \colhead{} & \colhead{}\\
\colhead{($\degr$,$\degr$)} & \colhead{(h m s)} & \colhead{($\degr$ $\arcmin$ $\arcsec$)} & \colhead{} & 
\colhead{(mK)} & \colhead{(mK km s$^{-1}$)} & \colhead{(km s$^{-1}$)} &
\colhead{(km s$^{-1}$)} & \colhead{(kpc)} & \colhead{(kpc)}& \colhead{} & \colhead{}\\
} 
\colnumbers
\startdata
G30.035$-$0.002 & 18:46:09.37 & $-$02:34:44.4 & 17/09/16 & 47 & 1.32(0.07) & 26.3(1.6) & 97.1(0.6) & 6.82  & 1.31  & PNe? & PN G030.0+00.0 / (K) G030.022 / IRAS 18436-0239\\
G78.911+0.792 & 20:29:08.19 & +40:15:26.5 & 16/09/16 & 18 & 0.40(0.04) & 20.8(2.9) & 10.7(1.3) & 2.72 & 0.97 & PNe? & PN Sd 1 / (K) G078.886+00.709\\
G78.931+0.722 & 20:29:29.59 & +40:13:55.0 & 16/09/16 & 32 & 1.05(0.07) & 30.7(2.2) & 17.2(1.0) & 3.07  & 0.87  & PNe? & PN Sd 1 / (K) G078.886+00.709\\
G84.913$-$3.505 & 21:07:00.13 & +42:13:01.8 & 14/12/17 & 35 & 2.24(0.05) & 58.7(1.5) & 25.5(0.7) & 1.28 & 0.07 & PNe & NGC 7027\\
G84.946$-$3.488 & 21:07:03.27 & +42:15:12.1 & 15/12/17 & 38 & 2.23(0.09) & 55.6(2.4) & 22.4(1.0) & 1.28 & 0.07 & PNe & NGC 7027\\
\enddata
\tablecomments{Characteristics of the potential PNe sample with (1) source name, (2)(3) the pointing center coordinates, (4) epoch of observations, (5) measured peak antenna temperature of a Gaussian fitting, (6) integral flux of the Gaussian fitting with error, (7) ({\tt\string FWHM}) of the Gaussian fitting with error, (8) radio defined LSR velocity at the peak value with error, (9)(10) Bayesian distance and error calculated following \citet{reid2016} with a default equal weighting ($P_{far}$ = 0.5) on near and far distance probability, (11) source type: ``PNe'' for sources associated explicitly with a PNe, ``PNe?'' for potential PNe sources, (12) other name(s) of the sources, for sources associated with an \citetalias{anderson2014} source, we label the type of the \citetalias{anderson2014} counterpart: ``K" for ``Known" sources, ``C" for ``candidate" sources, ``G" for ``Grouped" sources and ``Q" for ``Radio quiet" sources.\\}
\end{deluxetable*}
\end{longrotatetable}

\begin{longrotatetable}
\begin{deluxetable*}{p{1.9cm}p{1.7cm}p{1.6cm}p{1.2cm}p{0.7cm}p{1.7cm}p{1.7cm}p{1.7cm}p{0.6cm}p{0.6cm}p{0.8cm}p{3.5cm}}
\centering
\tablecaption{Characteristics of the SNR candidates\label{t7}}
\tablewidth{700pt}
\tabletypesize{\scriptsize}
\tablehead{
\colhead{Name} & \colhead{R.A.} & 
\colhead{Decl.} & \colhead{Epoch} & 
\colhead{T$_b$} & \colhead{S$_i$} & 
\colhead{$\Delta$V} & \colhead{$\nu$$_p$} & 
\colhead{d} & \colhead{$\sigma$$_d$} & \colhead{Notes} & \colhead{Other Name(s)}\\ 
\colhead{(l,b)} & \colhead{(J2000)} & \colhead{(J2000)} & \colhead{(dd/mm/yy)} & 
\colhead{} & \colhead{} & \colhead{} &
\colhead{} & \colhead{} & \colhead{}& \colhead{} & \colhead{}\\
\colhead{($\degr$,$\degr$)} & \colhead{(h m s)} & \colhead{($\degr$ $\arcmin$ $\arcsec$)} & \colhead{} & 
\colhead{(mK)} & \colhead{(mK km s$^{-1}$)} & \colhead{(km s$^{-1}$)} &
\colhead{(km s$^{-1}$)} & \colhead{(kpc)} & \colhead{(kpc)}& \colhead{} & \colhead{}\\
} 
\colnumbers
\startdata
G30.726+0.103 & 18:47:02.74 & $-$01:54:56.2 & 18/08/16 & 75 & 1.86(0.01) & 23.5(0.4) & 117.8(0.4) & 7.19 & 0.82 & SNR? & SNR G030.3+00.7 / (K) G030.796+00.183 / IRAS 18445-0158\\
 &  &  &  & 51 & 1.42(0.01) & 26.2(0.4) & 90.2(0.4) &  &  &  & \\
 &  &  &  & 36 & 0.91(0.01) & 23.7(0.4) & 42.0(0.4) &  &  &  & \\
G79.831+1.280 & 20:29:53.77 & +41:17:18.8 & 08/09/16 & 40 & 1.10(0.06) & 25.6(1.7) & -16.9(0.7) & 5.28 & 0.63 & SNR? & SNR G079.8+01.2 / (K) G080.362+01.212\\
G28.532+0.129 & 18:42:56.49 & $-$03:51:21.7 & 19/09/16 & 26 & 0.83(0.06) & 30.1(2.1) & 101.0(1.0) & 8.13 & 0.31 & SNR? & MAGPIS SNR? G28.5167+0.1333 / (K) G028.581+00.145\\
 &  &  &  & 19 & 1.45(0.08) & 70.2(4.6) & 26.0(2.0) &  &  &  & \\
G28.565+0.021 & 18:43:23.21 & $-$03:52:32.3 & 19/09/16 & 86 & 2.59(0.06) & 28.3(0.8) & 96.9(0.3) & 8.09 & 0.37 & SNR? & SNR G028.56+00.00 / (K) G028.607+00.019\\
 &  &  &  & 19 & 0.34(0.05) & 16.4(2.5) & 39.5(1.1) &  &  &  & \\
G27.102+0.024 & 18:40:41.50 & $-$05:10:32.5 & 09/09/16 & 42 & 1.41(0.07) & 31.1(1.8) & 92.9(0.7) & 8.55  & 0.39  & SNR? & MAGPIS SNR? G27.1333+0.0333\\
\enddata
\tablecomments{Characteristics of the potential SNR sample with (1) Source name, (2)(3) the pointing center coordinates, (4) epoch of observations, (5) measured peak antenna temperature of a Gaussian fitting, (6) integral flux of the Gaussian fitting with error, (7) ({\tt\string FWHM}) of the Gaussian fitting with error, (8) radio defined LSR velocity at the peak value with error, (9)(10) Bayesian distance and error calculated following \citet{reid2016} with a default equal weighting ($P_{far}$ = 0.5) on near and far distance probability, (11) source type, all the sources in this table are candidate sources thus are marked with ``SNR?'', (12) other name(s) of the sources, for sources associated with an \citetalias{anderson2014} source, we label the type of the \citetalias{anderson2014} counterpart: ``K" for ``Known" sources, ``C" for ``candidate" sources, ``G" for ``Grouped" sources and ``Q" for ``Radio quiet" sources. For multi-component sources, the components are sorted by peak line intensity.\\}
\end{deluxetable*}
\end{longrotatetable}

\appendix
\textbf{Appendix A:} RRL spectra for the HMSFR sample.\\

\begin{figure*}[ht]
\centering
\stackunder[5pt]{\includegraphics[width=2.33in]{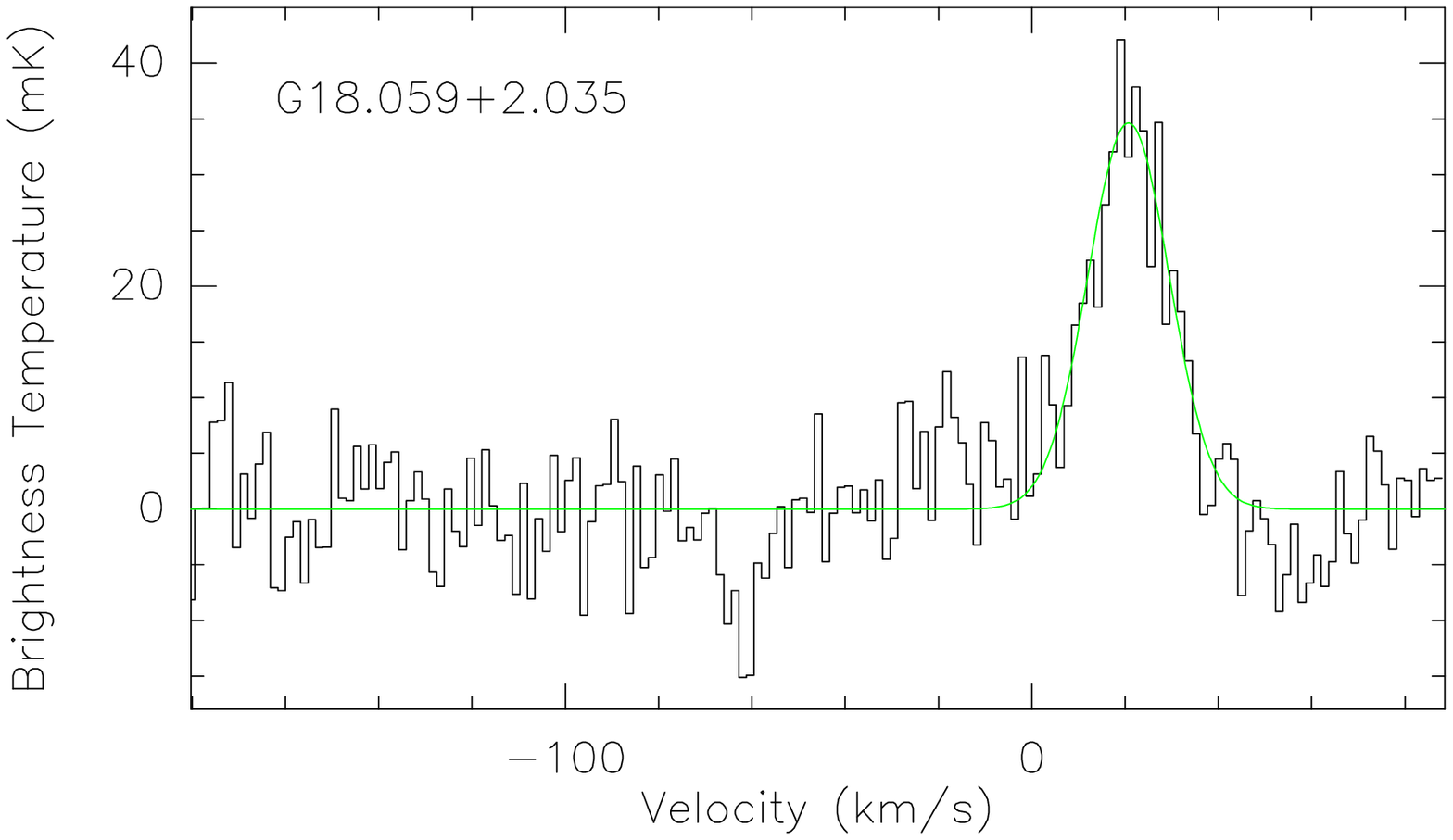}}{ }
\stackunder[5pt]{\includegraphics[width=2.33in]{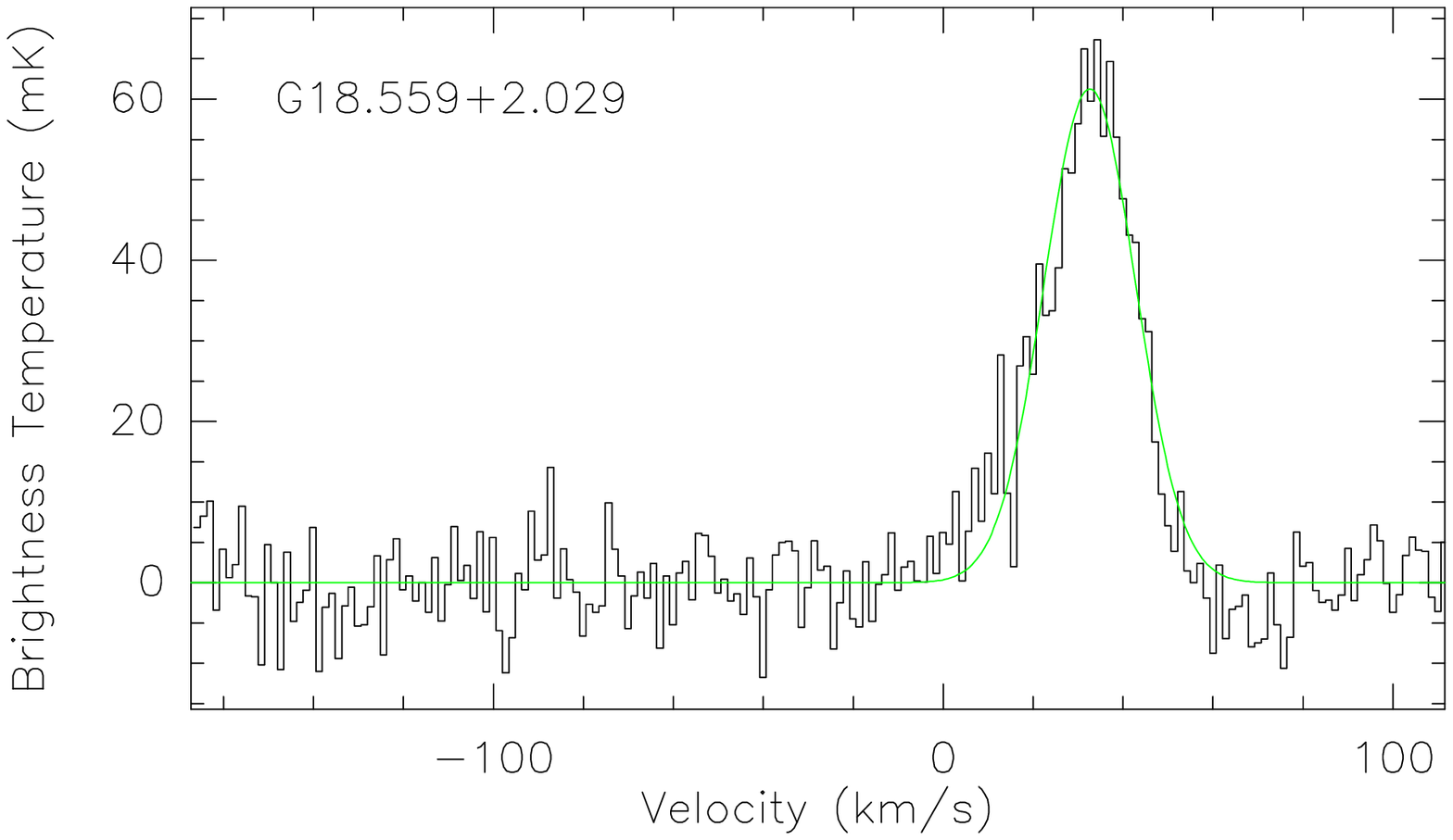}}{ }
\stackunder[5pt]{\includegraphics[width=2.33in]{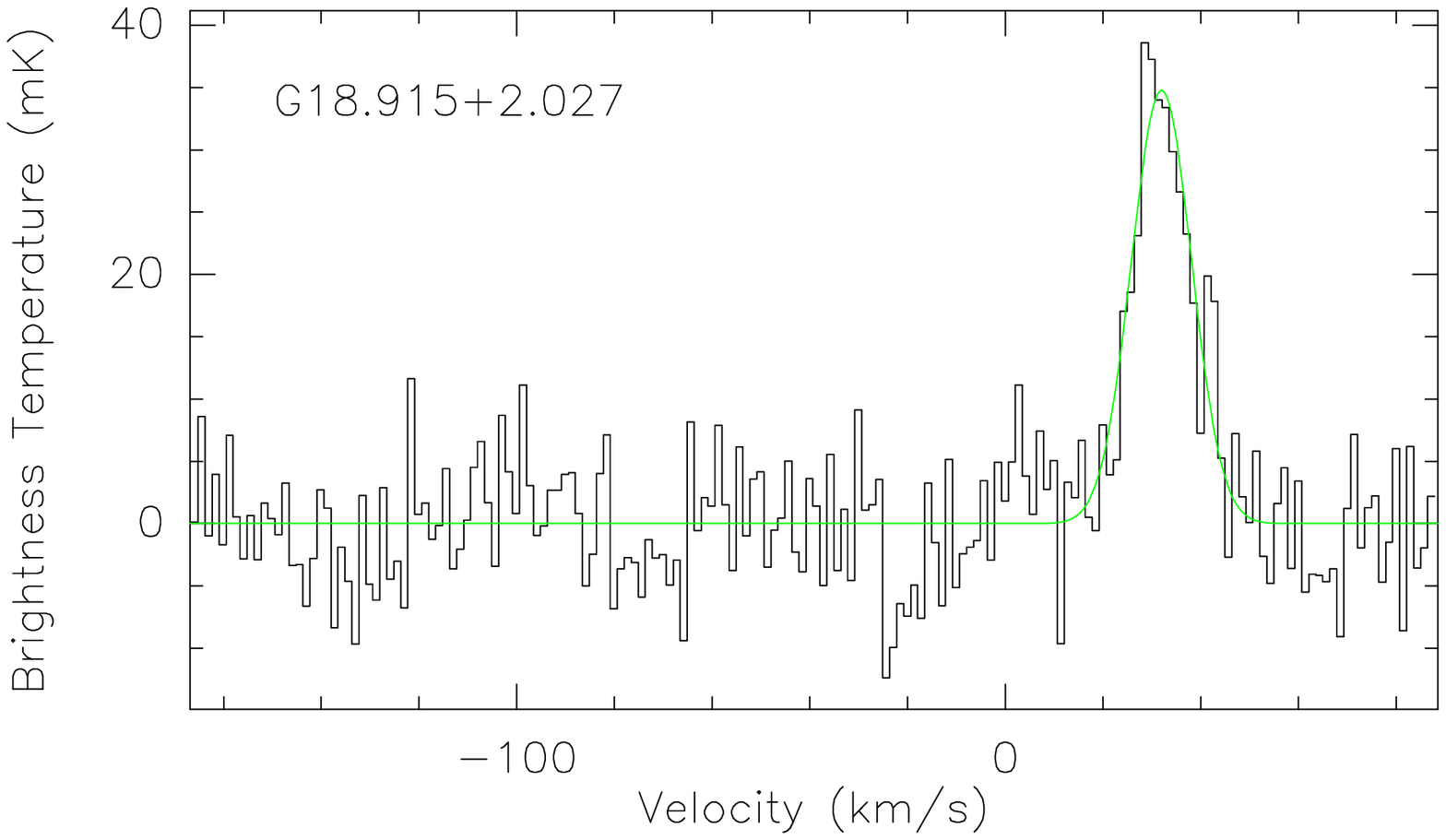}}{ }\\
\stackunder[5pt]{\includegraphics[width=2.33in]{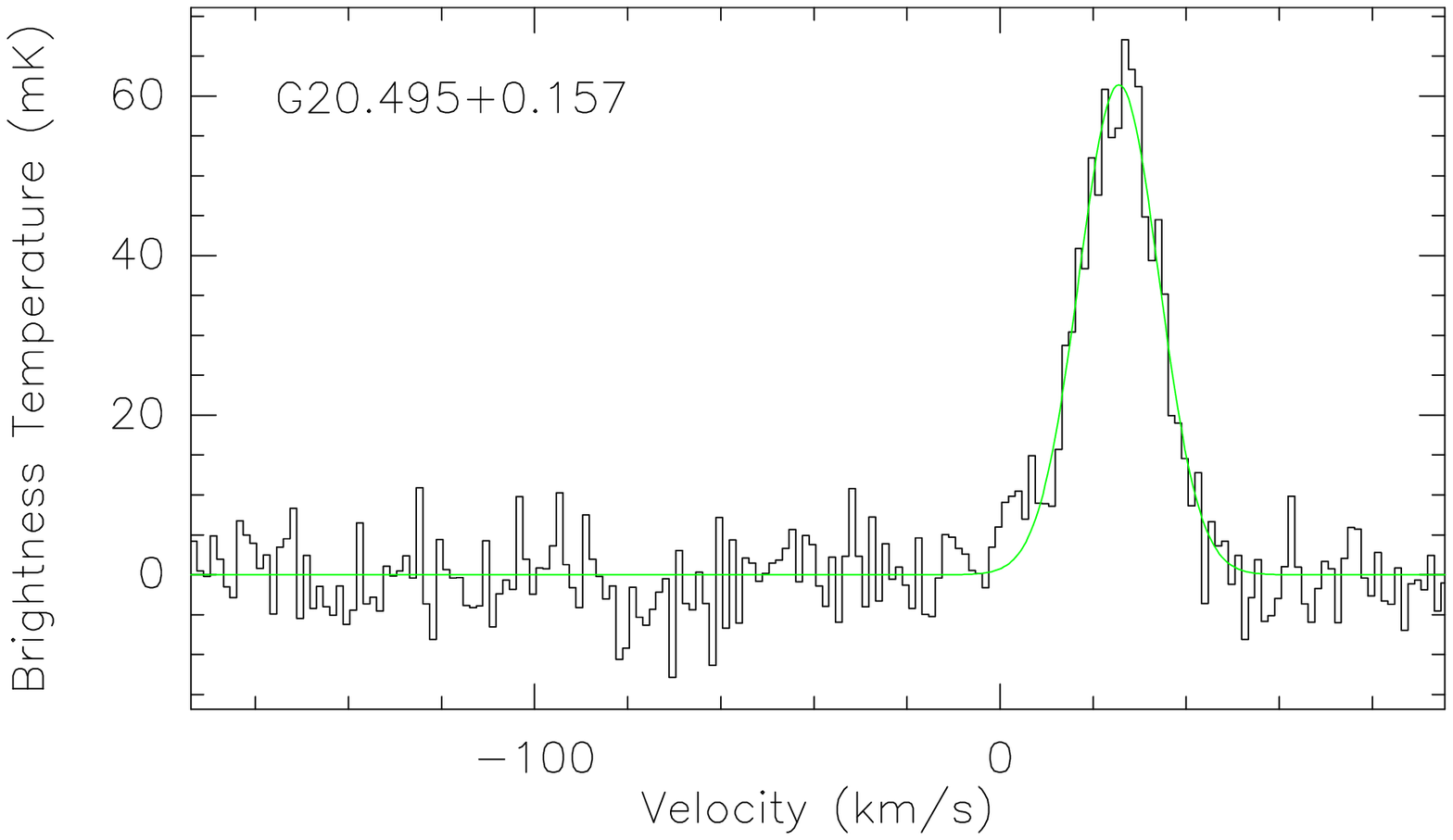}}{ }
\stackunder[5pt]{\includegraphics[width=2.33in]{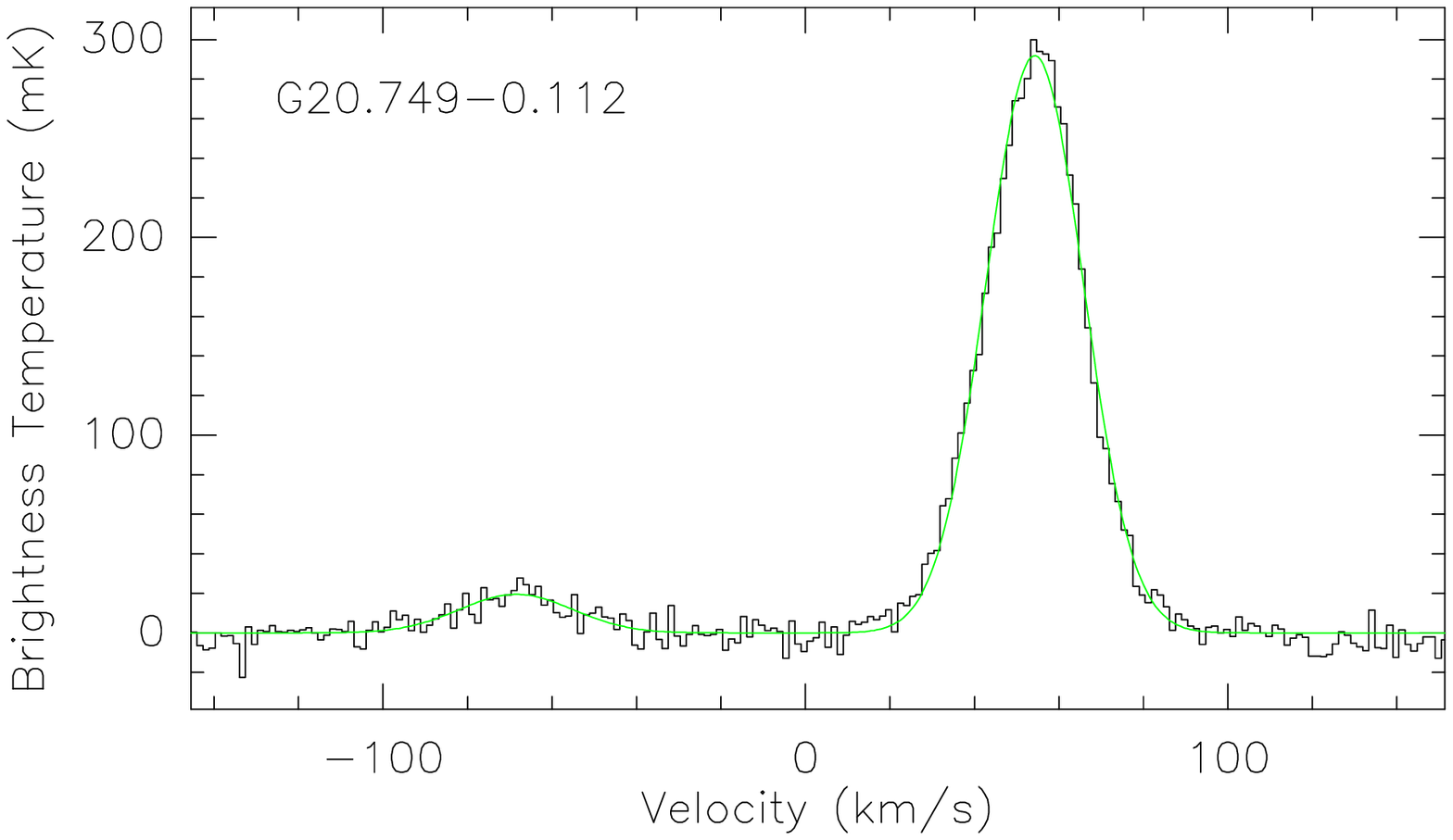}}{ }
\stackunder[5pt]{\includegraphics[width=2.33in]{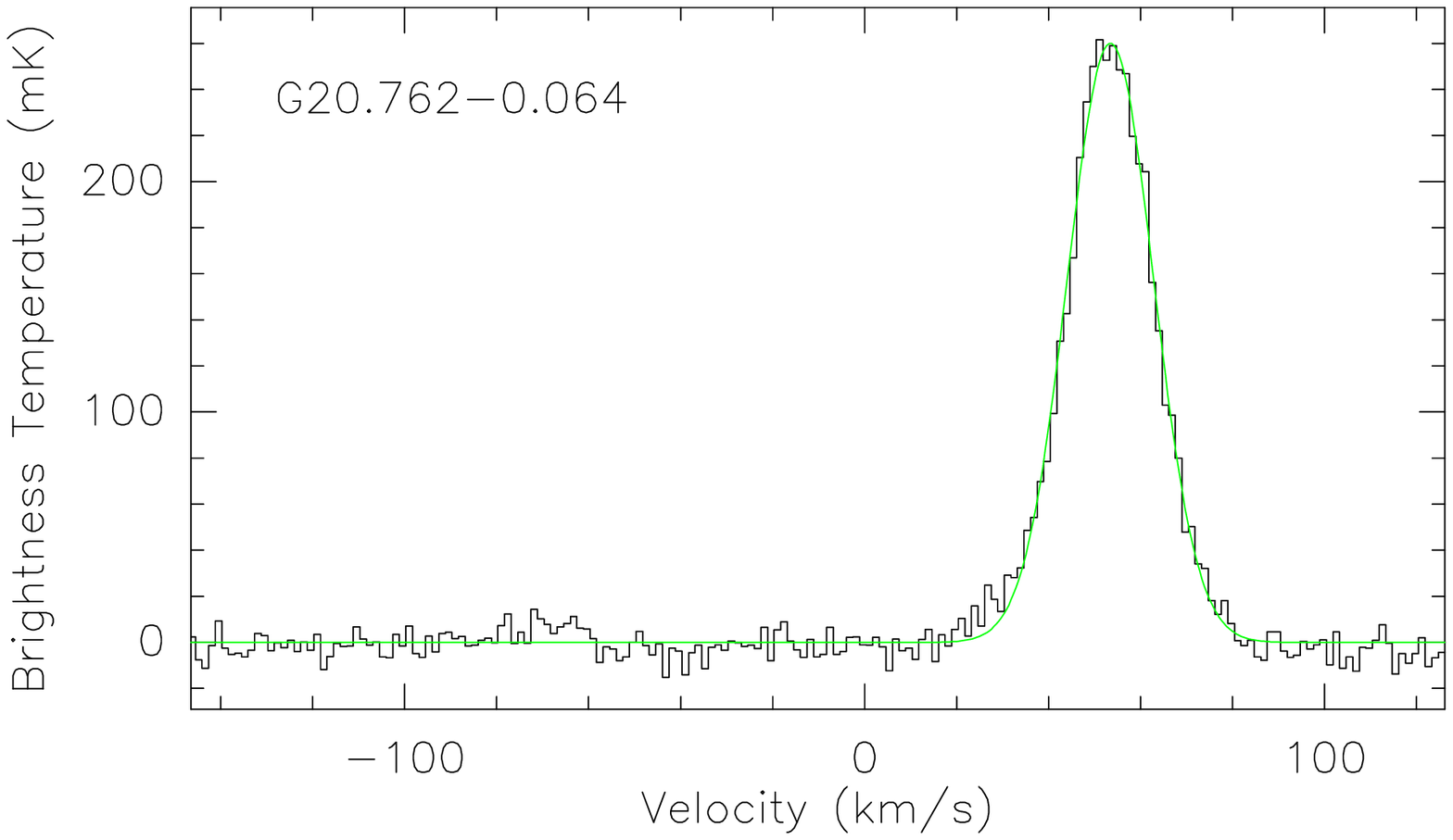}}{ }\\
\stackunder[5pt]{\includegraphics[width=2.33in]{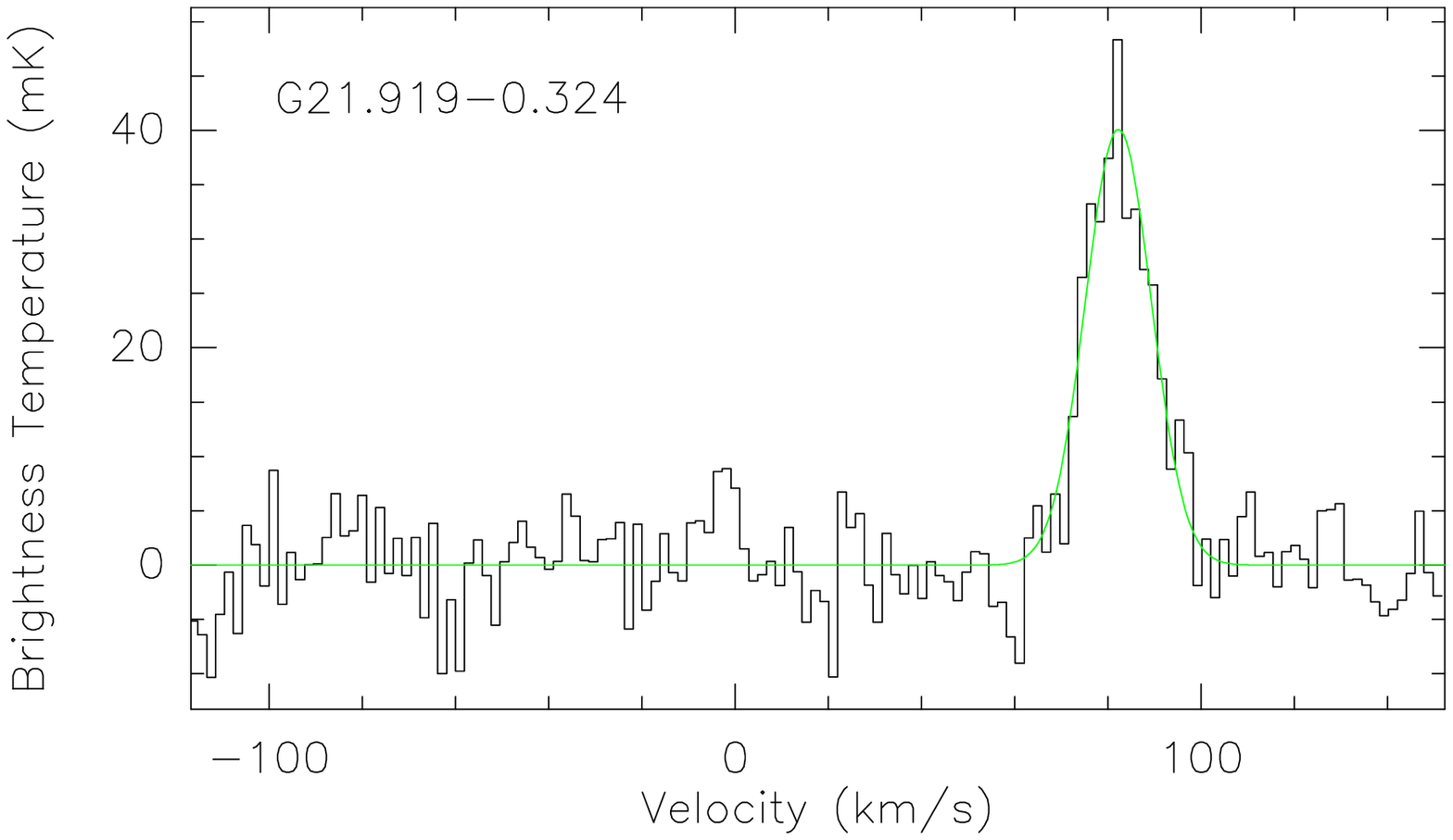}}{ }
\stackunder[5pt]{\includegraphics[width=2.33in]{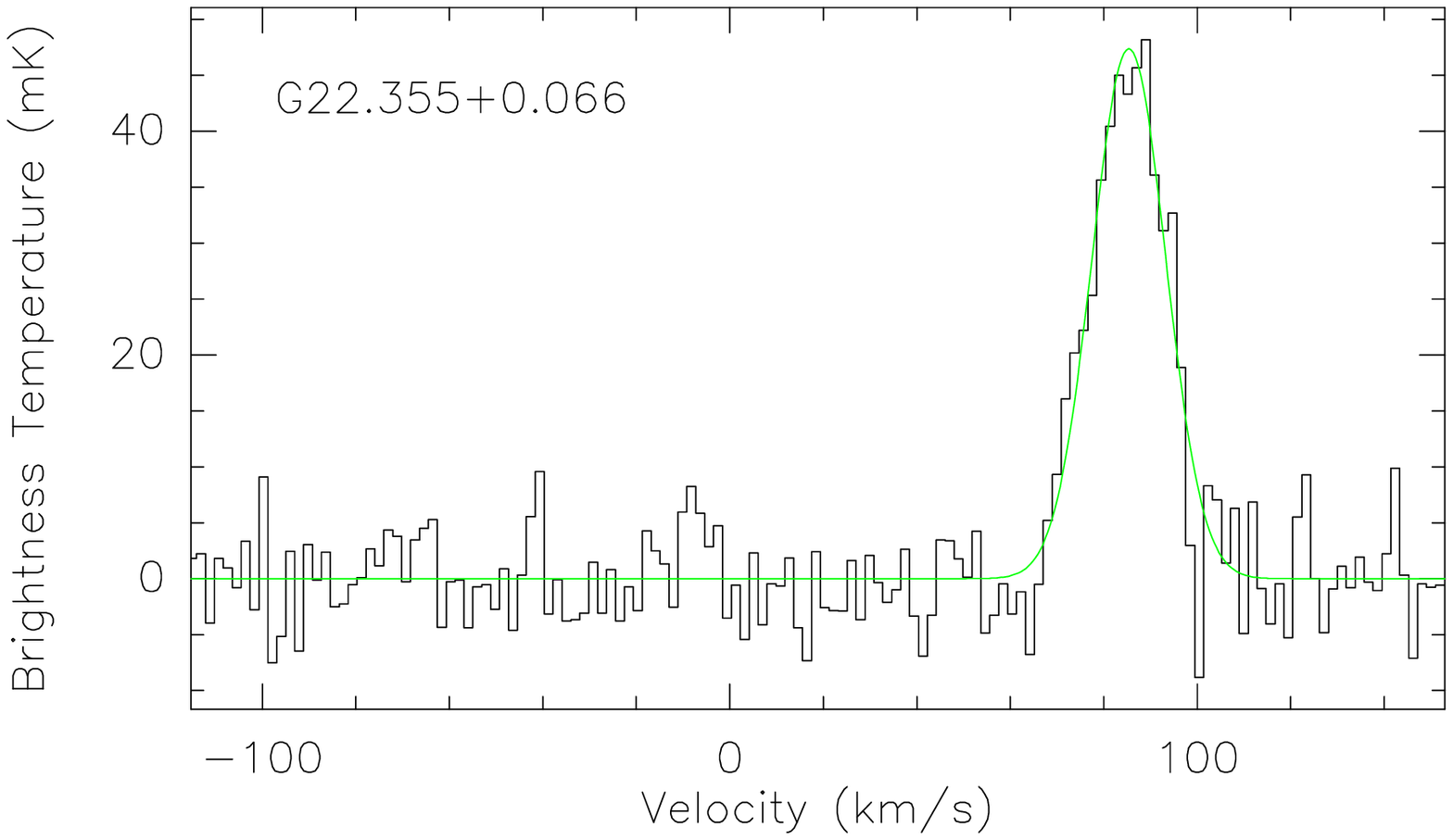}}{ }
\stackunder[5pt]{\includegraphics[width=2.33in]{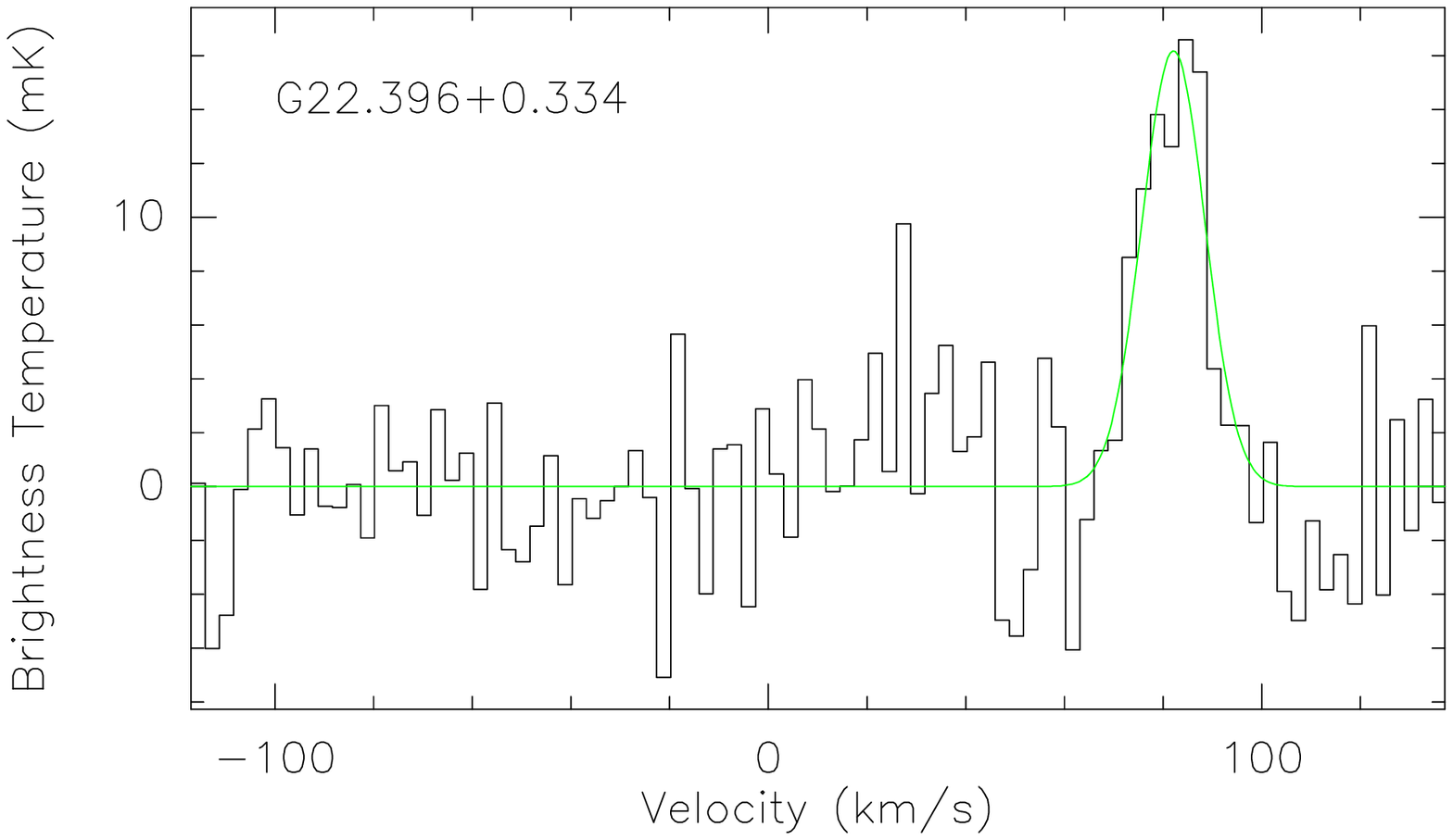}}{ }\\
\stackunder[5pt]{\includegraphics[width=2.33in]{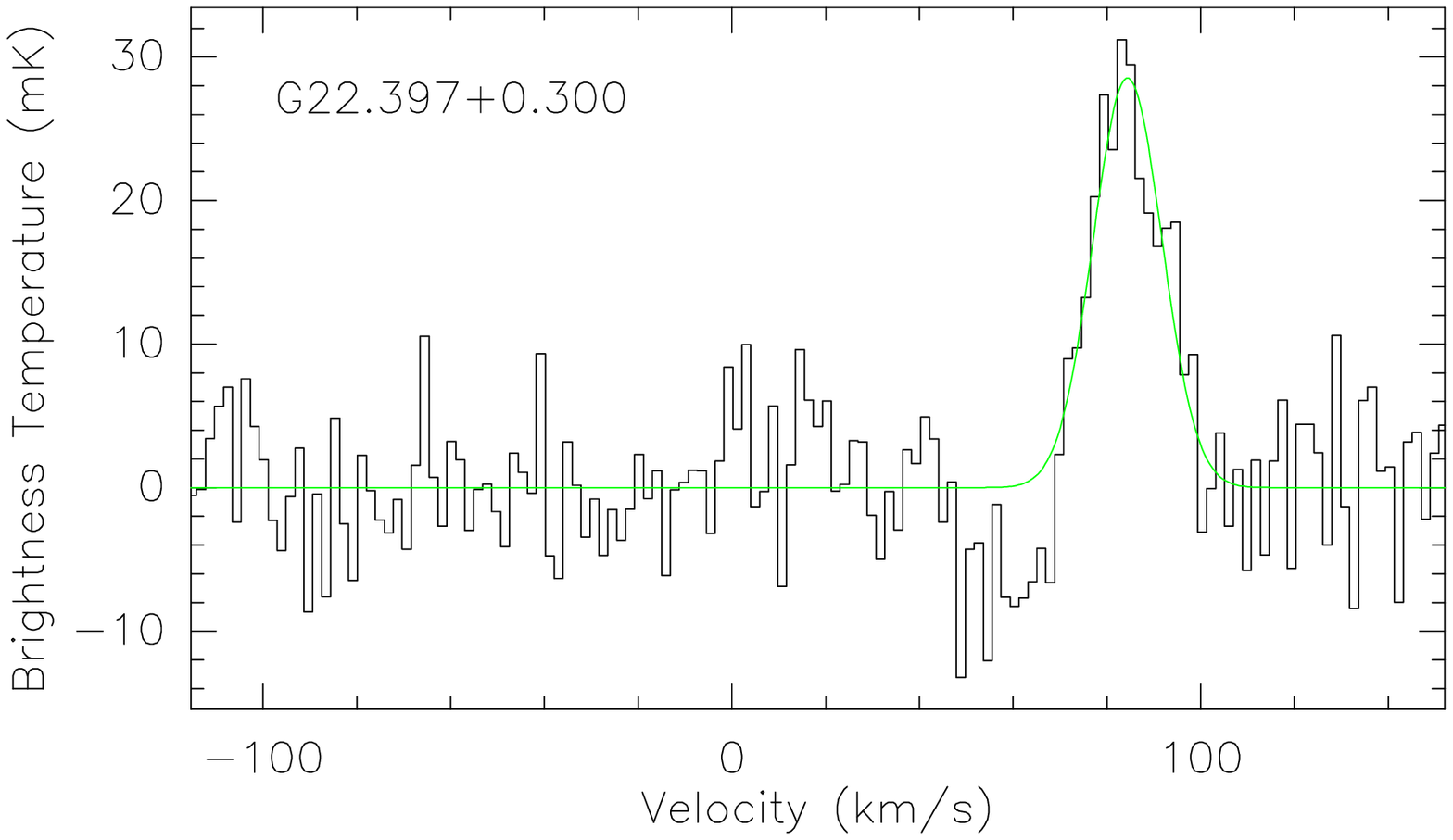}}{ }
\stackunder[5pt]{\includegraphics[width=2.33in]{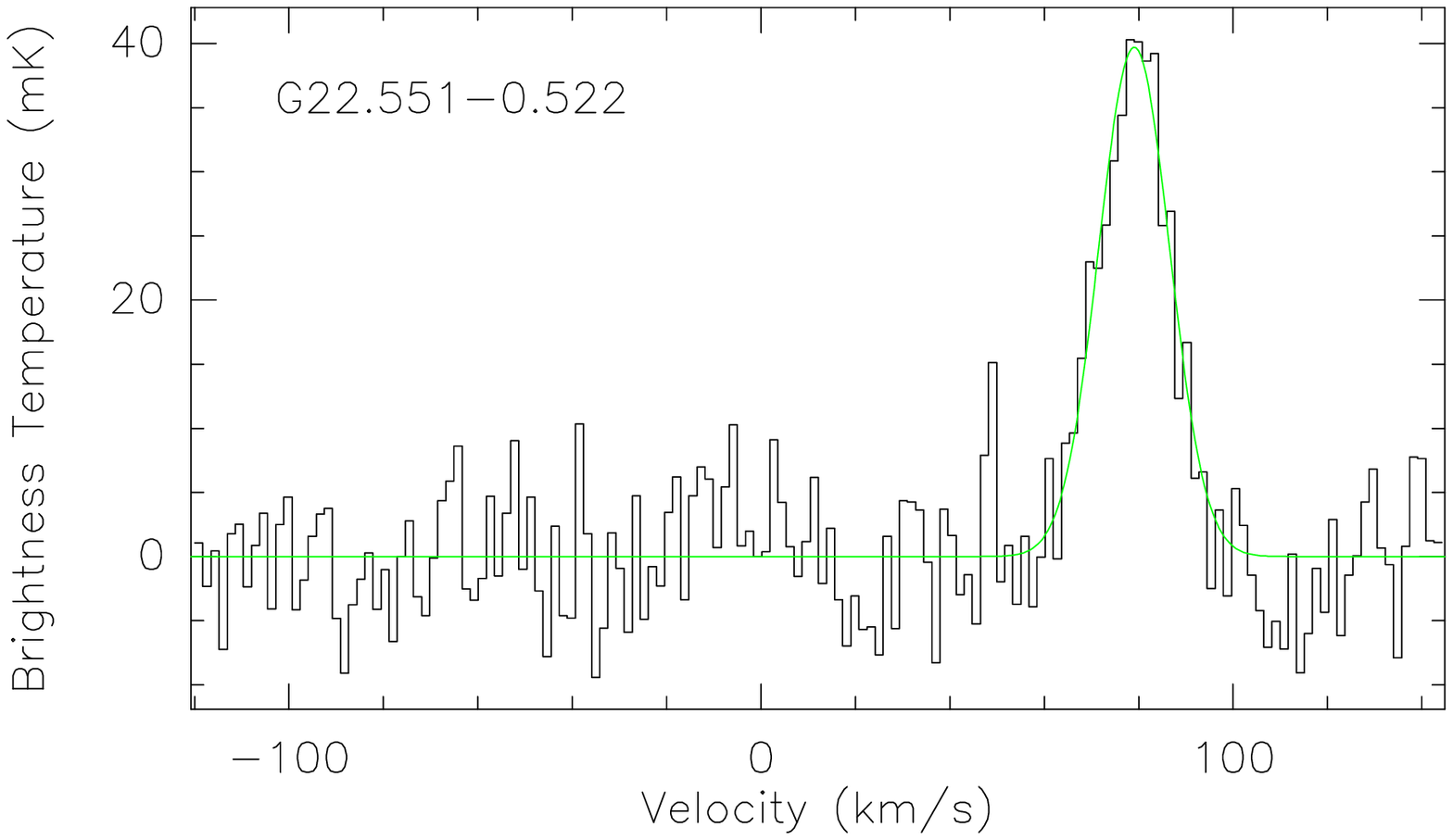}}{ }
\stackunder[5pt]{\includegraphics[width=2.33in]{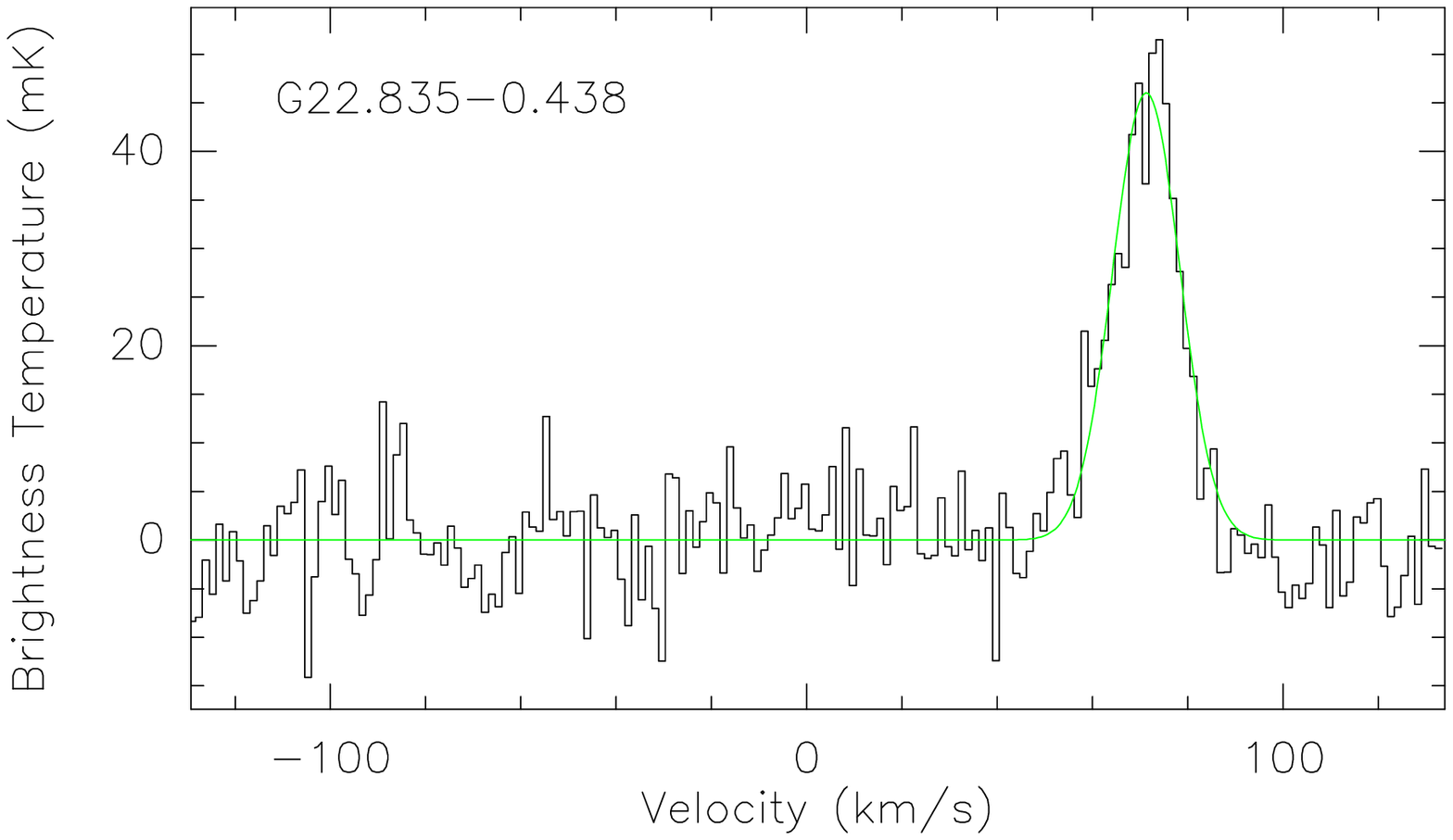}}{ }\\
\stackunder[5pt]{\includegraphics[width=2.33in]{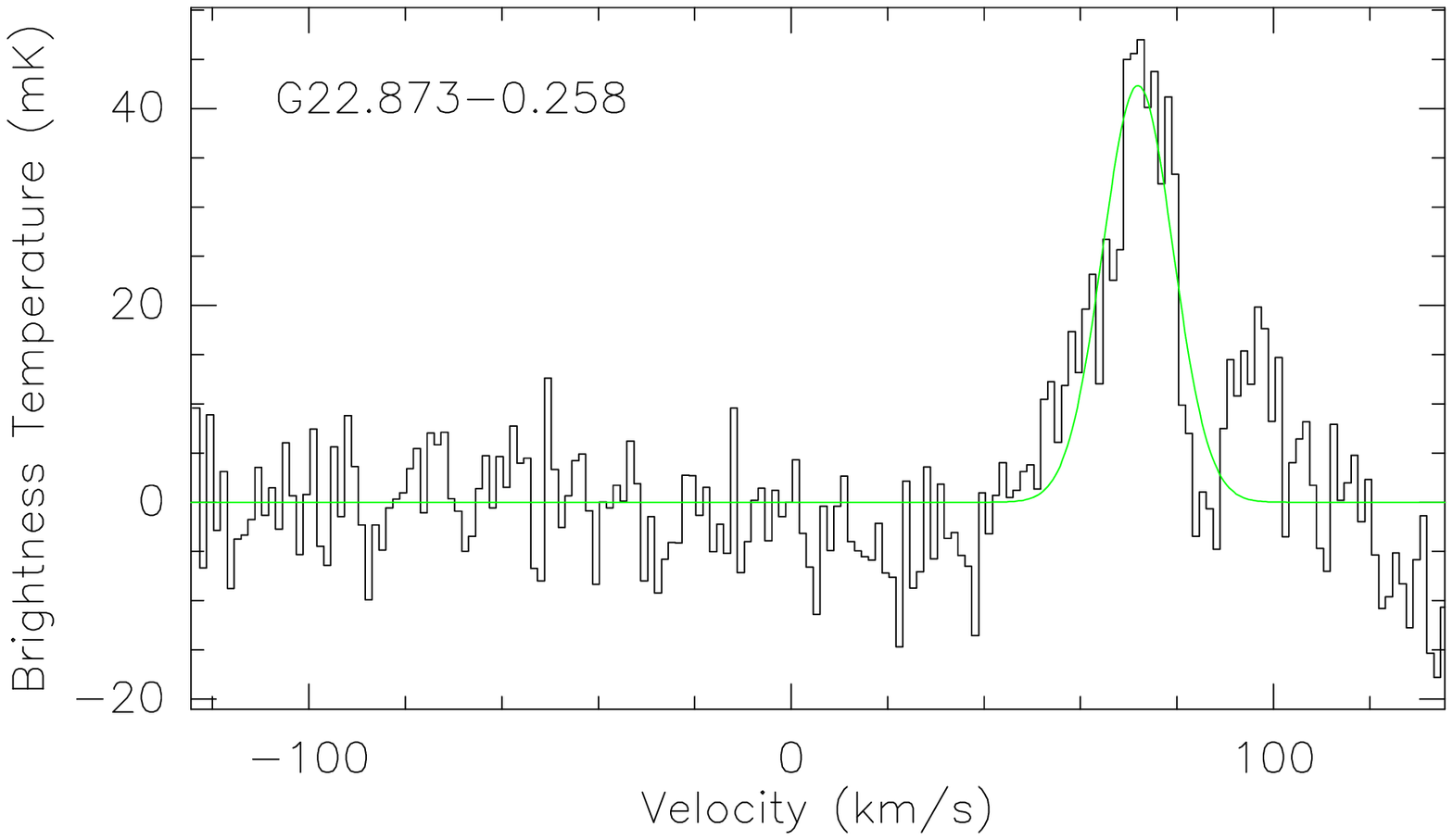}}{ }
\stackunder[5pt]{\includegraphics[width=2.33in]{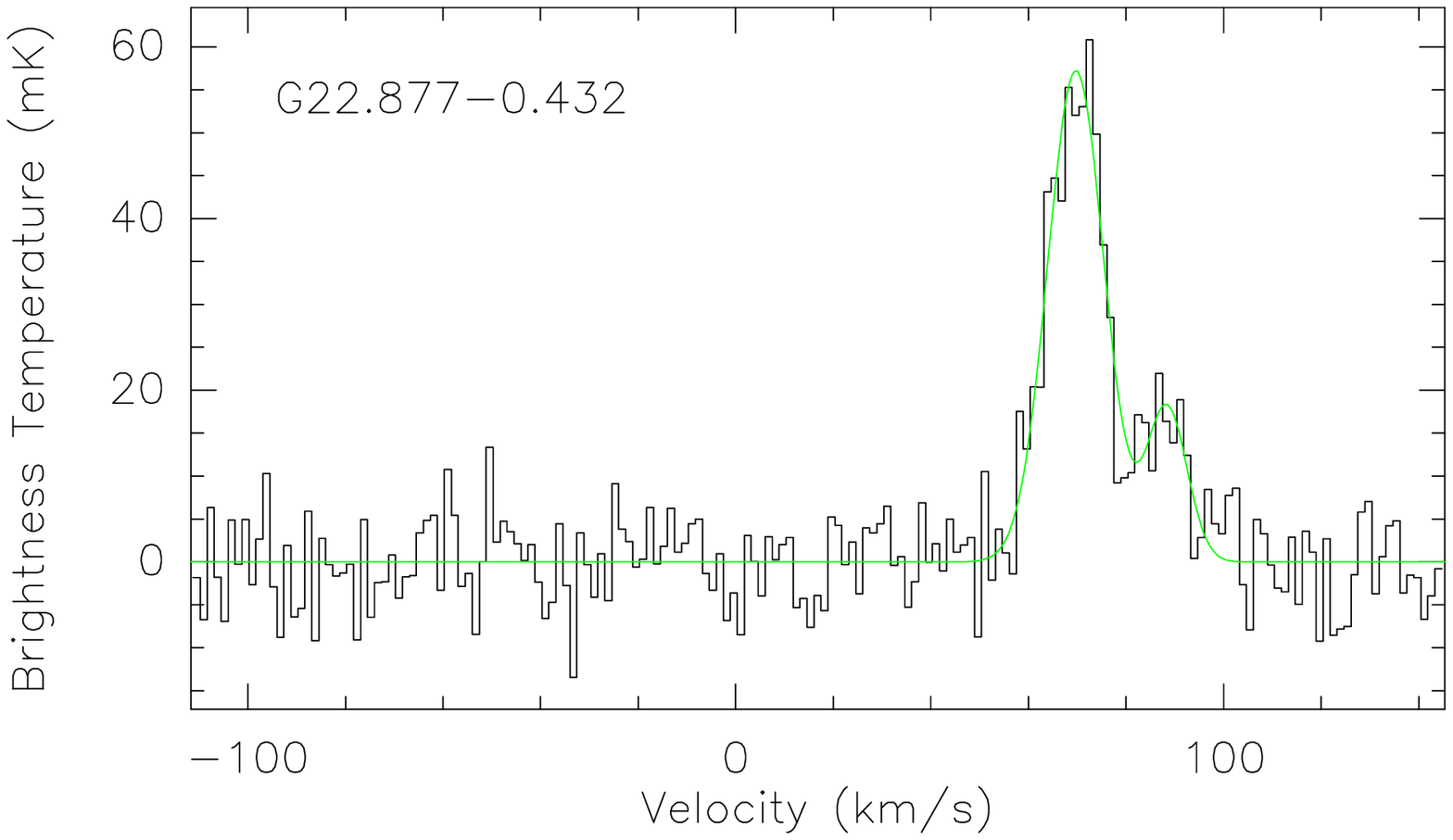}}{ }
\stackunder[5pt]{\includegraphics[width=2.33in]{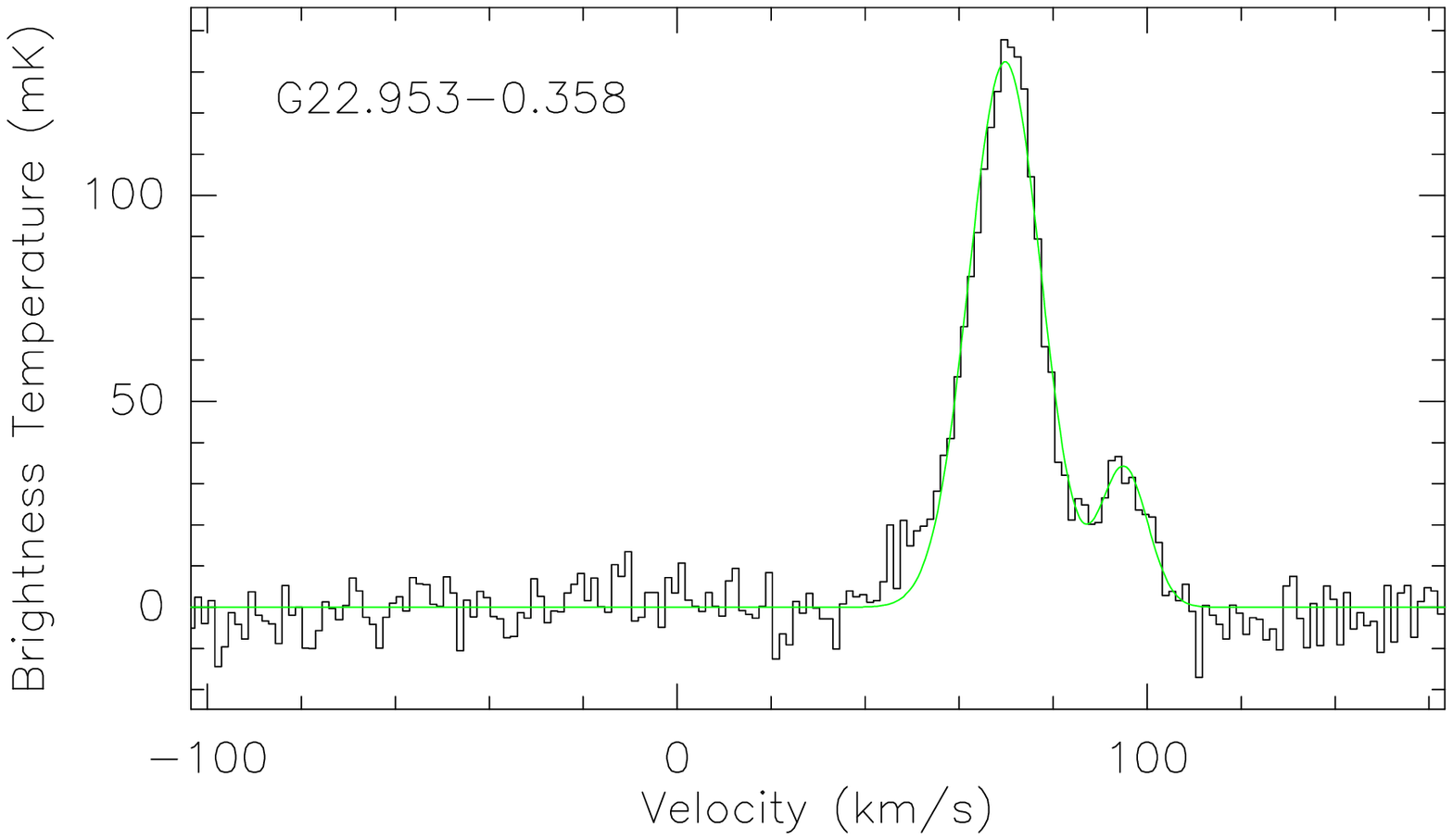}}{ }\\
\tablecomments{This figure is available in its entirety in the online journal. A portion is shown here for guidance regarding its form and content.}
\caption{RRL spectra for the HMSFRs.}
\end{figure*}

\clearpage
\textbf{Appendix B:} RRL spectra for the previously known and potential PNe sources.\\
\begin{figure*}[ht]
\centering
\stackunder[5pt]{\includegraphics[width=2.33in]{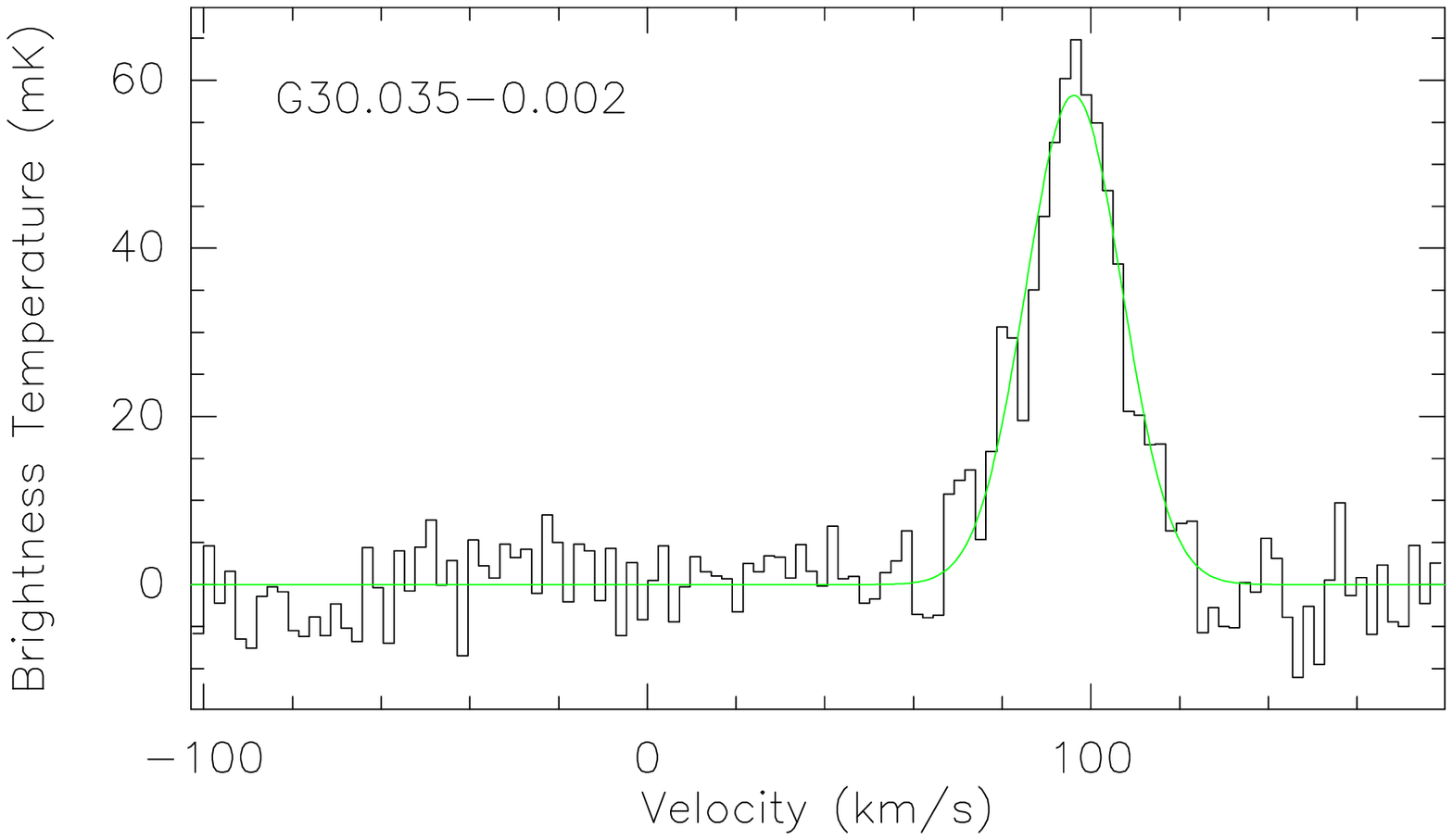}}{ }
\stackunder[5pt]{\includegraphics[width=2.33in]{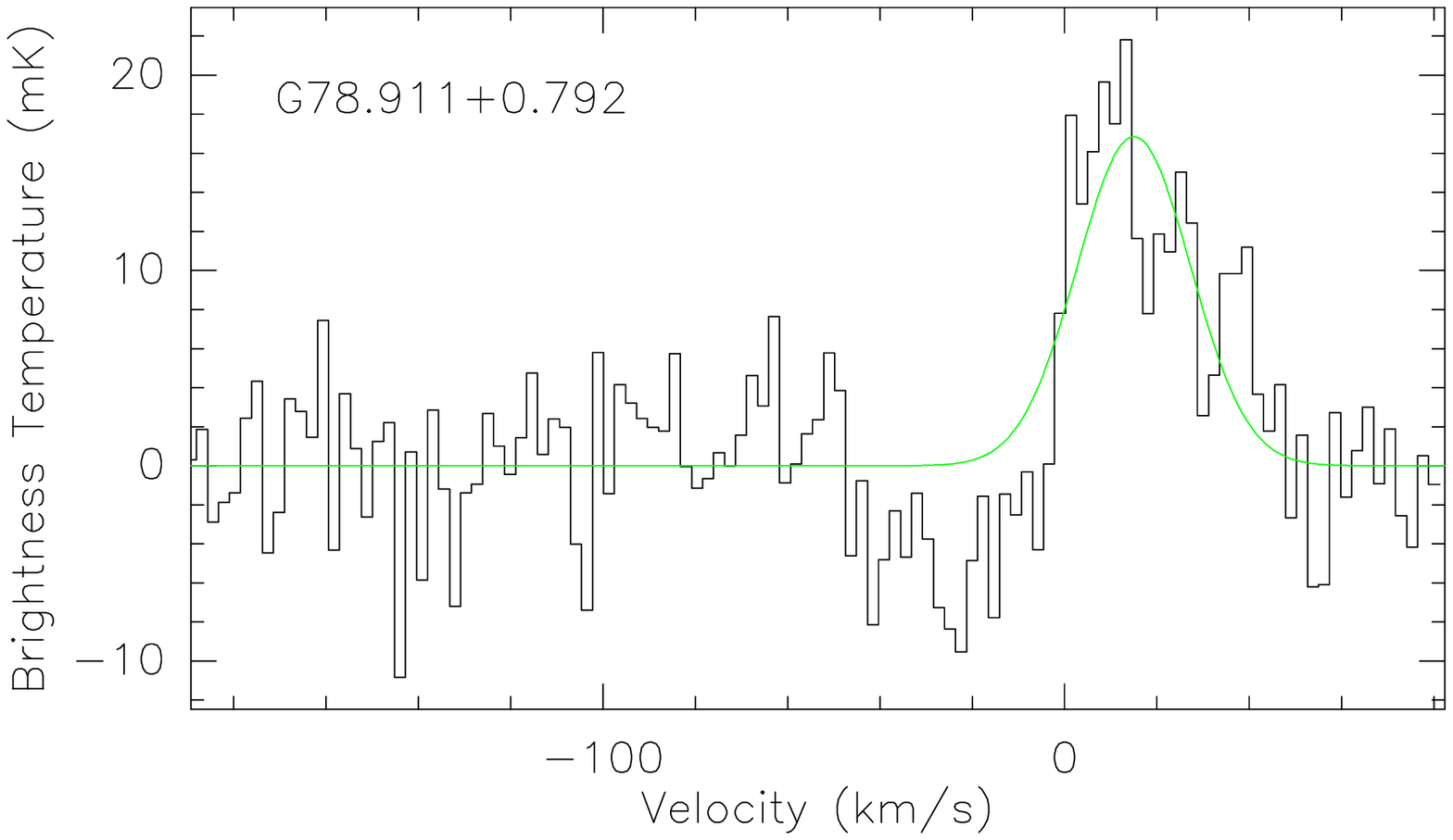}}{ }
\stackunder[5pt]{\includegraphics[width=2.33in]{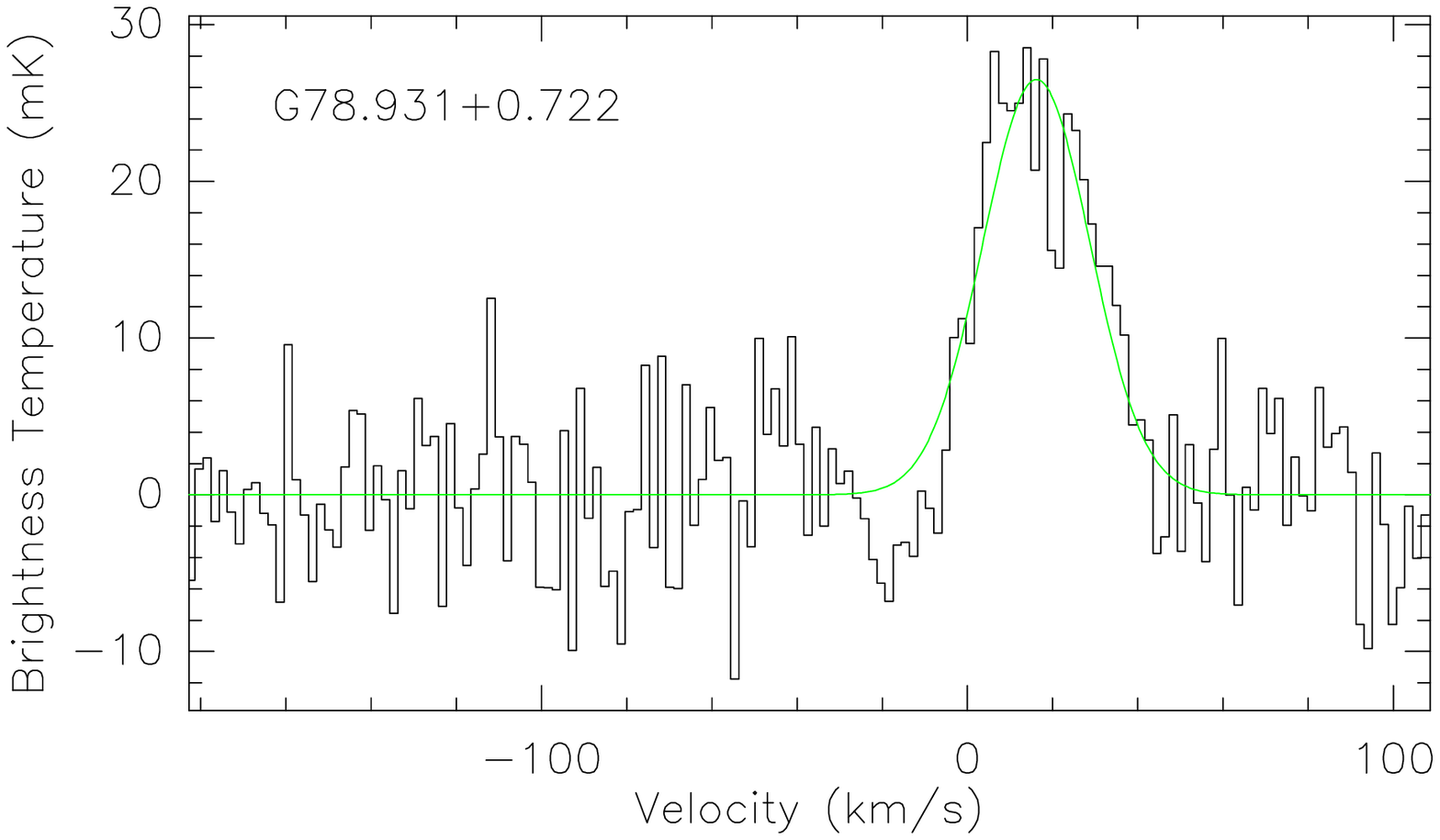}}{ }\\
\stackunder[5pt]{\includegraphics[width=2.33in]{J210700_13_G84_913-3_505.eps}}{ }
\stackunder[5pt]{\includegraphics[width=2.33in]{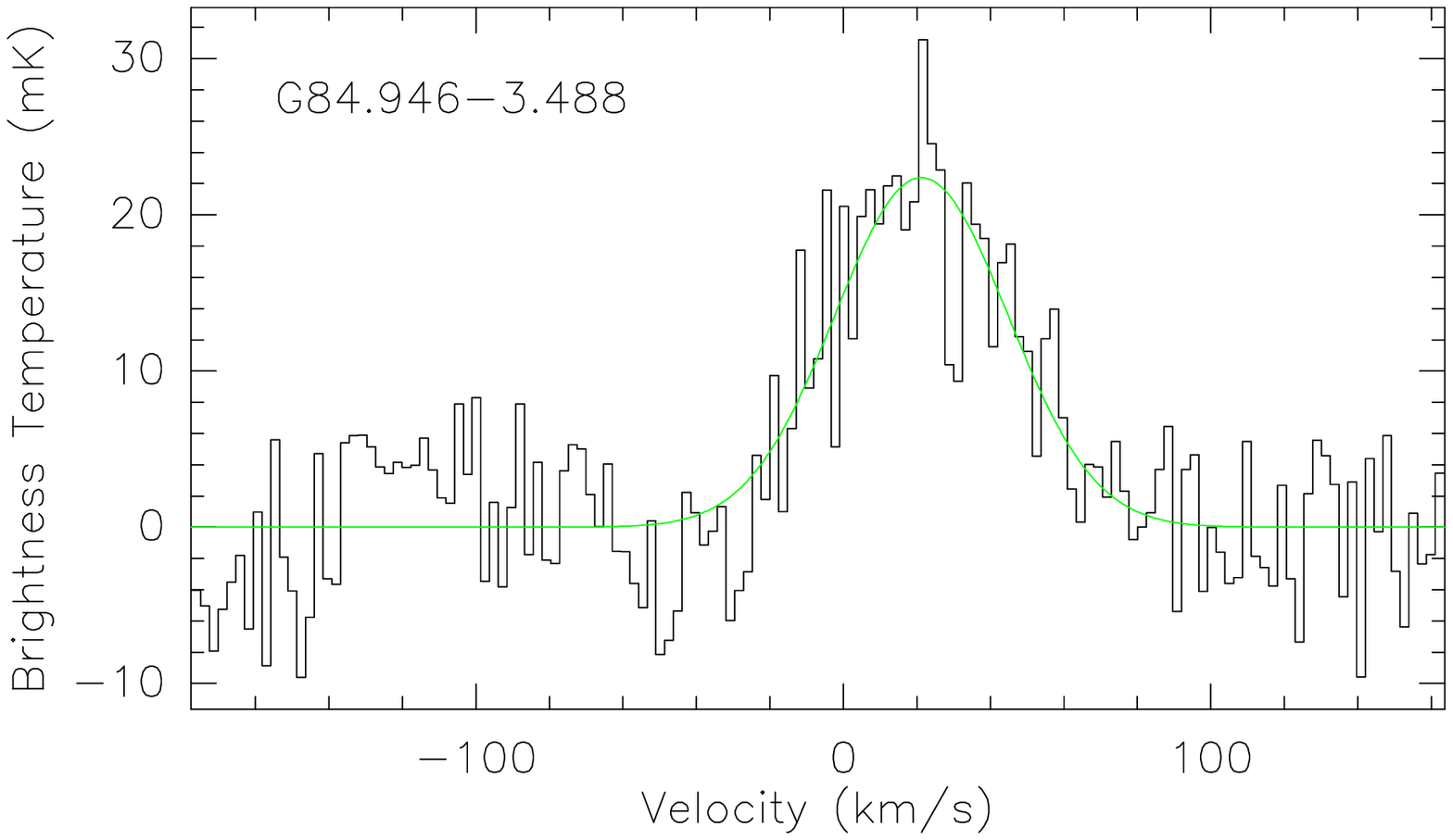}}{ }
\caption{RRL spectra for the candidate PNe sources.\\}
\end{figure*}

\vspace{3cm}

\textbf{Appendix C:} RRL spectra for the potential SNR sources.\\
\begin{figure*}[ht]
\centering
\stackunder[5pt]{\includegraphics[width=2.33in]{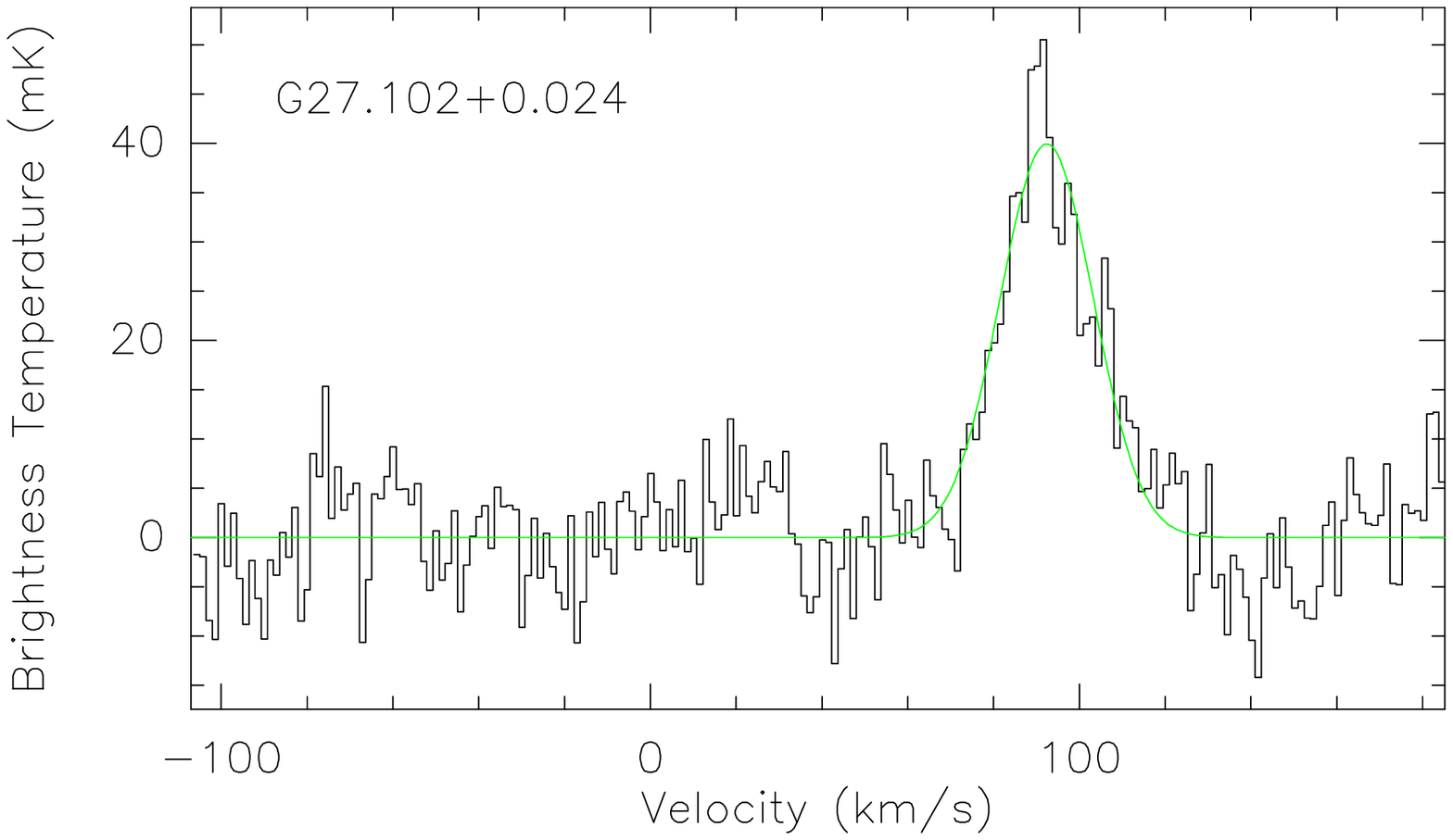}}{ }
\stackunder[5pt]{\includegraphics[width=2.33in]{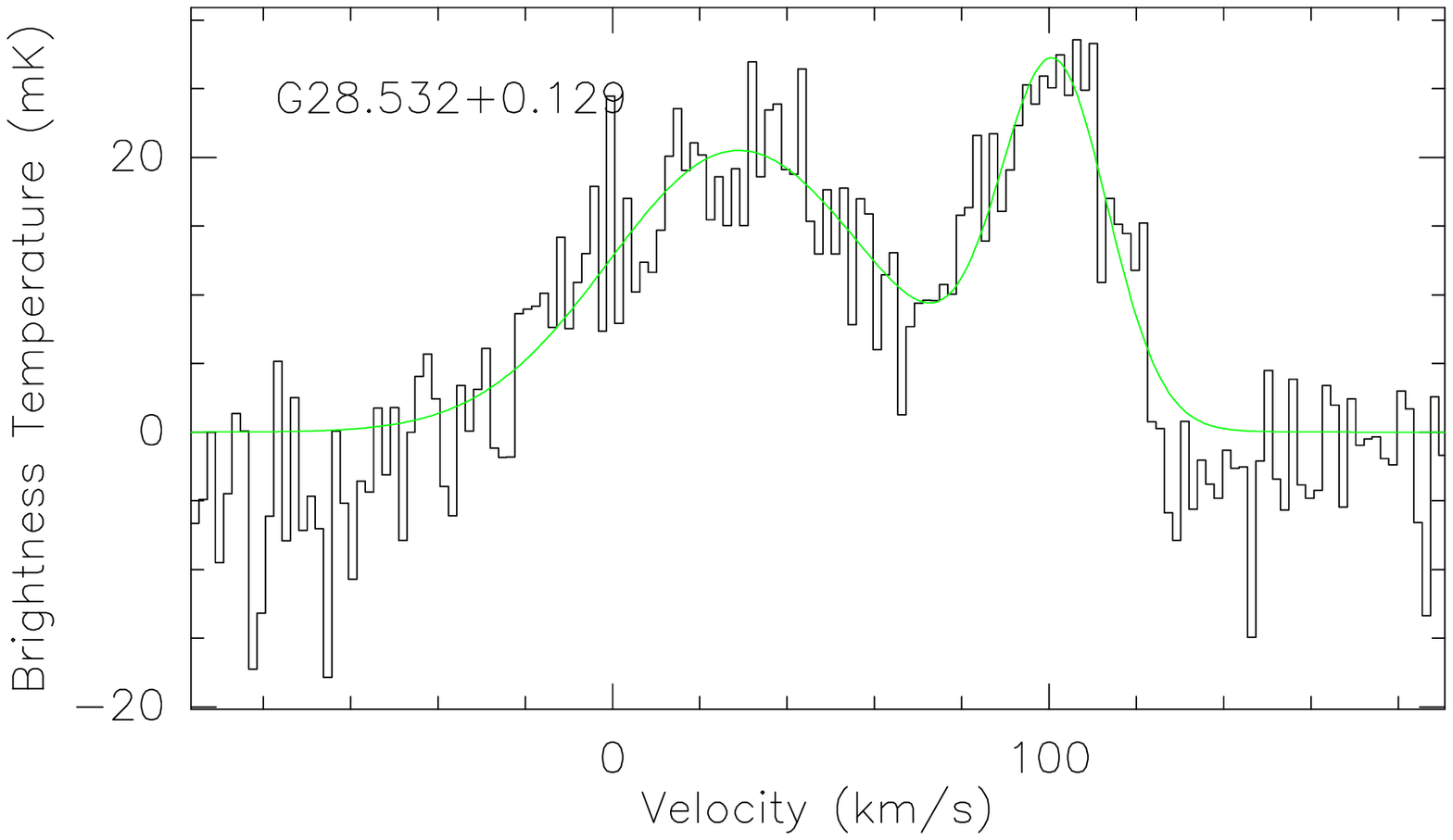}}{ }
\stackunder[5pt]{\includegraphics[width=2.33in]{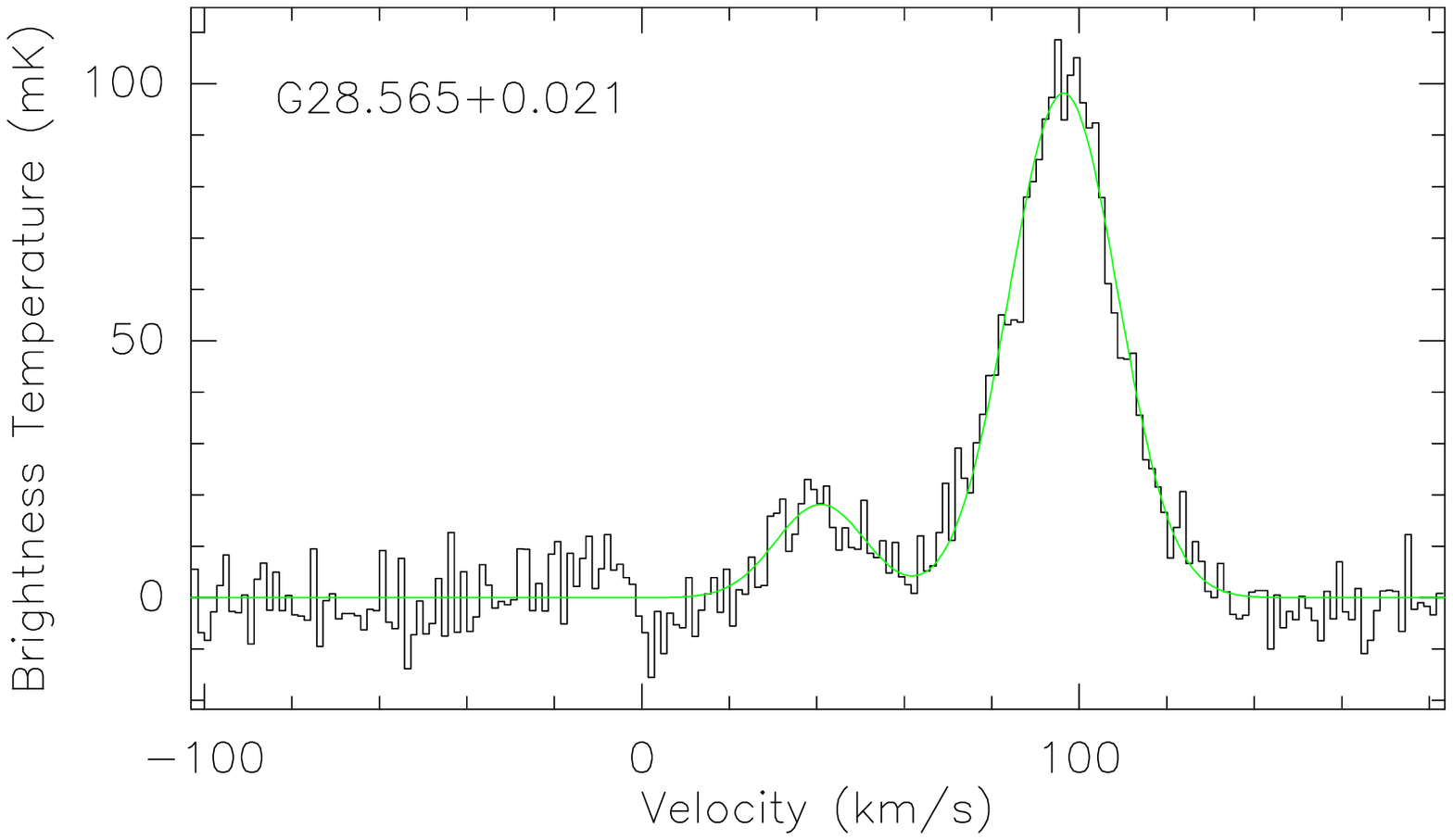}}{ }\\
\stackunder[5pt]{\includegraphics[width=2.33in]{J184702_73_G30_726+0_103.eps}}{ }
\stackunder[5pt]{\includegraphics[width=2.33in]{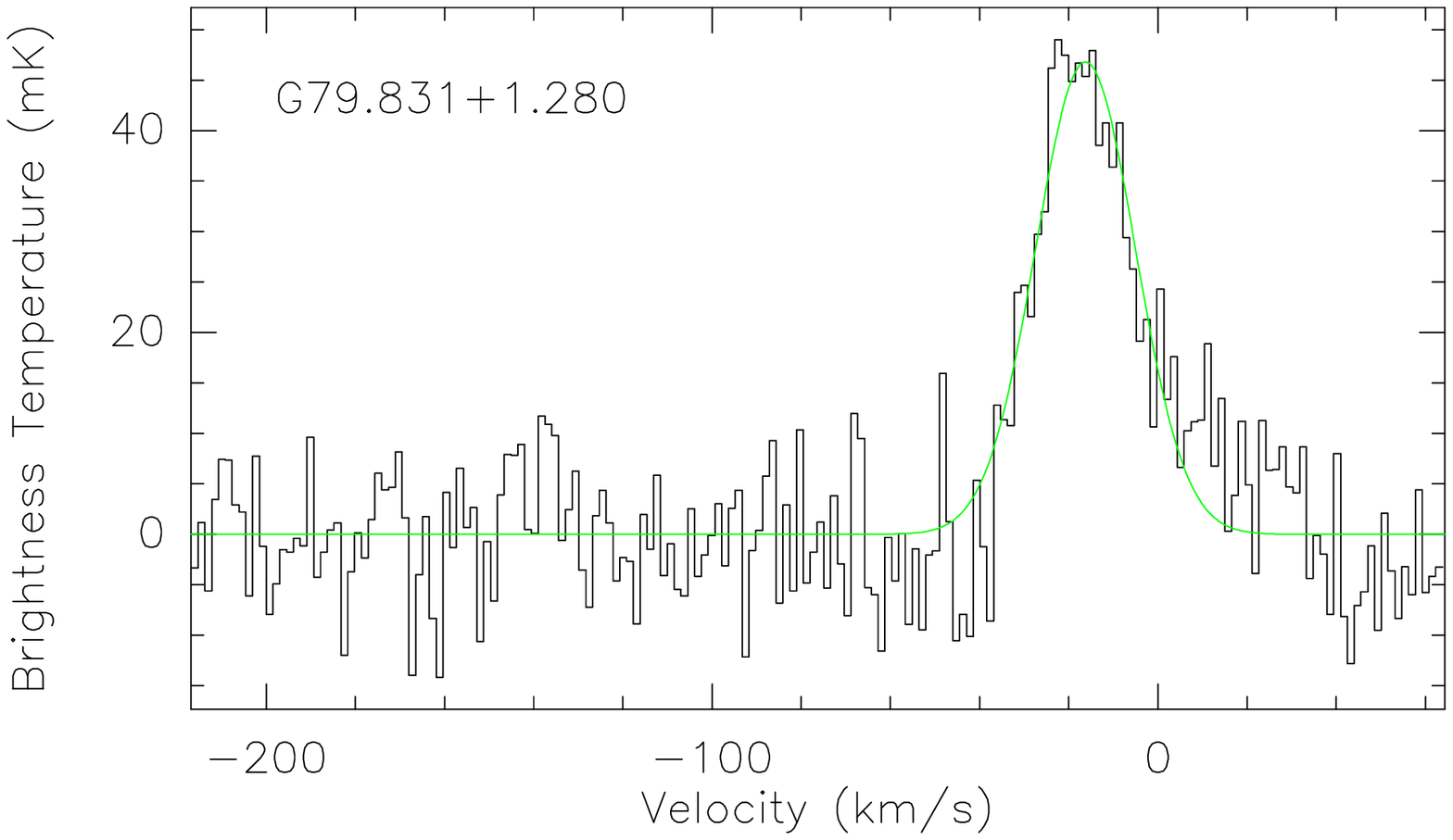}}{ }
\caption{RRL spectra for the candidate SNR sources.}
\end{figure*}

\end{document}